\documentclass[12pt]{article}

\usepackage{hyperref,bookmark} 
\hypersetup{colorlinks=true, allcolors=black} 
\usepackage{amsfonts,amsmath,amsxtra,amsthm}
\usepackage{amssymb}

\usepackage{bm}
\DeclareRobustCommand{\vect}[1]{\bm{#1}}
\usepackage[utf8]{inputenc}
\usepackage[english]{babel}
\usepackage{graphicx}

\usepackage{epsfig,epsf}
\usepackage{units}
\usepackage{amsmath}
\usepackage{amsthm}
\usepackage{latexsym}
\usepackage{amsfonts}
\usepackage{amssymb}
\usepackage{psfrag}
\usepackage{cite}
\usepackage{slashed}
\usepackage{dsfont}
\usepackage{amscd}
\usepackage{xcolor}

\newcommand{\be}{\begin{equation}}
\newcommand{\ee}{\end{equation}}
\newcommand{\bea}{\begin{eqnarray}}
\newcommand{\eea}{\end{eqnarray}}

\def\II{\hbox{{1}\kern-.25em\hbox{l}}}

\def\bg{\boldsymbol{\gamma}}

\def\beq{\begin{equation}}
\def\eeq{\end{equation}}
\def\beqq{\begin{equation*}}
\def\eeqq{\end{equation*}}

\def\bs{\begin{split}}
	\def\es{\end{split}}
\def\bg{\bm{\gamma}}
\def\bl{{\boldsymbol{\lambda}}}
\def\bbg{\underline{\bm{\gamma}}}
\def\bbl{\underline{\boldsymbol{\lambda}}}

\def\bo{\boldsymbol{\omega}}

\def\bx{\boldsymbol{x}}
\def\by{\boldsymbol{y}}

\def\bbx{\underline{\boldsymbol{x}}}
\def\bby{\underline{\boldsymbol{y}}}

\def\ch{\operatorname{ch}}

\def\cconst{\frac{2\pi i}{\o_1\o_2}}

\def\ee{\check{e}}

\def\Im{\operatorname{Im}}
\def\g{\gamma}

\def\j{\hat{j}}

\def\K{{K}}
\def\KK{\hat{K}}
\def\l{\lambda}

\def\mellin{:\stackrel{M}{=}}
\def\mellininverse{:\stackrel{M'}{=}}

\def\o{\omega}

\def\R{\mathbb{R}}
\def\Re{\mathrm{Re}\,}

\def\sh{\operatorname{sh}}

\newcommand{\rf}[1]{(\ref{#1})}

\def\LLL{\hat{\Lambda}}

\def\QQ{\hat{Q}}

\def\KKK{K}

\def\beq{\begin{equation}}
	\def\eeq{\end{equation}}
\def\beqq{\begin{equation*}}
	\def\eeqq{\end{equation*}}

\begin{document}

\vspace*{0.7cm}

\begin{center}
{\bf \large Baxter $Q$-operators in Ruijsenaars-Sutherland \\[3pt]hyperbolic systems: 
one- and two-particle cases}
\bigskip

{\bf N. Belousov$^{\dagger}$, S. Derkachov$^{\dagger}$, S. Kharchev$^{\times\ast}$, S. Khoroshkin$^{\circ\ast}$
}\medskip\\
$^\dagger${\it Steklov Mathematical Institute, Fontanka 27, St. Petersburg, 191023, Russia;}\smallskip\\
$^\times${\it National Research Center  “Kurchatov Institute”, Kurchatov Square 1, Moscow, 123182, Russia;}\smallskip\\
$^\circ${\it National Research University Higher School of Economics, Myasnitskaya 20, Moscow, 101000, Russia;}\smallskip\\
$^\ast${\it Institute for Information Transmission Problems RAS (Kharkevich Institute), Bolshoy Karetny per. 19, Moscow, 127994, Russia}
\end{center}

\setcounter{footnote}{0}
\setcounter{equation}{0}

\begin{abstract} \noindent
In these notes we review the technique of Baxter $Q$-operators in the Ruijsenaars-Sutherland hyperbolic systems in the cases of one and two particles. Using these operators we show in particular that eigenfunctions of these systems admit two dual integral representations and prove their orthogonality and completeness.
\end{abstract}

 \tableofcontents


\vspace{1cm}
\section{Introduction}

\paragraph{1.} In recent years, there has been a significant progress in the study of hyperbolic Ruijsenaars-Sutherland quantum integrable models. The study of these hyperbolic systems has gone along a path different from that used in the compact trigonometric versions of the corresponding models, where the main role is played by the theory of Macdonald polynomials and of the double affine Hecke  algebras by I. Cherednick. Instead, the cornerstone of constructions in hyperbolic systems is the so-called \textit{kernel function} by S. Ruijsenaars, first introduced and studied in \cite{R2}. The kernel function is factorized into a product of two-point functions and can be regarded as a counterpart of the reproducing kernel in the theory of Macdonald polynomials \cite{NS1,NS2}. With its use M. Halln\"as and S. Ruijsenaars constructed integral representations for the wave functions of the hyperbolic Ruijsenaars-Sutherland models \cite{HR1,HR2}.

In our recent papers \cite{BDKK1, BDKK2} we considered the Ruijsenaars system and used the kernel function to construct the commuting family of integral operators called \textit{Baxter $Q$-operators}. With these operators at hand we obtained another (dual) integral representation of the wave function, given by integrals over spectral parameters. We also proved that the wave function is symmetric with respect to both space and spectral variables and solves a bispectral problem, that is it is also an eigenfunction of difference operators acting on spectral variables. As it is shown in \cite{BDKK2}, all these properties stem from the commutativity of $Q$-operators, which was proven in \cite{BDKK1}. In a certain limit the Ruijsenaars model degenerates to the Sutherland model, and the whole construction can be repeated for it, as we will show in our future work.

The goal of these notes is to demonstrate basic ideas of $Q$-operator's technique in hyperbolic Ruijsenaars-Sutherland models in the cases of one and two particles, where many subtle points can be already visualized.

\paragraph{2.} Let us first demonstrate how $Q$-operators appear in these models.
Consider the hyperbolic Sutherland model with two particles. The Hamiltonian is given by
\begin{equation*}
H_{S} = - \partial_{x_1}^2 - \partial_{x_2}^2 + \frac{2g (g - 1)}{\sinh^2(x_1 - x_2)}.
\end{equation*}
Performing the similarity transformation
\begin{equation*}
H = \sinh^g|x_1 - x_2| \cdot H_{S} \cdot \frac{1}{\sinh^{g} |x_1 - x_2|}
\end{equation*}
we arrive at the operator
\begin{equation}\label{H}
H = - \partial_{x_1}^2 - \partial_{x_2}^2 - 2 g \coth(x_1 - x_2) ( \partial_{x_1} - \partial_{x_2} ) - 2g^2.
\end{equation}
It commutes with the ``full momentum'' operator
\begin{equation*}
P = - i \partial_{x_1} - i \partial_{x_2}.
\end{equation*}
With its use one can reduce the spectral problem for the operator $\rf{H}$ to the solution of ordinary differential equation, which coincides with classical hypergeometric equation over independent variable $t=e^{2(x_1-x_2)}$. Its solution can be represented by means of the Barnes type integral
\begin{equation}\label{Psi-1}
\begin{aligned}
\Psi_{\l_1, \l_2}(x_1,x_2) &= \frac{2^{2g - 3}}{ \pi \Gamma^2(g)} \int_\R d\gamma \, \Gamma \Bigl( \frac{i \l_1 - i \g + g}{2} \Bigr) \, \Gamma \Bigl( \frac{i \g - i \l_1 + g}{2} \Bigr) \\[3pt]
& \times \Gamma \Bigl( \frac{i \l_2 - i \g + g}{2} \Bigr) \, \Gamma \Bigl( \frac{i \g - i \l_2 + g}{2} \Bigr) \, e^{i (\l_1 + \l_2 - \g)x_2} \, e^{i \g x_1}
\end{aligned}
\end{equation}
so that
\begin{align*}
P \, \Psi_{\l_1, \l_2}(x_1,x_2) &= ( \l_1 + \l_2 ) \, \Psi_{\l_1, \l_2}(x_1,x_2),  \\[5pt]
H \, \Psi_{\l_1, \l_2}(x_1,x_2) &= ( \l_1^2 + \l_2^2 ) \, \Psi_{\l_1, \l_2}(x_1,x_2).
\end{align*}
The normalization constant behind the integral is chosen for the latter convenience. The solution of hypergeometric equation also admits Euler beta integral representation. To find it apply the Mellin transform to the integral \rf{Psi-1} (see Section \ref{sec:mell}), then \rf{Psi-1} takes the following form
\begin{equation}\label{Psi-2}
\Psi_{\l_1, \l_2}(x_1,x_2) = \int_\R dy \, \ch^{-g}(x_1 - y) \, \ch^{-g}(x_2 - y) \, e^{i \l_2 (x_1 + x_2 - y)} \, e^{i \l_1 y }.
\end{equation}
Clearly, this integral representation is symmetric with respect to $x_j$, and the first one \eqref{Psi-1} is symmetric with respect to $\lambda_j$
\begin{equation*}
\Psi_{\l_1, \l_2}(x_1,x_2) = \Psi_{\l_1, \l_2}(x_2,x_1)= 	\Psi_{\l_2, \l_1}(x_1,x_2).
\end{equation*}
Besides, one can check using the first representation \rf{Psi-1} that $\Psi_{\l_1, \l_2}(x_1,x_2)$ as a function of $\l_1,\l_2$ solves another (dual) spectral problem
\begin{align*}
\mathcal{P} \, \Psi_{\l_1, \l_2}(x_1,x_2) &= e^{2x_1 + 2x_2} \, \Psi_{\l_1, \l_2}(x_1,x_2), \\[6pt]
\mathcal{H} \, \Psi_{\l_1, \l_2}(x_1,x_2) &= \bigl( e^{2x_1} + e^{2x_2} \bigr) \, \Psi_{\l_1, \l_2}(x_1,x_2)
\end{align*}
with operators
\begin{equation}\label{dual-H}
\begin{aligned}
\mathcal{P} &= e^{-2i \partial_{\l_1} - 2i \partial_{\l_2}}, \\[6pt]
\mathcal{H} &= \frac{\l_1 - \l_2 + 2i(g - 1)}{\l_2 - \l_1} e^{-2i \partial_{\l_1}} + \frac{\l_2 - \l_1 + 2i(g - 1)}{\l_1 - \l_2} e^{-2i \partial_{\l_2}}.
\end{aligned}
\end{equation}

Now consider the asymptotic behavior of the eigenfunction $\Psi_{\l_1, \l_2}(x_1,x_2)$
as $x_2-x_1\to\infty$. Due to the structure of the Hamiltonian \rf{H} it should be given by a combination of plane waves. Its precise form can be obtained from the integral representation \eqref{Psi-1} by residue evaluation in the lower half-plane. The leading behavior of the residue series in the limit $x_2 - x_1 \rightarrow \infty$ is given by the closest to integration contour poles
\begin{equation}\label{Psi-as}
\begin{aligned}
&\Psi_{\l_1, \l_2}(x_1,x_2) = 2^{2g - 1} \Gamma^{-1}(g) \, e^{-g (x_2 - x_1)} \\[6pt]
&\times\Biggl[  \Gamma \Bigl( \frac{i \l_2 - i \l_1}{2} \Bigr) \, \Gamma \Bigl( \frac{i \l_1 - i \l_2}{2} + g \Bigr) \, e^{i \l_1 x_1 + i\l_2 x_2} \\[6pt]
& +  \Gamma \Bigl( \frac{i \l_1 - i \l_2}{2} \Bigr) \, \Gamma \Bigl( \frac{i \l_2 - i \l_1}{2} + g \Bigr) \, e^{i \l_2 x_1 + i\l_1 x_2} \,  \Biggr] + O\bigl(e^{-(g+2)(x_2-x_1)}\bigr).
\end{aligned}
\end{equation}
In order to separate the single plane wave in the asymptotics
we add an imaginary part $-i\varepsilon$ with $\varepsilon > 0$ to the variable $\l_2$ and consider the asymptotic behavior as $x_2 \rightarrow \infty$
\begin{equation}\label{Psi-as2}
\begin{aligned}
\Psi_{\l_1, \l_2 - i \varepsilon}(x_1,x_2) &= 2^{2g - 1} \Gamma^{-1}(g) \, \Gamma \Bigl( \frac{i \l_2 - i \l_1 + \varepsilon}{2} \Bigr) \, \Gamma \Bigl( \frac{i \l_1 - i \l_2 - \varepsilon}{2} + g \Bigr) \\[6pt]
&\times e^{i (\l_1 - i g) x_1 + i(\l_2 -i \varepsilon + ig) x_2}  + O\bigl(e^{-gx_2}\bigr).
\end{aligned}
\end{equation}
On the other hand, the same asymptotic of the integral \eqref{Psi-2} is given by
\begin{equation}\label{Psi-as3}
2^{g} \, e^{g x_1 + i(\l_2 - i \varepsilon + i g)x_2} \int_\R dy \, \ch^{-g}(x_1 - y) \, e^{i (\l_2 - i \varepsilon + ig)(x_1 - y)} \, e^{i \l_1 y}
\end{equation}
since
\begin{equation*}
\ch^{-g}(x_2-y) \sim 2^{g} \, e^{-g(x_2 - y)}, \qquad x_2\to\infty.
\end{equation*}
The comparison of \rf{Psi-as2} and \rf{Psi-as3} yields the integral identity
\begin{multline} \label{kh1}
\int_\R dy \, \ch^{-g}(x_1 - y)\, e^{i \l'_2 (x_1 - y)}e^{i\l_1 y} \\[6pt]
= 2^{g - 1}\Gamma^{-1}(g) \, \Gamma \Bigl( \frac{i \l'_2 - i \l_1 + g}{2} \Bigr) \, \Gamma \Bigl( \frac{i \l_1 - i \l'_2 +g}{2}  \Bigr)e^{i\l_1 x_1}
\end{multline}
with $\l_2' = \l_2 + i g - i \varepsilon$. It is equivalent to the classical beta integral.

The integrals \rf{Psi-2} and \rf{kh1} have similar structure. We can interpret both of them in terms of integral operators. The relation \rf{Psi-2} says that the two-particle wave function $\Psi_{\l_1,\l_2}(x_1,x_2)$ is obtained from one-particle function $\Psi_{\l_1}(x_1)=e^{i\l_1x_1}$ by means of the integral operator
\begin{equation*}
\bigl[ \Lambda(\l) f \bigr] (x_1, x_2) = \int_\R dy \, \ch^{-g}(x_1 - y) \, \ch^{-g}(x_2 - y) \, e^{i \l (x_1 + x_2 - y)} \, f(y),
\end{equation*}
which we call \textit{raising operator}. The relation \rf{kh1} says that the one-particle wave function $\Psi_{\l_1}(x_1)=e^{i\l_1x_1}$ is an eigenfunction of the integral operator
\begin{equation*}
\bigl[ Q(\l) f \bigr] (x_1) = \int_\R dy \, \ch^{-g}(x_1 - y)\, e^{i \l (x_1 - y)} \, f(y),
\end{equation*}
with the eigenvalue
\begin{equation}\label{kh2}
q(\l, \l_1) = 2^{g - 1}\Gamma^{-1}(g) \, \Gamma \Bigl( \frac{i\l - i\l_1 + g}{2} \Bigr) \, \Gamma \Bigl( \frac{i\l_1 - i\l + g}{2} \Bigr),
\end{equation}
which we call \textit{Baxter $Q$-operator}.

Analogously, the integral \rf{Psi-1} says that the two-particle wave function $\Psi_{\l_1,\l_2}(x_1,x_2)$ is obtained from one-particle function $\Psi_{\l_1}(x_1)=e^{i\l_1x_1}$ by means of the \textit{dual} raising integral operator
\begin{equation}\label{L2-dual}
\begin{aligned}
\bigl[ \hat{\Lambda}(x)  f \bigr] (\l_1, \l_2) &= \frac{2^{2g - 3}}{ \pi \Gamma^2(g)} \int_\R d\g \, \Gamma \Bigl( \frac{i\l_1 - i \g + g}{2} \Bigr) \, \Gamma \Bigl( \frac{i\g - i\l_1 + g}{2} \Bigr) \\[6pt]
&\times \Gamma \Bigl( \frac{i\l_2 - i \g + g}{2} \Bigr) \, \Gamma \Bigl( \frac{i\g - i\l_2 + g}{2} \Bigr) \, e^{i (\l_1 + \l_2 - \g)x_2} \, f(\g).
\end{aligned}
\end{equation}
Due to the structure of the dual Hamiltonian \eqref{dual-H} we can similarly consider the asymptotic behavior of the wave function as $\l_2\to \infty$ (with the help of Stirling formula for the gamma function). Then it gives the dual $Q$-operator
\begin{equation}\label{Q2-dual}
\bigl[ \hat{Q}(x)  f \bigr] (\l_1) = \frac{2^{g - 2}}{ \pi \Gamma(g)} \int_\R d\g \, \Gamma \Bigl( \frac{i\l_1 - i \g + g}{2} \Bigr) \, \Gamma \Bigl( \frac{i\g - i\l_1 + g}{2} \Bigr) \,  e^{i (\l_1 - \g)x_2} \, f(\g).
\end{equation}
Again its eigenfunctions are one-particle wave functions $\Psi_{\l_1}(x_1)=e^{i\l_1x_1}$ with the eigenvalue
\begin{equation}\label{kh3}
\hat{q}(x,x_1) = \ch^{-g}(x-x_1).
\end{equation}
The last statement amounts to the Fourier transform of the identity \eqref{kh1}.

Remarkably, these observations generalize to the $n$-particle case. In the work \cite{HR2} Halln\"as and Ruijsenaars showed that the $n$-particle eigenfunction of the Hamiltonian
\begin{equation}\label{H-n}
H_n = -\sum_{j = 1}^n \partial_{x_j}^2 - 2 g \sum_{\substack{j,k=1\\j<k}}^n \coth(x_j - x_k) (\partial_{x_j}-\partial_{x_k})
\end{equation}
can be obtained from the $(n-1)$-particle function
\begin{equation*}
\Psi_{\l_1, \dots, \l_n}(x_1, \dots, x_n) = \Lambda_n(\l_n) \, \Psi_{\l_1, \dots, \l_{n - 1}}(x_1, \dots, x_{n - 1})
\end{equation*}
using the integral operator
\begin{equation*}
\begin{aligned}
\bigl[ \Lambda_n(\l) f \bigr] (x_1, \dots,& x_n) = \int_{\R^{n - 1}} dy_1 \dots dy_{n - 1} \, \prod_{\substack{j,k=1 \\ j < k}}^{n - 1} \sh^{2g}|y_j - y_k| \\[6pt]
&\times e^{i \l \bigl( \sum\limits_{j = 1}^n x_j - \sum\limits_{j = 1}^{n - 1}y_j\bigr)}  \prod_{j = 1}^n \prod_{k = 1}^{n - 1 } \ch^{-g}(x_j - y_k) \,  f(y_1, \dots, y_{n - 1}).
\end{aligned}
\end{equation*}
As before, considering its asymptotic behavior as $x_n \rightarrow \infty$ we arrive at the integral operator
\begin{equation*}
\begin{aligned}
\bigl[ Q_{n - 1}(\l) f \bigr] (x_1, \dots, &x_{n - 1}) = \int_{\R^{n-1}} dy_1 \dots dy_{n-1} \, \prod_{\substack{j,k=1 \\ j < k}}^{n-1} \sh^{2g}|y_j - y_k| \\[6pt]
&\times e^{i \l \bigl( \sum\limits_{j = 1}^{n-1} x_j - \sum\limits_{j = 1}^{n-1}y_j\bigr)} \prod_{j,k = 1}^{n-1} \ch^{-g}(x_j - y_k) \,  f(y_1, \dots, y_{n-1}).
\end{aligned}
\end{equation*}
Due to the structure of the Hamiltonian \eqref{H-n} the $(n - 1)$-particle wave functions should be its eigenfunctions
\begin{multline}\label{Q-Psi}
Q_{n - 1}(\l) \, \Psi_{\l_1, \dots, \l_{n-1}}(x_1, \dots, x_{n-1}) \\
= q(\l, \l_1, \dots, \l_{n - 1}) \, \Psi_{\l_1, \dots, \l_{n-1}}(x_1, \dots, x_{n-1}).
\end{multline}
The same holds for the dual integral representation \cite{KK1} and dual operators $\hat{\Lambda}_n$, $\hat{Q}_{n - 1}$ generalizing \eqref{L2-dual}, \eqref{Q2-dual}. The equivalence of two integral representations in the case of $n$ particles is a nontrivial fact, but assuming it and comparing their asymptotics we notice that $Q$-operator's eigenvalue should be factorized into functions \eqref{kh2}
\begin{equation}\label{q-fact}
q(\l, \l_1, \dots, \l_{n - 1}) = (n-1)! \prod_{j = 1}^{n - 1} q(\l, \l_j).
\end{equation}
This can be also guessed from the general principle of $S$-matrix factorization in the theory of integrable models.

The proof of the $Q$-operator diagonalization \eqref{Q-Psi} in the case of $n$-particles Sutherland model will be given in our future work. In these notes we trace the path for the first nontrivial case --- the operator $Q_2(\l)$. The key ingredient of the proof is the commutativity of $Q$-operators
\begin{equation*}
Q_n(\l) \, Q_n(\rho) = Q_n(\rho) \, Q_n(\l).
\end{equation*}
In the certain limit this identity degenerates to exchange relation between $Q$ and $\Lambda$ operators, which in turn implies the statement \eqref{Q-Psi} together with factorization \eqref{q-fact}.

One more remarkable feature of the above calculation is the form of the eigenvalues \rf{kh2} and \rf{kh3} of operators $Q(\l)$ and $\hat{Q}(x)$: the eigenvalue \rf{kh2} is the constituent of the kernel of dual $Q$-operator $\hat{Q}(x)$, and  the eigenvalue \rf{kh3} is the constituent of the kernel of ${Q}(\l)$. Using this property one can write down one more integral representation for the wave function
\begin{equation*}
\Psi_{\l_1, \dots, \l_n}(x_1, \dots, x_n) = e^{i \lambda_n x_n} \, Q_{n - 1}(\l_n) \, \hat{Q}_{n - 1}(x_n) \, \Psi_{\l_1, \dots, \l_{n - 1}}(x_1, \dots, x_{n - 1})
\end{equation*}
and prove the equality of two mentioned integral representations found in \cite{HR2,KK1}.

\paragraph{3.} The described program is implemented for the Ruijsenaars hyperbolic system in \cite{BDKK2}. This system is governed by commuting difference operators
\begin{equation}\label{Macd}
M_k = \sum_{\substack{A\subset \{1, \dots, n\} \\ |A|=k}} \,
\prod_{\substack{a\in A \\ b\notin A}}
\frac{\sh\frac{\pi}{\o_1}\left(x_a-x_b-i g\right)}
{\sh\frac{\pi}{\o_1}\left(x_a-x_b\right)}
\prod_{a \in A} e^{-i \o_2 \partial_{x_a}}, \qquad k = 1, \dots, n.
\end{equation}
Here and in what follows we assume that $g, \omega_1, \omega_2$ are positive constants, such that
\begin{equation*}
0 < g < \omega_1 + \omega_2.
\end{equation*}
In \cite{BDKK2} a more general case of complex constants is considered.
The Sutherland system can be obtained after scaling $g \rightarrow g \omega_2$ in the limit $\omega_2 \rightarrow 0$.

Here is a short list of some results from \cite{HR1,BDKK1,BDKK2}. The building blocks of the construction are the measure function $\mu_g(\bx_n)$ and the kernel function $K_g(\bx_n,\by_m)$, which are defined for tuples
$$\bx_n=(x_1,\ldots,x_n),\qquad \by_m=(y_1,\ldots, y_m)$$
via products
\begin{align*}
\mu_g(\bx_n)=\prod_{\substack{i,j=1 \\ i\not=j}}^n\mu_g(x_i-x_j), \qquad K_g(\bx_n,\by_m)=\prod_{i=1}^n \prod_{j = 1}^m K_g(x_i-y_j)
\end{align*}
where
\begin{equation}
\begin{aligned}\label{mera1}
\mu_g(x)&=S_2(i x|\bo) S_2(g-i x|\bo), \\[6pt] 
K_g(x)&=S_2^{-1}\Bigl( \frac{g}{2}+i x\Big|\bo\Bigr)S_2^{-1}\Bigl(\frac{g}{2}-i x\Big|\bo\Bigr).
\end{aligned}
\end{equation}
Here $S_2(z|\bo)$ is the double sine function, see Appendix \ref{AppA}. Denote also
\begin{equation*}
g^\ast=\o_1+\o_2-g.
\end{equation*}
In these notations the Baxter operator is the integral operator
\begin{equation*}
\bigl[ Q_{n}(\lambda | \bo) f\bigr] (\bm{x}_n) =  d_n(g) \, \int_{\mathbb{R}^n} d\bm{y}_n \, Q(\bm{x}_n, \bm{y}_n; \lambda) f(\bm{y}_n)
\end{equation*}
with the kernel
\begin{equation*}
Q(\bx_n,\by_{n};\l)= e^{\cconst \l(\bbx_n-\bby_n)}
\, \K_{g^*}(\bx_n,\by_{n}) \,\mu_{g^*}(\by_{n})
\end{equation*}
and normalizing constant
\begin{equation*}
d_n(g) = \frac{1}{n!} \left[ \sqrt{\omega_1 \omega_2} S_2(g|\bo) \right]^{-n}.
\end{equation*}
Here and below for a tuple $\bx_n=(x_1,\ldots,x_n)$ we use the notation $\bbx_n$ for the sum of components
\begin{equation*}
\bbx_n=x_1+\ldots+x_n.
\end{equation*}
The raising operator $\Lambda_{n}(\l)$ is a similar integral operator
\begin{equation*}
\bigl[\Lambda_{n}(\l | \bo)f\bigr](\bx_n)=d_{n - 1}(g) \int_{\R^{n-1}}d\by_{n-1} \, \Lambda(\bx_n,\by_{n-1};\l) f(\by_{n-1})
\end{equation*}
with the kernel
\begin{equation*}
\Lambda(\bx_n,\by_{n-1};\l)= e^{\cconst \l(\bbx_n-\bby_{n-1})}
\, \K_{g^*}(\bx_n,\by_{n-1}) \, \mu_{g^*} (\by_{n-1}).
\end{equation*}
It was proved in \cite{HR1} that the function
\begin{equation}\label{I12}
\Psi_{\bl_n}(\bx_n | \bo) = \Lambda_{n}(\l_n | \bo) \, \Lambda_{n-1}(\l_{n-1} | \bo)\cdots \Lambda_{2}(\l_2 | \bo) \, e^{\cconst \l_1x_1}
\end{equation}
is a joint eigenfunction of the operators \rf{Macd}
\begin{equation*}
M_k\,\Psi_{\bl_n}(\bx_n | \bo)= e_k \bigl(e^{\frac{2\pi \lambda_1}{\o_1}}, \dots, e^{\frac{2\pi \lambda_n}{\o_1}}\bigr)  \Psi_{\bl_n}(\bx_n | \bo), \qquad k = 1, \dots, n
\end{equation*}
under the condition $g < \omega_1$. Here $e_k$ are the elementary symmetric functions. Furthermore, in \cite {BDKK2} we showed that it is also an eigenfunction of the Baxter operators $Q_n(\l | \bo)$
\begin{equation*}
Q_{n}(\lambda | \bo) \, \Psi_{ \bm{\lambda}_n }(\bm{x}_n | \bo) = \prod_{j = 1}^n K_g(\lambda-\lambda_j) \, \Psi_{ \bm{\lambda}_n }(\bm{x}_n | \bo).
\end{equation*}
In a similar manner dual Baxter and raising operators are integral operators
\begin{equation*}
\begin{aligned}
\bigl[ \QQ_n(x | \bo) f\bigr] (\bm{\l}_n)& = d_n({g}^*)\, \int_{\mathbb{R}^n} d\bm{\gamma}_n \, \QQ(\bm{\l}_n, \bm{\gamma}_n; x) f(\bm{\gamma}_n),\\[7pt]
\bigl[\LLL_{n}(x | \bo)f\bigr](\bl_n)&=d_{n - 1}({g}^*) \int_{\R^{n-1}}d\bg_{n-1} \, \LLL(\bl_n,\bg_{n-1};x) f(\bg_{n-1})
\end{aligned}
\end{equation*}
with the kernels
\begin{equation*}
\begin{split} 
\QQ(\bl_n,\bg_{n};x)&= e^{\cconst x (\bbl_n-\bbg_n)}	 \, K_g(\bl_n,\bg_{n}) \,\mu_g(\bg_{n}),\\[7pt]
\LLL(\bl_n,\bg_{n-1};x)&= e^{\cconst x (\bbl_n-\bbg_{n-1} )} K_g(\bl_n,\bg_{n-1}) \, \mu_g (\bg_{n-1}).
\end{split}
\end{equation*}
The duality property established in \cite{BDKK2} implies that the wave function admits along with~\rf{I12} another integral representation
\begin{equation*}
\Psi_{\bl_n}(\bx_n | \bo)=\LLL_{n}(x_n | \bo) \, \LLL_{n-1}(x_{n-1} | \bo)\cdots \LLL_{2}(x_2 | \bo) \, e^{\cconst \l_1x_1},
\end{equation*}
so that it solves the spectral problem for dual Macdonald operators as well, and it is also an eigenfunction of dual Baxter operators
\begin{equation*}
\QQ_{n}(x| \bo) \, \Psi_{ \bm{\lambda}_n }(\bm{x}_n | \bo) = \prod_{j = 1}^n K_{g^*}(x-x_j) \, \Psi_{ \bm{\lambda}_n }(\bm{x}_n | \bo). 
\end{equation*}
We remark that operators defined here are different from the ones in \cite{BDKK1, BDKK2} by rescaling of spectral parameters $\lambda_j \rightarrow \lambda_j/ \o_1 \o_2$. Such rescaling simplifies formulas in the case of real constants $\omega_i$, and oppositely complicates matters in the case of complex $\omega_i$ considered in \cite{BDKK1, BDKK2}.

After rescaling $g \rightarrow g \omega_2$ in the limit $\omega_2 \rightarrow 0$ the Ruijsenaars system degenerates to the Sutherland system so that the measure and kernel functions \eqref{mera1} turn into
\begin{align*}
&\mu(x)=\sh^g |x|,&& \hat{\mu}(\l)= 2^{1 - g} \Gamma(g) \,\Gamma^{-1}\Bigl(\frac{i\lambda}{2}+g \Bigr)\Gamma^{-1}\Bigl(-\frac{i\lambda}{2} \Bigr),\\[6pt]
&K(x)=\ch^{-g}(x),&&
\KK(\l)=2^{g - 1} \Gamma^{-1}(g) \, \Gamma\Bigl(\frac{g+i\lambda}{2} \Bigr)\Gamma\Bigl(\frac{g-i\lambda}{2} \Bigr),
\end{align*}
see Section \ref{sec:reduc}.
With these functions Baxter and raising operators and their duals are defined by the same formulas. In the future work we will show that all statements listed above remain valid for Sutherland system as well.

\paragraph{4.} The plan of the paper is as follows. In Section 2 we collect all required calculations for $n=1$ case. Basic integral identities for $n=1$ are Euler beta integral and its hyperbolic generalization.
There are two possible points of view on non-relativistic model. 
If one has all needed relativistic formulas, one can 
try to obtain non-relativistic analogs by the appropriate reduction.
On the other hand one can try to work out all needed non-relativistic formulas independently. We test both points of view, that is we reduce the appearing operator identities to the classical integrals independently and also trace the degeneration procedure of these identities from Ruijsenaars to Sutherland model. Although one-particle wave functions are just plane waves and their completeness and orthogonality is well-known, in this section we present a proof suitable to generalization for $n>1$. 

Section~3 is devoted to more nontrivial $n=2$ case. Here the basic operator relations are given by the integral identities which implicitly appeared in \cite{HR4}. 
The presentation in the case $n=2$ is close to the one given for $n=1$. We prove the equivalence of two integral representations for the eigenfunction
in non-relativistic case by different methods, one of them admits natural generalization to arbitrary $n$.   
The calculation of the scalar product between eigenfunctions of the operator $Q_2(\lambda)$ 
is also performed by two methods. The first one is a standard method from textbooks, yet we do not know its generalization to the case of arbitrary $n$. Due to this reason we present calculation of the scalar product using $Q$-operator, which works for all constructed eigenfunctions and admits natural generalization to arbitrary~$n$. In the last subsection we discuss the completeness of eigenfunctions. Due to the remarkable property of self-duality in relativistic case the completeness relation is in fact equivalent to the orthogonality relation up to the change $g\rightleftarrows g^*$. In non-relativistic case the completeness relation for the eigenfunctions of $Q$-operator is equivalent to the orthogonality relation for eigenfunctions of the dual $\QQ$-operator.

\newpage
\setcounter{footnote}{0}

\section{Case $n=1$}

In this section we demonstrate
the main relations using the simplest example $n=1$.
The plan step by step by subsections is the following:
\begin{enumerate}
\item Description of all $Q$-operators as integral operators and proof of their commutativity.
\item Diagonalization of $Q$-operators in a straightforward way. Calculation of the eigenvalues is based on the beta integral and its generalization.
\item Reductions that relate different $Q$-operators. Derivation of commutation relations between $Q$-operators and $\Lambda$-operators from the commutation relations between $Q$-operators.
\item Calculation of the scalar product for $Q$-operator eigenfunctions. For $n = 1$ it is reduced to the proof of the
standard formula for the plane waves. For illustration we prove this formula
using two nonstandard regularizations which are very useful in the general case.
\end{enumerate}

\subsection{Definitions of $Q$-operators and commutativity}

We have three one-parametric families of commuting operators
\begin{equation} \label{com1}
\begin{aligned}
Q(\lambda)Q(\mu) &= Q(\mu)Q(\lambda), \\[6pt]
\QQ(x)\QQ(y) &= \QQ(y)\QQ(x), \\[6pt]
\QQ(x|\bo)\QQ(y|\bo) &= \QQ(y|\bo)\QQ(x|\bo).
\end{aligned}
\end{equation}
The $Q$-operator $Q(\lambda)$ depends on the spectral parameter $\lambda$
and acts on the functions of variable $x$:
it is the integral operator
\begin{align*}
&[Q(\lambda)\Psi](x) = \int\limits_{-\infty}^{+\infty}
d y\, e^{i \lambda\left(x-y\right)} K(x-y)\, \Psi(y)
\end{align*}
with the kernel $e^{i \lambda\left(x-y\right)} K(x-y)$.
For the dual $\QQ$-operator $\QQ(x)$ the roles of variable $x$ and spectral parameter $\lambda$ interchange: the variable $x$ plays the role of the
spectral parameter and $\QQ $-operator acts on the functions of variable $\lambda$. It is the integral operator
\begin{align*}
&[\QQ(x)\Psi](\lambda) = \int\limits_{-\infty}^{+\infty}
\frac{d \gamma}{2\pi}\, e^{i x\left(\lambda-\gamma\right)} \KK(\lambda-\gamma)\, \Psi(\gamma)
\end{align*}
with the kernel $e^{i x\left(\lambda-\gamma\right)} \KK(\lambda-\gamma)$. The explicit expressions for $K(x)$ and $\KK(\lambda)$ are given by the formulas
\begin{align*}
K(x) = \frac{1}{\mathrm{ch}^g(x)}, \qquad  \KK(\lambda) = \frac{\Gamma\left(\frac{g+i\lambda}{2}\right)
\Gamma\left(\frac{g-i\lambda}{2}\right)}{2^{1-g}\Gamma(g)}
\end{align*}
and in explicit form we have
\begin{align*}
&[Q(\lambda)\Psi](x) =
\int\limits_{-\infty}^{+\infty}
d y\,
\frac{e^{i \lambda\left(x-y\right)}}{\mathrm{ch}^g(x-y)}\,\Psi(y)\,,\\
&[\QQ(x)\Psi](\lambda) =
\int\limits_{-\infty}^{+\infty}
\frac{d \gamma}{2\pi}\,
e^{i x\left(\lambda-\gamma\right)}\frac{\Gamma\left(\frac{g+i(\lambda-\gamma)}{2}\right)
\Gamma\left(\frac{g-i(\lambda-\gamma)}{2}\right)}{2^{1-g}\Gamma(g)}\,
\Psi(\gamma)\,.
\end{align*}
On the higher relativistic level we have the integral operator which we supply in this section by index $\bo$ to recall its dependence on periods $\bo=(\o_1,\o_2)$.
\begin{align*}
[\QQ(x|\bo)\Psi](\lambda) = \int\limits_{-\infty}^{+\infty}
d \gamma \, e^{\frac{2\pi i }{\omega_1\omega_2}x
\left(\lambda-\gamma\right)} \KKK_g(\lambda-\gamma)\, \Psi(\gamma)
\end{align*}
with the kernel $e^{\frac{2\pi i }{\omega_1\omega_2}x
\left(\lambda-\gamma\right)} \KKK_g(\lambda-\gamma)$ where
\begin{align*}
\KKK_g(\lambda) = \frac{S\left(\omega_1+\omega_2+i\lambda-\frac{g}{2}\right)}
{S\left(i\lambda+\frac{g}{2}\right)} =
\frac{1}
{S\left(\frac{g}{2}+i\lambda\right)\,S\left(\frac{g}{2}-i\lambda\right)}.
\end{align*}
Here $S(z) : = S_2(z | \bo)$ is the double sine function, see its properties in Appendix \ref{AppA}. In explicit form we have
\begin{align*}
[\QQ(x|\bo)\Psi](\lambda) = \int\limits_{-\infty}^{+\infty}
d \gamma\, e^{\frac{2\pi i}{\omega_1\omega_2}
x\left(\lambda-\gamma\right)}\,
\frac{S\left(\omega_1+\omega_2+i(\lambda-\gamma)-\frac{g}{2}\right)}
{S\left(i(\lambda-\gamma)+\frac{g}{2}\right)}\, \Psi(\gamma).
\end{align*}

Relations \eqref{com1} are proved in all cases uniformly.
Let us consider for definiteness the first relation
$Q(\lambda)Q(\mu) = Q(\mu)Q(\lambda)$.
We have to prove that integral kernels of operators in both sides
of equality coincide or, equivalently, that integral
kernel of the operator $Q(\lambda) Q(\mu)$ is symmetric under
exchange $\lambda \rightleftarrows \mu$.
The corresponding kernel is given by the convolution of the kernels of operators
$Q(\lambda)$ and $Q(\mu)$. The needed symmetry is proved by the change of integration
variable $s = z+x-t$
\begin{multline*}
\int\limits_{-\infty}^{+\infty}
d s\,e^{i \lambda\left(x-s\right)}\,K(x-s)\,
e^{i \mu\left(s-z\right)}\,K(s-z) \\
 =
\int\limits_{-\infty}^{+\infty}
d t\,e^{i \lambda\left(t-z\right)}\,K(t-z)\,
e^{i \mu\left(x-t\right)}\,K(x-t).
\end{multline*}

\subsection{Eigenfunctions and beta integrals}
All $Q$-operators commute with operator of translation $T^{a}f(x) = f(x+a)$ or, equivalently,
integral kernels of $Q$-operators depend only on difference of coordinates.
Eigenfunctions of the operator $T^{a}$ are usual plane waves so that we expect
the same for all $Q$-operators.  It is indeed the fact 
\begin{align*}
Q(\lambda)\,e^{i\lambda_1 x} &= q(\lambda,\lambda_1)\,e^{i\lambda_1 x}, \\[6pt]
\QQ(x)\,e^{ix_1 \lambda} &= \hat{q}(x,x_1)\,e^{ix_1 \lambda}, \\[6pt]
\QQ(x|\bo)\,
e^{\frac{2\pi i }{\omega_1\omega_2}x_1\lambda} &= \hat{q}(x,x_1|\bo)\,
e^{\frac{2\pi i }{\omega_1\omega_2}x_1\lambda}.
\end{align*}
and eigenvalues coincide with the Fourier transformation of the integral kernel
\begin{align*}
q(\lambda,\lambda_1) &= \KK(\lambda-\lambda_1),\\[6pt]
\hat{q}(x,x_1) &= K(x-x_1),\\[6pt]
\hat{q}(x,x_1|\bo) &= \sqrt{\omega_1\omega_2}\,S(g^*)\,\KKK_{g^*}(x-x_1).
\end{align*}
The proof is straightforward. Let us start from the operator $Q(\lambda)$
\begin{align*}
Q(\lambda)\,e^{i\lambda_1 x} =
\int\limits_{-\infty}^{+\infty}
d y\,e^{i \lambda\left(x-y\right)}\,K(x-y)\, e^{i\lambda_1 y} =
e^{i\lambda_1 x}\,\int\limits_{-\infty}^{+\infty}
d z\,e^{i \left(\lambda-\lambda_1\right)z}\,K(z).
\end{align*}
After change of variables $y = x-z$ in initial integral the $x$-dependence is
factorized out in the form $e^{i\lambda_1 x}$. The remaining integral does not
depend on $x$ and gives the explicit expression for the corresponding eigenvalue.
Functions $K(z)$ and $\KK(\lambda)$ are connected by Fourier transformation
\begin{align}
\label{F1}
\int\limits_{-\infty}^{+\infty}
d z\,e^{i \lambda z}\,K(z) = \KK(\lambda), \qquad
\int\limits_{-\infty}^{+\infty}
\frac{d \lambda}{2\pi}\,e^{-i \lambda z}\,\KK(\lambda) = K(z)
\end{align}
so that one obtains
\begin{align*}
Q(\lambda)\,e^{i\lambda_1 x} = \KK(\lambda-\lambda_1)\,e^{i\lambda_1 x}\,.
\end{align*}
The first relation in \eqref{F1} in explicit form looks as follows
\begin{align}\label{beta0}
\int\limits_{-\infty}^{+\infty}
d z\,
\frac{e^{i \lambda z}}{\mathrm{ch}^g (z)} =
\frac{\Gamma\left(\frac{g+i\lambda}{2}\right)
\Gamma\left(\frac{g-i\lambda}{2}\right)}{2^{1-g}\Gamma(g)}.
\end{align}
It is equivalent to the Euler beta integral in the form
\begin{align*}
\int_0^\infty\frac{x^a}{(1+x)^{a+b}}\frac{dx}{x}=
\frac{\Gamma(a)\Gamma(b)}{\Gamma(a+b)}.
\end{align*}
Indeed, changing the variable $x = e^{2z}$ in \eqref{beta0} we have
\begin{equation*}
\begin{aligned}
\int\limits_{-\infty}^{+\infty}
d z\,
\frac{e^{i\lambda z}}{\mathrm{ch}^g (z)}
&=
2^g\,\int\limits_{-\infty}^{+\infty}
d z\,
\frac{e^{i(\lambda-ig) z}}{\left(1+e^{2z}\right)^g }  \\[6pt]
&= 2^{g-1}\,\int\limits_{0}^{+\infty}
\frac{d x}{x}\,
\frac{x^{\frac{g+i\lambda}{2}}}{\left(1+x\right)^g } =
\frac{\Gamma\left(\frac{g+i\lambda}{2}\right)
\Gamma\left(\frac{g-i\lambda}{2}\right)}{2^{1-g}\Gamma(g)}.
\end{aligned}
\end{equation*}
The second relation in \eqref{F1} explicitly looks as follows
\begin{align}\label{beta00}
\int\limits_{-\infty}^{+\infty}
\frac{d\lambda}{2\pi}\,
e^{- i \lambda z}\,\frac{\Gamma\left(\frac{g+i\lambda}{2}\right)
\Gamma\left(\frac{g-i\lambda}{2}\right)}
{2^{1-g}\Gamma(g)} =
\frac{1}{\mathrm{ch}^g(z)}.
\end{align}
It is simply the formula of the inverse Fourier transformation.

Previous calculations can be almost literally repeated for the dual $\QQ$-operator
\begin{align*}
\QQ(x)\,e^{ix_1 \lambda} &= \int\limits_{-\infty}^{+\infty}
\frac{d \gamma}{2\pi}\, e^{i x\left(\lambda-\gamma\right)} \KK(\lambda-\gamma)\,
e^{ix_1 \gamma} \\[6pt]
&= e^{ix_1 \lambda}\,
\int\limits_{-\infty}^{+\infty}
\frac{d \gamma}{2\pi}\, e^{i (x-x_1)\gamma} \KK(\gamma)
= K(x-x_1)\,e^{ix_1 \lambda}\,,
\end{align*}
and for the remaining higher level $\QQ(x|\bo)$-operator
\begin{equation*}
\begin{aligned}
\QQ(x|\bo)\,
&e^{\frac{2\pi i }{\omega_1\omega_2}x_1\lambda} =
\int\limits_{-\infty}^{+\infty}
d \gamma\, e^{\frac{2\pi i }{\omega_1\omega_2}x
	\left(\lambda-\gamma\right)} \KKK_g(\lambda-\gamma)\,
e^{\frac{2\pi i }{\omega_1\omega_2}x_1\gamma} \\[6pt]
= \,&e^{\frac{2\pi i }{\omega_1\omega_2}x_1\lambda}\,
\int\limits_{-\infty}^{+\infty}
d \gamma\, e^{\frac{2\pi i }{\omega_1\omega_2}\gamma
	\left(x-x_1\right)} \KKK_g(\gamma) =
e^{\frac{2\pi i }{\omega_1\omega_2}x_1\lambda}\,
\sqrt{\omega_1\omega_2}\,S(g^*)\,\KKK_{g^*}(x-x_1)
\end{aligned}
\end{equation*}
where $g^* = \omega_1+\omega_2-g$ and on the last step we used the generalization of the
beta integral in the form
\begin{align}\label{betah}
\int\limits_{-\infty}^{+\infty}
d z\,e^{\frac{2\pi i }{\omega_1\omega_2}x z}\,
K_g(z)  = \sqrt{\omega_1\omega_2}\,S(g^*)\,\KKK_{g^*}(x),
\end{align}
or explicitly (see \cite{PT,FKV})
\begin{align*}
\int\limits_{-\infty}^{+\infty}
d z\,
\frac{e^{\frac{2\pi i }{\omega_1\omega_2}x z}}
{S\left(\frac{g}{2}+iz\right)S\left(\frac{g}{2}-iz\right)} =
\frac{\sqrt{\omega_1\omega_2}\,S(g^*)}
{S\left(\frac{g^*}{2}+ix\right)
S\left(\frac{g^*}{2}-ix\right)}\,.
\end{align*}

\subsection{Reductions} \label{sec:reduc}

Now we demonstrate that formulas from the higher level are in some sense universal and all formulas for operators $Q(\lambda)$ and $\QQ(x)$ can be obtained by simple reductions from the corresponding formulas for the operator $\QQ(x|\bo)$.

The needed reduction is based on the leading asymptotics as $\omega_2 \to 0$
\begin{align*}
K_{g\omega_2}(\lambda\omega_2) &\to
\frac{2^{1-g}\Gamma(g)}{2\pi}\,
\left(\frac{2\pi\omega_2}{\omega_1}\right)^{g-1}
\KK(2\lambda),\\[8pt]
K_{\omega_1+\omega_2-g\omega_2}(\lambda\omega_2) &\to
2^{g}\,K\left(\frac{\pi \lambda}{\omega_1}\right).
\end{align*}
In Appendix \ref{AppA} we derive these formulas and demonstrate that
beta integral \eqref{betah} reduces in the corresponding
asymptotic regimes to the relations \eqref{F1}.

\subsubsection{Commutation relations between $Q$-operators} \label{QQ1-red}

We start from the commutation relation for $Q$-operators in relativistic model
\begin{align}\label{QQ_*}
\QQ(x|\bo)\QQ(y|\bo) = \QQ(y|\bo)\QQ(x|\bo)
\end{align}
and derive the commutation relations
\begin{align*}
Q(\lambda)Q(\mu) = Q(\mu)Q(\lambda), \qquad
\QQ(x)\QQ(y) = \QQ(y)\QQ(x)
\end{align*}
by appropriate reductions.
In integral form the relation \eqref{QQ_*} looks as follows
\begin{multline}\label{KK_*}
\int\limits_{-\infty}^{+\infty}
d \gamma\,e^{\frac{2\pi i x (\lambda-\gamma)}{\omega_1\omega_2}}
K_g(\lambda-\gamma)\,e^{\frac{2\pi i y (\gamma-\mu)}{\omega_1\omega_2}}
K_g(\gamma-\mu) \\
=
\int\limits_{-\infty}^{+\infty}
d \gamma\,e^{\frac{2\pi i y (\lambda-\gamma)}{\omega_1\omega_2}}
K_g(\lambda-\gamma)\,e^{\frac{2\pi i x (\gamma-\mu)}{\omega_1\omega_2}}
K_g(\gamma-\mu).
\end{multline}
In this identity we rescale
\begin{equation*}
g \to g\omega_2, \quad \lambda \to \frac{\lambda\omega_2}{2}, \quad \gamma \to \frac{\gamma\omega_2}{2}, \quad \mu \to \frac{\mu\omega_2}{2}, \quad x \to \frac{\omega_1 x}{\pi}, \quad  y \to \frac{\omega_1 y}{\pi}
\end{equation*}
and then using the leading asymptotic as $\omega_2\to 0$
\begin{align*}
K_{g\omega_2}(\lambda\omega_2) \to
\frac{2^{1-g}\Gamma(g)}{2\pi}\,
\left(\frac{2\pi\omega_2}{\omega_1}\right)^{g-1}
\KK(2\lambda), \qquad
\KK(\lambda) = \frac{\Gamma\left(\frac{g+i\lambda}{2}\right)
\Gamma\left(\frac{g-i\lambda}{2}\right)}{2^{1-g}\Gamma(g)}
\end{align*}
we obtain
\begin{multline*}
\int\limits_{-\infty}^{+\infty}
d \gamma\,e^{i x (\lambda-\gamma)}
\KK(\lambda-\gamma)\,e^{i y (\gamma-\mu)}
\KK(\gamma-\mu) \\
= \int\limits_{-\infty}^{+\infty}
d \gamma\,e^{i y (\lambda-\gamma)}\KK(\lambda-\gamma)\,e^{i x (\gamma-\mu)}
\KK(\gamma-\mu)\,.
\end{multline*}
It is precisely the integral form of the relation
$\QQ(x)\QQ(y) = \QQ(y)\QQ(x)$.

For the second reduction we start from the relation
\eqref{KK_*} for the dual coupling constant, that is $g \to \omega_1+\omega_2-g$. Again rescale
\begin{equation*}
g \to g\omega_2, \quad \lambda \to \frac{\lambda\,\omega_1\omega_2}{\pi}, \quad
\gamma \to \frac{\gamma\,\omega_1\omega_2}{\pi}, \quad \mu \to \frac{\mu\,\omega_1\omega_2}{\pi}, \quad x \to \frac{x}{2}, \quad y \to \frac{y}{2}
\end{equation*}
and using the leading asymptotic as $\omega_2\to 0$
\begin{align*}
K_{\omega_1+\omega_2-g\omega_2}(\lambda\omega_2) \to
2^{g}\,K\left(\frac{\pi \lambda}{\omega_1}\right), \qquad
K(x) = \frac{1}{\mathrm{ch}^g(x)}
\end{align*}
we arrive at
\begin{multline*}
\int\limits_{-\infty}^{+\infty}
d \gamma\,e^{i x (\lambda-\gamma)}
K(\lambda-\gamma)\,e^{i y (\gamma-\mu)}
K(\gamma-\mu) \\
=
\int\limits_{-\infty}^{+\infty}
d \gamma\,e^{i y (\lambda-\gamma)}
K(\lambda-\gamma)\,e^{i x (\gamma-\mu)}
K(\gamma-\mu)\,,
\end{multline*}
which is the integral form of the relation
$Q(x)Q(y) = Q(y)Q(x)$ (modulo simple renaming of the spectral and function variables $x\,,y \rightleftarrows \lambda\,,\mu$).

\subsubsection{$Q\Lambda$-commutation relations}

In the general case $Q\Lambda$-commutation relations are used to diagonalize $Q$-operators and can be derived from the $QQ$-commutation relations.
Now we are going to demonstrate how it works in the simplest
example $n=1$ for all kinds of operators. Reduction considered in this section
can be viewed as a preparation for similar calculations in the case $n=2$.

The $n=1$ example is in some sense degenerate because
$\Lambda$-operators are operators of multiplication
by plane waves. In fact $Q\Lambda$-commutation
relations in this case are equivalent to the statement that plane waves are
eigenfunctions of the $Q$-operators, which has been already checked.

First of all we perform all calculations in relativistic case and
on the next stage repeate the same procedure at lower level,
i.e. in non-relativistic situation.
We start from the commutation relation
\begin{align}\label{QQ1}
\QQ(x|\bo)\QQ(y|\bo) = \QQ(y|\bo)\QQ(x|\bo)
\end{align}
and derive the relation
\begin{align}\label{QL1}
\QQ(x|\bo)\,\hat{\Lambda}\left(y-\textstyle\frac{ig^*}{2}|\bo\right) =
\hat{\Lambda}\left(y-\textstyle\frac{ig^*}{2}|\bo\right)\,
\hat{q}\left(x\,,y-\textstyle\frac{ig^*}{2}|\bo\right)
\end{align}
All operators act on the functions of the variable $\lambda$ and
operator $\hat{\Lambda}(y|\bo)$
is the operator of multiplication by $e^{\frac{2\pi i \lambda}{\omega_1\omega_2} y}$.
Note that starting from $QQ$-commutation
relations one obtains $\Lambda$-operators in $Q\Lambda$-commutation
relations with shifted spectral parameters $y \to y- i g^*/2$.
The shift is universal for all $n$ so that it is instructive to fix it
in the simplest case $n=1$.

In integral form the commutation relation \eqref{QQ1} looks as follows
\begin{multline}\label{KKbo}
\int\limits_{-\infty}^{+\infty}
d \gamma\,e^{\frac{2\pi i x (\lambda-\gamma)}{\omega_1\omega_2}}
K_g(\lambda-\gamma)\,e^{\frac{2\pi i y (\gamma-\mu)}{\omega_1\omega_2}}
K_g(\gamma-\mu) \\
=
\int\limits_{-\infty}^{+\infty}
d \gamma\,e^{\frac{2\pi i y (\lambda-\gamma)}{\omega_1\omega_2}}
K_g(\lambda-\gamma)\,e^{\frac{2\pi i x (\gamma-\mu)}{\omega_1\omega_2}}
K_g(\gamma-\mu)\,
\end{multline}
and we consider reduction of this identity as $\mu \to \infty$ using
the following leading asymptotic
\begin{align*}
K_g(\lambda)  \to
e^{\frac{2\pi i}{\omega_1\omega_2}\,\lambda\,\frac{i g^{*}}{2}}, \qquad \lambda \to \infty
\end{align*}
see Appendix \ref{AppA-as2}.
In the left hand side of relation \eqref{KKbo} we have for $\mu \to \infty$
\begin{align*}
K_g(\gamma-\mu) = K_g(\mu - \gamma)\to e^{\frac{2\pi i}{\omega_1\omega_2}\,(\mu-\gamma)\,\frac{i g^{*}}{2}}.
\end{align*}
To obtain the same asymptotic in the right hand side we have to shift $\gamma \to \gamma+\mu$ and
then send $\mu \to \infty$ using
\begin{align*}
K_g(\lambda-\gamma-\mu) \to e^{\frac{2\pi i}{\omega_1\omega_2}\,(\mu + \gamma-\lambda)\,\frac{i g^{*}}{2}}.
\end{align*}
So, the leading asymptotics of both sides give
\begin{multline*}
\int\limits_{-\infty}^{+\infty}
d \gamma\,e^{\frac{2\pi i x (\lambda-\gamma)}{\omega_1\omega_2}}
K_g(\lambda-\gamma)\,e^{\frac{2\pi i y (\gamma-\mu)}{\omega_1\omega_2}}
e^{\frac{2\pi i}{\omega_1\omega_2}\,(\mu-\gamma)\,\frac{i g^{*}}{2}} \\
=
\int\limits_{-\infty}^{+\infty}
d \gamma\,e^{\frac{2\pi i y (\lambda-\mu-\gamma)}{\omega_1\omega_2}}
e^{\frac{2\pi i}{\omega_1\omega_2}\,(\mu+\gamma-\lambda)\,\frac{i g^{*}}{2}}\,e^{\frac{2\pi i x \gamma}{\omega_1\omega_2}}
K_g(\gamma)\,,
\end{multline*}
so that $\mu$-dependent contributions can be canceled and
after all we obtain the identity
\begin{multline*}
\int\limits_{-\infty}^{+\infty}
d \gamma\,e^{\frac{2\pi i x (\lambda-\gamma)}{\omega_1\omega_2}}
K(\lambda-\gamma)\,
e^{\frac{2\pi i \gamma}{\omega_1\omega_2}\left(y-\frac{ig^*}{2}\right)} \\
=
e^{\frac{2\pi i \lambda}{\omega_1\omega_2}\left(y-\frac{ig^*}{2}\right)}\,
\int\limits_{-\infty}^{+\infty}
d \gamma\,e^{\frac{2\pi i \gamma}{\omega_1\omega_2}
\left(x-y+\frac{ig^*}{2}\right)}
K(\gamma)\,,
\end{multline*}
which is exactly the relation \eqref{QL1} in explicit form.

Now we repeat similar reduction starting
from commutation relation
\begin{align}\label{QQh}
\QQ (x)\QQ (y) = \QQ (y)\QQ (x)
\end{align}
and derive the relation
\begin{align}\label{QLh}
\QQ(x)\,\hat{\Lambda}\left(y-\textstyle\frac{i\pi}{2}\right) =
\hat{\Lambda}\left(y-\textstyle\frac{i\pi}{2}\right)\,
\hat{q}\left(x\,,y-\textstyle\frac{i\pi}{2}\right).
\end{align}
All operators act on the functions of the variable $\lambda$ and $\hat{\Lambda}(y)$ is the operator of multiplication by $e^{i \lambda y}$.
Note the same rule as in relativistic case: starting from $QQ$-commutation
relations one obtains $\Lambda$-operators with shifted spectral parameters $y \to y-i \pi/2$.
Again the shift is universal for all $n$ and it is instructive to fix it
in the simplest case $n=1$.

The commutation relation \eqref{QQh} in integral form looks as follows
\begin{multline*}
\int\limits_{-\infty}^{+\infty}
\frac{d \gamma}{2\pi}\,e^{i x (\lambda-\gamma)}
\KK (\lambda-\gamma)\,e^{i y (\gamma-\mu)}
\KK (\gamma-\mu) \\
=
\int\limits_{-\infty}^{+\infty}
\frac{d \gamma}{2\pi}\,e^{i y (\lambda-\gamma)}
\KK (\lambda-\gamma)\,e^{i x (\gamma-\mu)}
\KK (\gamma-\mu)\,.
\end{multline*}
The leading asymptotic of $\KK (\gamma-\mu)$ for
$\mu \to \infty$ has the following form (see Appendix \ref{AppA})
\begin{align*}
\KK (\gamma-\mu) =
\frac{\Gamma\left(\frac{g+i(\gamma-\mu)}{2}\right)
\Gamma\left(\frac{g-i(\gamma-\mu)}{2}\right)}{2^{1-g}\Gamma(g)} \to
\frac{2\pi}{\Gamma(g)}\,\mu^{g-1}\,e^{\frac{\pi}{2}(\gamma-\mu)}.
\end{align*}
The whole reduction is the same as in previous case: in the left hand side we simply send $\mu \to \infty$, but in the right hand side we shift $\gamma \to \gamma+\mu$ and then send $\mu \to \infty$
\begin{multline*}
\int\limits_{-\infty}^{+\infty}
\frac{d \gamma}{2\pi}\,e^{i x (\lambda-\gamma)}
\KK (\lambda-\gamma)\,e^{i y (\gamma-\mu)}\,
\mu^{g-1}\,e^{\frac{\pi}{2}(\gamma-\mu)} \\
=
\int\limits_{-\infty}^{+\infty}
\frac{d \gamma}{2\pi}\,e^{i y (\lambda-\gamma-\mu)}\,\mu^{g-1}\,
e^{\frac{\pi}{2}(\gamma-\lambda-\mu)}\,e^{i x \gamma}
\KK (\gamma).
\end{multline*}
Again $\mu$-dependent contributions can be canceled and one obtains
\begin{align*}
\int\limits_{-\infty}^{+\infty}
\frac{d \gamma}{2\pi}\,e^{i x (\lambda-\gamma)}
\KK (\lambda-\gamma)\,e^{i \gamma\left(y-\frac{i\pi}{2}\right)} =
e^{i \lambda\left(y-\frac{i\pi}{2}\right)}
\int\limits_{-\infty}^{+\infty}
\frac{d \gamma}{2\pi}\,e^{i \gamma \left(x-y+\frac{i\pi}{2}\right)}
\KK (\gamma)\,,
\end{align*}
which is exactly the relation \eqref{QLh} in an explicit form.

The last relation
\begin{align}\label{QL0}
Q(\lambda)\,\Lambda\left(\mu-ig\right) = \Lambda\left(\mu-ig\right)\,q(\lambda\,,\mu-ig)
\end{align}
is derived from the commutativity relation
$$
Q(\lambda)\,Q(\mu) = Q(\mu)\,Q(\lambda)\,
$$
in the same way. More general reduction is
presented for the case $n=2$ below.

\subsection{Regularization and the scalar product}

The orthogonality relation
\begin{align*}
\langle\Psi_{\mu}|\Psi_{\lambda}\rangle =
\int\limits_{-\infty}^{+\infty}
d x\, \overline{\Psi_{\mu}(x)}\,\Psi_{\lambda}(x) =
2\pi\,\delta(\lambda-\mu)
\end{align*}
and completeness relation
\begin{align*}
\int\limits_{-\infty}^{+\infty}
\frac{d\lambda}{2\pi}\,
\Psi_{\lambda}(x)\,\overline{\Psi_{\lambda}(y)} = \delta(x-y)
\end{align*}
for the $Q$-operators' eigenfunctions $\Psi_{\lambda}(x) = e^{i\lambda x}$
are reduced to the standard integral
\begin{align}\label{standard}
\int\limits_{-\infty}^{+\infty}
d x\, e^{i\lambda x} = 2\pi\,\delta(\lambda).
\end{align}
Below we demonstrate on this simplest example the method of
calculation of the scalar product which we shall use in general
case of arbitrary $n$.

The integral \eqref{standard} is ill-defined and
we shall understand this integral as an appropriate
limit of the regularized integral. Regularization should have
two natural properties:
\begin{itemize}
\item regularized integrals must be convergent;
\item regularized integrals must be calculable in an explicit form.
\end{itemize}

\subsubsection{$Q$-operator regularization}

In this subsection we introduce regularization such that the regularized integral gives precisely
the action of the $Q$-operator on eigenfunction.
It appears that in general case it is possible to introduce
similar regularization preserving the same property:
regularized integral contains action of
$Q$-operator on eigenfunction and can be calculated in explicit form.

Let us regularize integral adding the external
point $z_0$ and inserting additional $\varepsilon >0 $ in exponent
\begin{align*}
&\int\limits_{-\infty}^{+\infty}
d x \, e^{i \lambda x} =
\frac{1}{2^{g}}\lim _{z_0 \to +\infty}\,
\lim _{\varepsilon \to 0}\,
\int\limits_{-\infty}^{+\infty}
d x \,
\frac{e^{g(z_0 - x)}\,e^{\varepsilon x}\,\,
e^{i\lambda x}}
{\mathrm{ch}^g(x-z_0)}\,.
\end{align*}
To prove that in the limit $z_0\to +\infty$ the initial integral is reproduced
we use the following asymptotic
\begin{align*}
\frac{1}{2^{g}}\frac{1}{\mathrm{ch}^g(x-z_0)} =
\frac{e^{-g(z_0-x)}}
{\left(1+e^{-2(z_0-x)}\right)^g} \to
e^{-g(z_0-x)}, \qquad z_0 \rightarrow + \infty.
\end{align*}
For $x \to +\infty$ the integrand behaves as $e^{2gz_0}\,e^{(-2g + i\lambda +\varepsilon)x}$, so that
decreasing factor $e^{-2g x}$ guarantees the convergence.
For $x \to -\infty$ the integrand behaves as $e^{i\lambda x +\varepsilon x}$ and
decreasing factor $e^{\varepsilon x}$ guarantees
the convergence for $\varepsilon>0$.
The calculation of the integral follows the same line as calculation
of $Q$-operator eigenvalue and everything is reduced to the beta integral
\begin{align*}
\int\limits_{-\infty}^{+\infty}
d x \,
\frac{e^{g(z_0 - x)}\,e^{\varepsilon x}\,\,
e^{i\lambda x}}
{\mathrm{ch}^g(x-z_0)} &= \{x \to x+z_0\} =
e^{(\varepsilon+i\lambda) z_0}\,\int\limits_{-\infty}^{+\infty}
d x \,
\frac{e^{i(\lambda-i\varepsilon +ig) x}}
{\mathrm{ch}^g x} \\[6pt]
&=e^{(\varepsilon+i\lambda) z_0}\,
\frac{\Gamma\left(\frac{g+i(\lambda-i\varepsilon +ig)}{2}\right)
\Gamma\left(\frac{g-i(\lambda-i\varepsilon +ig)}{2}\right)}
{2^{1-g}\Gamma(g)} \\[6pt]
&= e^{(\varepsilon+i\lambda) z_0}\,
\frac{\Gamma\left(\frac{i(\lambda-i\varepsilon)}{2}\right)
\Gamma\left(g -\frac{i(\lambda-i\varepsilon)}{2}\right)}
{2^{1-g}\Gamma(g)}.
\end{align*}
This formula shows that order of the limits is fixed.
If we shall fix $\varepsilon>0$ then $\lim _{z_0 \to +\infty}$ does
not exist due to the factor $e^{\varepsilon z_0}$.
It is possible to argue that the prescribed order is the
right one without explicit calculation. Let us introduce slightly
different regularization
\begin{align*}
&\int\limits_{-\infty}^{+\infty}
d x \, e^{i \lambda x} =
\frac{1}{2^{g}}\lim _{z_0 \to +\infty}\,
\lim _{\varepsilon \to 0}\,
\int\limits_{-\infty}^{+\infty}
d x \,
\frac{e^{g(z_0 - x)}\,e^{-\varepsilon |x|}\,\,
e^{i\lambda x}}
{\mathrm{ch}^g(x-z_0)}\,.
\end{align*}
Due to the factor $e^{-\varepsilon |x|}$ integral is
convergent at $x\to \pm\infty$. Hence, $z_0$-regularization is not needed and the order
of the limits is not important.
But in prescribed order it is possible to change
$e^{-\varepsilon |x|} \to e^{\varepsilon x}$
\begin{align*}
\int\limits_{-\infty}^{+\infty}
d x \, e^{i \lambda x} &=
\frac{1}{2^{g}}\lim _{z_0 \to +\infty}\,
\lim _{\varepsilon \to 0}\,
\int\limits_{-\infty}^{+\infty}
d x \,
\frac{e^{g(z_0 - x)}\,e^{-\varepsilon |x|}\,\,
e^{i\lambda x}}
{\mathrm{ch}^g(x-z_0)} \\
&=
\frac{1}{2^{g}}\lim _{z_0 \to +\infty}\,
\lim _{\varepsilon \to 0}\,
\int\limits_{-\infty}^{+\infty}
d x \,
\frac{e^{g(z_0 - x)}\,e^{\varepsilon x}\,\,
e^{i\lambda x}}
{\mathrm{ch}^g(x-z_0)}\,.
\end{align*}
because at $x \to +\infty$ everything is regularized by
external $z_0$.

Using prescribed order of limits we reproduce the standard answer
\begin{equation*}
\begin{aligned}
\int\limits_{-\infty}^{+\infty}
d x \, e^{i \lambda x} &= \lim _{z_0 \to +\infty}\,
\lim _{\varepsilon \to 0}\, e^{(\varepsilon+i\lambda) z_0}\,
\frac{\Gamma\left(\frac{i(\lambda-i\varepsilon)}{2}\right)
	\Gamma\left(g -\frac{i(\lambda-i\varepsilon)}{2}\right)}
{2^g\,2^{1-g}\Gamma(g)} \\[8pt]
& = \frac{2}{i}\frac{\Gamma\left(1+\frac{i\lambda}{2}\right)
	\Gamma\left(g -\frac{i\lambda}{2}\right)}
{2\,\Gamma(g)} \lim _{z_0 \to +\infty}\,\lim _{\varepsilon \to 0}\,
\frac{e^{i\lambda z_0}}{\lambda-i\varepsilon} \\[8pt]
& = \frac{2}{i}\frac{\Gamma\left(1+\frac{i\lambda}{2}\right)
	\Gamma\left(g -\frac{i\lambda}{2}\right)}
{2\,\Gamma(g)}\,2\pi i \delta(\lambda) = 2\pi \delta(\lambda)
\end{aligned}
\end{equation*}
where we used the formula
\begin{align}\label{g=1}
\lim _{z_0 \to +\infty}\,\lim _{\varepsilon \to 0}\,
\frac{e^{i\lambda z_0}}{\lambda-i\varepsilon} = 2\pi i \delta(\lambda).
\end{align}
It is a relative of the formula
\begin{align*}
\lim _{z_0 \to +\infty}\,
\frac{\sin(\lambda z_0)}{\lambda} = \pi\delta(\lambda)
\end{align*}
and the possible proof is the following.
Let us consider the integral with the test function and divide it on
two parts: the first
integral can be calculated by residues and due
to cancelation of singularity at $\lambda=0$ it is possible to put
$\varepsilon \to 0 $ in the second part
\begin{equation*}
\begin{aligned}
\int_{\mathbb{R}} f(\lambda)\,
\frac{e^{i\lambda z_0}}{\lambda-i\varepsilon}\,d\lambda &=
f(0)\,\int_{\mathbb{R}}\,
\frac{e^{i\lambda z_0}}{\lambda-i\varepsilon}\,d\lambda  +
\int_{\mathbb{R}} \frac{f(\lambda)-f(0)}
{\lambda-i\varepsilon}\,e^{i\lambda z_0}\,d\lambda \\[6pt]
&\xrightarrow{\varepsilon\to 0} 2\pi i\,f(0)  +
\int_{\mathbb{R}} \frac{f(\lambda)-f(0)}
{\lambda}\,e^{i\lambda z_0}\,d\lambda.
\end{aligned}
\end{equation*}
Due to the Riemann-Lebesgue lemma the second contribution
tends to zero in the limit $z_0\to\infty$, so that we obtain
after removing both regularizations
\begin{align*}
\int_{\mathbb{R}} f(\lambda)\,
\frac{e^{i\lambda z_0}}{\lambda -i\varepsilon}\,d\lambda \to 2\pi i\,f(0),
\end{align*}
or equivalently
\begin{align*}
\frac{e^{i\lambda z_0}}{\lambda -i\varepsilon}
\to 2\pi i\,\delta(\lambda).
\end{align*}

\subsubsection{$\hat{Q}$-operator regularization}

There exists a second variant of regularization.
It is possible to introduce regularization, such that resulting integral coincides with
the action of the $\hat{Q}$-operator on eigenfunction.
In general case it is possible to introduce
a similar regularization preserving the same property:
regularized integral gives action of
$\hat{Q}$-operator on eigenfunction and can be calculated in explicit form.

To regularize the integral we introduce two external parameters $\gamma_0$ and $\varepsilon >0 $
\begin{equation}
\begin{aligned}
\int\limits_{-\infty}^{+\infty}
\frac{d \lambda}{2\pi} \,
e^{ix\lambda}\, =
\frac{\Gamma(g)}{2\pi}\,
\lim _{\gamma_0 \to +\infty}\,
\lim _{\varepsilon \to 0}\,\gamma_0^{1-g}\,
&\int\limits_{-\infty}^{+\infty}
\frac{d \lambda}{2\pi}\,
e^{-\frac{\pi}{2}(\lambda-\gamma_0)}
\,e^{\varepsilon\lambda}
\,e^{i x\lambda} \\[6pt]
&\times 
\frac{\Gamma\left(\frac{g+i(\lambda-\gamma_0)}{2}\right)
	\Gamma\left(\frac{g-i(\lambda-\gamma_0)}{2}\right)}{2^{1-g}\Gamma(g)}.
\end{aligned}
\end{equation}
To prove that in the limit $\gamma_0\to +\infty$ we reproduce
the initial integral we use the following asymptotic
as $\gamma_0\to +\infty$
\begin{align*}
\frac{\Gamma\left(\frac{g+i(\lambda-\gamma_0)}{2}\right)
\Gamma\left(\frac{g-i(\lambda-\gamma_0)}{2}\right)}{2^{1-g}\Gamma(g)} \to
\frac{2\pi}{\Gamma(g)}\,\gamma_0^{g-1}\,e^{\frac{\pi}{2}(\lambda-\gamma_0)}\,.
\end{align*}
To check convergence we test behaviour in two regions:
$\lambda\to +\infty$ and $\lambda \to -\infty$.
For $\lambda \to +\infty$ the integrand behaves as $e^{(-\pi + ix +\varepsilon)\lambda}$, so that
integral converges in this region due to decreasing factor $e^{-\pi \l}$. For $\lambda \to -\infty$
we have $e^{( i x +\varepsilon)\lambda}$, so that for $\varepsilon>0$
integral converges in this region too.
The $\lambda$-integral can be calculated explicitly
(it is easy to recognize in this integral the
action of the $\QQ$-operator on eigenfunction)
\begin{multline*}
\int\limits_{-\infty}^{+\infty}
\frac{d \lambda}{2\pi}\,
e^{-\frac{\pi}{2}(\lambda-\gamma_0)}
\,e^{\varepsilon\lambda}
\,e^{i x \lambda}\,
\frac{\Gamma\left(\frac{g+i(\lambda-\gamma_0)}{2}\right)
	\Gamma\left(\frac{g-i(\lambda-\gamma_0)}{2}\right)}{2^{1-g}\Gamma(g)} \\[6pt]
= e^{i\gamma_0(x-i\varepsilon)}\,
\int\limits_{-\infty}^{+\infty}
\frac{d \lambda}{2\pi}
\,e^{i\lambda(x+\frac{i\pi}{2}-i\varepsilon)}\,
\frac{\Gamma\left(\frac{g+i\lambda}{2}\right)
	\Gamma\left(\frac{g-i\lambda}{2}\right)}{2^{1-g}\Gamma(g)} \\[6pt]
= \frac{e^{i\gamma_0(x-i\varepsilon)}}{\ch^g(x+\frac{i\pi}{2}-i\varepsilon)}=
\frac{e^{i\gamma_0(x-i\varepsilon)}\,e^{-i\frac{\pi}{2}g}}{
	\sh^g(x-i\varepsilon)}
\end{multline*}
so that we have
\begin{multline*}
\int\limits_{-\infty}^{+\infty}
\frac{d \lambda}{2\pi} \,
e^{i x \lambda}\, = \frac{\Gamma(g)}{2\pi}\,e^{-i\frac{\pi}{2}g}\,
\lim_{\gamma_0 \to +\infty}\,
\lim_{\varepsilon \to 0}\,
\gamma_0^{1-g}\,\frac{e^{i\gamma_0(x-i\varepsilon)}}{
	\sh^g(x-i\varepsilon)} \\[6pt]
=\frac{\Gamma(g)}{2\pi}\,e^{-i\frac{\pi}{2}g}\,
\lim _{\varepsilon \to 0}\,
\frac{(x-i\varepsilon)^g}{\sh^g(x-i\varepsilon)}
\lim _{\gamma_0 \to +\infty}\,
\lim _{\varepsilon \to 0}
\frac{\gamma_0^{1-g}\,e^{i x \gamma_0}}{
	(x-i\varepsilon)^g} \\[8pt]
=
\lim_{\varepsilon \to 0}\,
\frac{(x-i\varepsilon)^g}{\sh^g(x-i\varepsilon)}\,\,\delta(x) = \delta(x).
\end{multline*}
In the last line we used the formula
\begin{align}\label{g}
\lim _{\gamma_0 \to +\infty}\,\lim _{\varepsilon \to 0}\,
\frac{\gamma_0^{1-g}\,e^{i x \gamma_0}}
{(x-i\varepsilon)^g} =
\frac{2\pi}{\Gamma(g)}\,e^{i\frac{\pi}{2}g}\,\delta(x).
\end{align}
Note that this formula is reduced to \eqref{g=1} in the case $g=1$.
The possible proof of this formula is the following.
Let us consider the integral with the test function and
divide this integral into three parts ($\delta>0$)
\begin{multline*}
\gamma_0^{1-g}\,\int\limits_{-\infty}^{+\infty} d x\,
\frac{e^{i x \gamma_0}}{(x-i\varepsilon)^g}\,f(x) =
\,\gamma_0^{1-g}\,\int\limits_{-\delta}^{+\delta} d x\,
\frac{e^{i x \gamma_0}}{(x-i\varepsilon)^g}\,f(x)\\[6pt]
+\gamma_0^{1-g}\,\int\limits_{-\infty}^{-\delta} d x\,
\frac{e^{i x \gamma_0}}{(x-i\varepsilon)^g}\,f(x)+
\gamma_0^{1-g}\,\int\limits_{\delta}^{+\infty} d x\,
\frac{e^{i x \gamma_0}}{(x-i\varepsilon)^g}\,f(x).
\end{multline*}
The last two terms do not contain singularity in integration domains, so that it is possible to put $\varepsilon = 0$
and due to the Riemann-Lebesgue lemma both contributions tend to zero in the limit $\gamma_0\to+\infty$. In the first integral we perform the
change of variables $y = \gamma_0 x$ and then use the standard formula for the Fourier transformation of the generalised function $(y-i0)^{-g}$
\begin{multline*}
\lim_{\gamma_0 \to +\infty}\,
\lim_{\varepsilon \to 0}\,\gamma_0^{1-g}\,
\int\limits_{-\delta}^{+\delta} d x\,
\frac{e^{i x \gamma_0}}{(x-i\varepsilon)^g}\,f(x) =
\lim _{\gamma_0 \to +\infty}\,
\lim _{\varepsilon \to 0}\,
\int\limits_{-\delta\gamma_0}^{+\delta\gamma_0} d y\,
\frac{e^{i y }}{(y-i\varepsilon)^g}\,f\left(\frac{y}{\gamma_0}\right)  \\[6pt]
= f(0)\,\int\limits_{-\infty}^{+\infty} d y\,
\frac{e^{i y }}{(y-i0)^g} = f(0)\,\frac{2\pi}{\Gamma(g)}\,e^{i\frac{\pi}{2}g}.
\end{multline*}
The used Fourier transformation formula has the following form \cite{GS}
\begin{align*}
\int\limits_{-\infty}^{+\infty} d y\,
(y-i0)^{-g}\,e^{i y p} = \frac{2\pi}{\Gamma(g)}\,e^{i\frac{\pi}{2}g}\,
p_{+}^{g-1}.
\end{align*}

\newpage

\section{Case $n=2$}

The plan step by step by subsections is the following:
\begin{enumerate}
\item Description of all $Q$-operators and $\Lambda$-operators as integral operators.
\item Derivation of the commutation relations between $Q$-operators and
$\Lambda$-operators from commutativity of $Q$-operators.
\item Construction of eigenfunctions. Demonstration of the equivalence of Mellin-Barnes and Hallnas-Ruijsenaars representations.
\item Calculation of the scalar product between eigenfunctions of all kinds using $Q$-operator regularizations.
\end{enumerate}

\subsection{$Q$-operators and $\Lambda$-operators as integral operators}

Let us define $Q$-operators in the case $n=2$ as integral operators by the formulas
\begin{align*}
[Q_2(\lambda)\Psi](x_1,x_2) &=
\int\limits_{-\infty}^{+\infty}
d y_1 d y_2\,\sh^{2g}|y_1-y_2|\,
e^{i\lambda\left(x_1+x_2-y_1-y_2\right)} \\[6pt]
& \qquad  \times  \prod_{i,k=1,2}K(x_i-y_k)\,\Psi(y_1, y_2), \\[6pt]
[\QQ_2(x)\Psi](\lambda_1,\lambda_2) &=
\int\limits_{-\infty}^{+\infty}
\frac{d \gamma_1}{2\pi}\,\frac{d \gamma_2}{2\pi}\,
\mu(\gamma_1,\gamma_2)\,e^{i x\left(\lambda_1+\lambda_2-\gamma_1-\gamma_2\right)} \\[6pt]
& \qquad   \times \prod_{i,k=1,2}\hat{K}(\lambda_i-\gamma_k)\,\Psi(\gamma_1, \gamma_2), \\[6pt]
[\QQ_2(x|\bo)\Psi](\lambda_1,\lambda_2) &=
\int\limits_{-\infty}^{+\infty}
d \gamma_1\,d \gamma_2\,\mu_g(\gamma_1,\gamma_2)\,
e^{\frac{2\pi i}{\omega_1\omega_2} x\left(\lambda_1+\lambda_2-\gamma_1-\gamma_2\right)} \\[6pt]
& \qquad  \times \prod_{i,k=1,2}K_g(\lambda_i-\gamma_k)
\,\Psi(\gamma_1,\gamma_2).
\end{align*}
The building blocks are the same as in the case $n=1$
\begin{gather*}
K(x) = \frac{1}{\mathrm{ch}^g(x)}, \qquad  \KK(\lambda) = \frac{\Gamma\left(\frac{g+i\lambda}{2}\right)
\Gamma\left(\frac{g-i\lambda}{2}\right)}{2^{1-g}\Gamma(g)}, \\[8pt]
\KKK_g(\lambda) =
\frac{1}
{S\left(\frac{g}{2}+i\lambda\right) S\left(\frac{g}{2}-i\lambda\right)},
\end{gather*}
except the nontrivial measure: $\sh^{2g}|y_1-y_2|$ in the simplest case and
\begin{align}\label{mu}
\mu(\gamma_1,\gamma_2) &= \frac{[2^{1-g}\Gamma(g)]^2}
{\Gamma\left(g\pm\frac{i(\gamma_1-\gamma_2)}{2}\right)
\Gamma\left(\pm\frac{i(\gamma_1-\gamma_2)}{2}\right)}, \\[6pt]
\label{mug}
\mu_g(\gamma_1,\gamma_2) &= S(g\pm i(\gamma_1-\gamma_2))S(\pm i(\gamma_1-\gamma_2)).
\end{align}
Here and in what follows we use notation for the products of functions
\begin{equation*}
f(a\pm b) = f(a+b) f(a-b).
\end{equation*}
\textbf{Remark.} Note that in relativistic case Fourier transformation
of the function $K_g$ gives the function $K_{g^*}$ \eqref{betah}.
This means that the transition to the dual $Q$-operator is reduced to the change $g \to g^*$ and renaming of arguments $x \rightleftarrows \lambda$
\begin{multline}
\label{Q2bod}
[Q_2(\lambda|\bo)\Psi](x_1,x_2) =
\int\limits_{-\infty}^{+\infty}
d y_1\,d y_2\,\mu_{g^*}(y_1\,,y_2)\,
e^{\frac{2\pi i}{\omega_1\omega_2} \lambda\left(x_1+x_2-y_1-y_2\right)} \\[6pt]
\times 
\prod_{i,k=1,2}K_{g^*}(x_i-y_k)
\,\Psi(y_1,y_2).
\end{multline}
To avoid the simple duplication of formulas we state
all the results only for the operator $Q_2(\lambda|\bo)$ or
$\QQ_2(x|\bo)$.

Commutation relation
\begin{align}\label{QQbo}
\QQ_2(x|\bo)\,\QQ_2(y|\bo) = \QQ_2(y|\bo)\,\QQ_2(x|\bo)
\end{align}
is proven by residue calculation method in \cite{BDKK1}.
The same method (in fact, in a much simpler form) can be used for the proof
of the commutation relation
\begin{align*}
\QQ_2(x)\,\QQ_2(y) = \QQ_2(y)\,\QQ_2(x).
\end{align*}
Unfortunately we do not have at
the moment any direct proof of the commutation relation
\begin{align*}
Q_2(\lambda)\,Q_2(\rho) = Q_2(\rho)\,Q_2(\lambda),
\end{align*}
except the case $g = 1$, see Appendix \ref{App-g=1}. However, the last two commutation relations can be deduced from the first one \eqref{QQbo} by
the appropriate reduction in full analogy with $n=1$ case, see Section \ref{QQ1-red}.

The closest relatives of $Q$-operators are integral $\Lambda$-operators defined as
\begin{align*}
&[\Lambda_2(\lambda)\Psi](x_1,x_2) =
\int\limits_{-\infty}^{+\infty}
d y\,
e^{i\lambda\left(x_1+x_2-y\right)}\,
K(x_1-y)\,K(x_2-y)\,\Psi(y),\\[6pt]
&[\hat{\Lambda}_2(x)\Psi](\lambda_1,\lambda_2) =
\int\limits_{-\infty}^{+\infty}
\frac{d \gamma}{2\pi}\,e^{i x\left(\lambda_1+\lambda_2-\gamma\right)}
\,\hat{K}(\lambda_1-\gamma)\,\hat{K}(\lambda_2-\gamma)\,\Psi(\gamma),\\[6pt]
&[\hat{\Lambda}_2(x|\bo)\Psi](\lambda_1,\lambda_2) =
\int\limits_{-\infty}^{+\infty}
d \gamma\,
e^{\frac{2\pi i}{\omega_1\omega_2} x\left(\lambda_1+\lambda_2-\gamma\right)}\,
K_g(\lambda_1-\gamma)\,K_g(\lambda_2-\gamma)
\,\Psi(\gamma).
\end{align*}
%

\subsection{$Q\Lambda$-commutation relations}
In this section we explain how commutation relations between $Q$-operators
\begin{align*}
Q_2(\l) \, Q_2(\rho) &= Q_2(\rho) \, Q_2(\l), \\[6pt]
\hat{Q}_2(x) \, \hat{Q}_2(y) &= \hat{Q}_2(y) \, \hat{Q}_2(x), \\[6pt]
\QQ_2(x|\bo) \, \QQ_2(y|\bo) &= \QQ_2(y|\bo) \, \QQ_2(x|\bo)
\end{align*}
degenerate to the relations between $Q$ and $\Lambda$-operators
\begin{align}\label{QL-1}
Q_2(\l) \, \Lambda_2(\rho) &= 2\, q(\l,\rho) \, \Lambda_2(\rho) \, Q_1(\l), \\[6pt] \notag
\hat{Q}_2(x) \, \hat{\Lambda}_2(y) &= 2\,\hat{q}(x,y)\, \hat{\Lambda}_2(y) \, \hat{Q}_1(x), \\[6pt] \notag
\QQ_2(x|\bo) \, \hat{\Lambda}_2(y|\bo) &=  2\, \hat{q}(x,y|\bo) \,
\hat{\Lambda}_2(y|\bo) \, Q_1(x|\bo).
\end{align}
where
\begin{gather*}
q(\l,\rho) = \hat{K}(\l - \rho), \qquad \hat{q}(x,y) = K(x - y), \\[6pt]
 \hat{q}(x,y|\bo) = \sqrt{\omega_1\omega_2}\,S(g^*)\,\KKK_{g^*}(x-y).
\end{gather*}
For simplicity we consider the case of operators with hyperbolic functions $Q_2(\l)$ and $\Lambda_2(\l)$. The derivation for the rest two cases follows the same steps.

Baxter and raising operators of all kinds have similar kernels. Furthermore, their building blocks $K(x)$ and $\mu(x)$ have nice asymptotics as $|x| \rightarrow \infty$ and the corresponding bounds for $x \in \mathbb{R}$. In the simplest case of hyperbolic functions
\begin{equation*}
K(x) = \ch^{-g}(x), \qquad \mu(x) = \sh^g|x|
\end{equation*}
we have exponential asymptotics and bounds
\begin{align} \notag
&K(x) \sim 2^{g}\, e^{-g|x|}, && \hspace{-2cm} \mu(x) \sim 2^{-g} \, e^{g |x| }, \qquad \hspace{1cm} |x| \rightarrow \infty, \\[8pt]
\label{Kmu-b}
&|K(x)| \leq e^{-g|x|}, && \hspace{-2cm} |\mu(x)| \leq 2^{-g} \, e^{g |x| }, \qquad \hspace{0.9cm} x \in \mathbb{R}.
\end{align}
The case of double sines differs only by constants, and the case of gamma functions differs by constants and subleading polynomial growth. Nevertheless, in all three cases these asymptotics and bounds allow to degenerate the kernel of $Q$-operator to the kernel of $\Lambda$-operator in some limit.

For $n = 2$ in the case of hyperbolic functions the kernels are
\begin{align*}
Q(x_1, x_2, y_1, y_2; \l) &= e^{i \l(x_1 + x_2 - y_1 - y_2)}  \ch^{-g}(x_1 - y_1)  \ch^{-g}(x_2 - y_1) \\[6pt]
& \times  \ch^{-g}(x_1 - y_2)  \ch^{-g}(x_2 - y_2)  \sh^{2g}|y_2 - y_1|, \\[8pt]
\Lambda(x_1, x_2, y_1; \l) &= e^{i \l(x_1 + x_2 - y_1)}  \ch^{-g}(x_1 - y_1)  \ch^{-g}(x_2 - y_1).
\end{align*}
They are related in the limit $y_2 \rightarrow \infty$
\begin{equation}\label{Q-lim}
\lim_{y_2 \rightarrow \infty} e^{gy_1 + i \lambda y_2} \, Q(x_1, x_2, y_1, y_2; \l) = \Lambda(x_1, x_2, y_1; \l - ig).
\end{equation}
Note the shift by $-ig$ in the variable $\lambda$ from the right.

The same reduction can be applied to the commutation relation of $Q$-operators
\begin{equation*}
Q_2(\l) \, Q_2(\rho) = Q_2(\rho) \, Q_2(\l).
\end{equation*}
Denote by $\mathcal{Q}$ the kernel of two $Q$-operators product
\begin{equation*}
\mathcal{Q}(x_1, x_2, z_1, z_2; \l, \rho) = \int_{\mathbb{R}^2} dy_1 dy_2 \, Q(x_1, x_2, y_1, y_2; \l) \, Q(y_1, y_2, z_1, z_2; \rho).
\end{equation*}
Then the commutativity is equivalent to the integral identity
\begin{equation}\label{QQ-ker}
\mathcal{Q}(x_1, x_2, z_1, z_2; \l, \rho) = \mathcal{Q}(x_1, x_2, z_1, z_2; \rho, \l).
\end{equation}
In complete analogy with the limit \eqref{Q-lim}, we consider the limit of the left-hand side of identity~\eqref{QQ-ker}
\begin{equation}\label{QQ-lim}
\lim_{z_2 \rightarrow \infty} e^{gz_1 + i \rho z_2} \, \mathcal{Q}(x_1, x_2, z_1, z_2; \l, \rho) = \mathcal{L}(x_1, x_2, z_1; \l, \rho - ig)
\end{equation}
and arrive at the integral $\mathcal{L}$ that represents the kernel of the product $Q_2(\l) \, \Lambda_2(\rho - ig)$
\begin{equation*}
\mathcal{L}(x_1, x_2, z_1; \l, \rho - ig) = \int_{\mathbb{R}^2} dy_1 dy_2 \, Q(x_1, x_2, y_1, y_2; \l) \, \Lambda(y_1, y_2, z_1; \rho - ig).
\end{equation*}
Note again the same shift by $-ig$. The only subtle point is the interchange of the limit and the integral. To justify it, use bounds \eqref{Kmu-b} and dominated convergence theorem. We also remark that if the original integral $\mathcal{Q}$ is absolutely convergent for $\Im(\l - \rho) \in (-2g, 2g)$, the integral $\mathcal{L}$ is absolutely convergent for $\Im(\l - \rho) \in (-2g, 0)$, so that we assume this condition when performing the limit.

The limit of the right-hand side of \eqref{QQ-ker} is more involved. After multiplying by the same function $e^{g z_1 + i \rho z_2}$, as in the limit \eqref{QQ-lim}, we have
\begin{equation}\label{QQ-r}
\begin{aligned}
e^{gz_1 + i \rho z_2} \, &\mathcal{Q}(x_1, x_2, z_1, z_2; \rho, \l) = \int\limits_{\mathbb{R}^2} dy_1 dy_2 \, e^{i \rho(x_1 + x_2 + z_2 - y_1 - y_2 ) + g z_1 + i \l(y_1 + y_2 - z_1 - z_2)} \\[6pt]
& \times \ch^{-g}(x_1 - y_1) \ch^{-g}(x_2 - y_1) \ch^{-g}(x_1 - y_2) \ch^{-g}(x_2 - y_2) \\[6pt]
& \times \ch^{-g}(y_1 - z_1) \ch^{-g}(y_2 - z_1) \ch^{-g}(y_1 - z_2) \ch^{-g}(y_2 - z_2) \\[6pt]
& \times \sh^{2g}|y_1 - y_2| \,  \sh^{2g}|z_1 - z_2|.
\end{aligned}
\end{equation}
The integrand contains exponent $e^{i (\rho - \l)z_2}$ and therefore doesn't have pointwise limit as $z_2 \rightarrow \infty$. However, this exponent can be canceled by the shift of variable $y_2 \rightarrow y_2 + z_2$. Besides, to use dominated convergence theorem we should have the integrand which can be bounded by integrable function independent of $z_2$ (for big enough $z_2$). To end up with such bound we notice that the integrand before the shift \eqref{QQ-r}, say $F(y_1, y_2)$, is symmetric with respect to $y_1, y_2$, and therefore the integration domain can be reduced to $y_2 > y_1$
\begin{equation*}
\int_{\mathbb{R}^2} dy_1 dy_2 \, F(y_1, y_2) = 2 \int_{y_2 > y_1} dy_1 dy_2 \, F(y_1, y_2).
\end{equation*}
Therefore, after the shift the integral can be rewritten as follows
\begin{equation*}
\begin{aligned}
e^{gz_1 + i \rho z_2} & \, \mathcal{Q}(x_1, x_2, z_1, z_2; \rho, \l) = 2 \hspace{-0.4cm} \int\limits_{y_2 + z_2 > y_1} \hspace{-0.4cm} dy_1 dy_2 \, e^{i \rho(x_1 + x_2  - y_1 - y_2 ) + g z_1 + i \l(y_1 + y_2 - z_1)} \\[6pt]
& \times \ch^{-g}(x_1 - y_1) \ch^{-g}(x_2 - y_1) \ch^{-g}(x_1 - y_2 - z_2) \ch^{-g}(x_2 - y_2 - z_2) \\[6pt]
& \times \ch^{-g}(y_1 - z_1) \ch^{-g}(y_2 + z_2 - z_1) \ch^{-g}(y_1 - z_2) \ch^{-g}(y_2) \\[6pt]
& \times \sh^{2g}|y_1 - y_2 - z_2| \,  \sh^{2g}|z_1 - z_2|.
\end{aligned}
\end{equation*}
Now due to the condition $y_2 + z_2 > y_1$ we, in particular, have the bound
\begin{equation*}
\sh^{2g}|y_1 - y_2 - z_2| \leq 2^{-2g} \, e^{ 2g (y_2 + z_2 - y_1)}
\end{equation*}
for all $y_1, y_2$ in the integration domain. It allows to obtain the uniform in $z_2$ bound for the integrand (using again \eqref{Kmu-b}). Note that in the limit $z_2 \rightarrow \infty$ the domain of integration becomes the whole $\mathbb{R}^2$. Then applying dominated convergence theorem
\begin{equation*}
\lim_{z_2 \rightarrow \infty} e^{gz_1 + i \rho z_2} \, \mathcal{Q}(x_1, x_2, z_1, z_2; \rho, \l) = \mathcal{R}(x_1, x_2, z_1; \rho - ig, \l)
\end{equation*}
in the limit we obtain the function $\mathcal{R}$ given by two separated integrals
\begin{multline*}
\mathcal{R}(x_1, x_2, z_1; \rho - ig, \l) = 2 \int_{\mathbb{R}} dy_2 \, e^{i(\l - \rho + ig) y_2} \, \ch^{-g}(y_2) \\[6pt]
\times \int_{\mathbb{R}} dy_1 \, e^{i(\rho - ig)(x_1 + x_2 - y_1) + i \l (y_1 - z_1)} \\[6pt]
\times \ch^{-g}(x_1 - y_1) \ch^{-g}(x_2 - y_1) \ch^{-g}(y_1 - z_1) .
\end{multline*}
The first one is just the beta integral \eqref{beta0}
\begin{align*}
\int_{\mathbb{R}} dy_2 \, e^{i(\l - \rho + ig) y_2} \, \ch^{-g}(y_2) = 
\frac{\Gamma(\frac{g + i(\l - \rho + ig)}{2})\,
\Gamma(\frac{g - i(\l - \rho + ig)}{2})}{2^{1-g} \Gamma(g)} =
q(\l,\rho - ig).
\end{align*}
The second integral coincides with the kernel of the operator $\Lambda_2(\rho - ig) \, Q_1(\l)$.

In the limit we have the equivalence of the kernels from the left and right-hand sides $\mathcal{L} = \mathcal{R}$ and therefore the equivalence of operators
\begin{equation*}
Q_2(\l) \, \Lambda_2(\rho - ig) = 2\, q(\l,\rho - ig)\, \Lambda_2(\rho - ig) \, Q_1(\l).
\end{equation*}
Shifting $\rho \rightarrow \rho + i g$ we arrive at the stated identity \eqref{QL-1}.

\subsection{Eigenfunctions}

The $Q\Lambda$-commutation relation allows to construct
eigenfunction of the $Q$-operator using corresponding $\Lambda$-operator.
Let us consider the example from the previous section.
We have two commutation relations
\begin{align}\label{QL2}
Q_2(\l) \, \Lambda_2(\lambda_2) &= 2\, q(\l,\lambda_2) \, \Lambda_2(\lambda_2) \, Q_1(\lambda), \\[6pt] \notag
Q_1(\lambda) \, \Lambda_1(\lambda_1) &= q(\l,\lambda_1) \, \Lambda_1(\lambda_1)
\end{align}
where the second one is the relation \eqref{QL0} from $n=1$ example.
Note that in the section $n=1$ for simplicity we omit index $n=1$
in notations of all operators but now we have to restore it.

Let us consider the following function
\begin{equation}\label{Psi2}
\begin{aligned}
\Psi_{\lambda_1,\lambda_2}(x_1, x_2) &= \Lambda_2(\lambda_2)\,\Lambda_1(\lambda_1) \\[6pt]
&= \int\limits_{-\infty}^{+\infty} d t\,
e^{i\lambda_2\left(x_1+x_2-t\right)}\,K(x_1-t)\,K(x_2-t)\,
e^{i\lambda_1 t}.
\end{aligned}
\end{equation}
It is easy to show using \eqref{QL2} that $\Psi_{\lambda_1\lambda_2}(x_1, x_2)$ is an eigenfunction of the operator $Q_2(\lambda)$ and calculate the corresponding eigenvalue. Indeed we have
\begin{align*}
Q_2(\lambda)\,\Lambda_2(\lambda_2)\,\Lambda_1(\lambda_1) & =
2\, q(\l,\lambda_2) \, \Lambda_2(\lambda_2) \, Q_1(\lambda)\,\Lambda_1(\lambda_1) \\[6pt]
& = 2\, q(\l,\lambda_2) \,q(\l,\lambda_1)\, \Lambda_2(\lambda_2) \,\Lambda_1(\lambda_1),
\end{align*}
or in more detailed notation
\begin{align}
[Q_2(\lambda)\,\Psi_{\lambda_1,\lambda_2}](x_1, x_2) =
2\, q(\l,\lambda_2) \,q(\l,\lambda_1)\,
\Psi_{\lambda_1,\lambda_2}(x_1, x_2).
\end{align}
It is clear that it is possible to repeat everything almost literally
and construct in the same way eigenfunctions of all $Q$-operators.
In this way we obtain eigenfunctions of the operator $\hat{Q}_2(x)$
\begin{align}\nonumber
&\Phi_{x_1, x_2}(\lambda_1, \lambda_2)  =
\hat{\Lambda}_2(x_2)\,\hat{\Lambda}_1(x_1) \\[6pt] \nonumber
& \hspace{2.3cm} = \int\limits_{-\infty}^{+\infty} \frac{d \gamma}{2\pi}\,
e^{i x_2\left(\lambda_1+\lambda_2-\gamma\right)}\,
\hat{K}(\lambda_1-\gamma)\,\hat{K}(\lambda_2-\gamma)\,
e^{i x_1 \gamma}, \\[6pt]
\label{Psi2h}
& [\hat{Q}_2(x)\,\Phi_{x_1, x_2}](\lambda_1, \lambda_2) =
2\,\hat{q}(x,x_2) \,\hat{q}(x,x_1)\,
\Phi_{x_1, x_2}(\lambda_1, \lambda_2)
\end{align}
and eigenfunctions of the operator $\hat{Q}_2(x|\bo)$
\begin{align}\notag
&\Phi_{x_1, x_2}(\lambda_1, \lambda_2|g,\bo) =
\hat{\Lambda}_2(x_2|\bo)\,\hat{\Lambda}_1(x_1|\bo) \\[6pt]\notag
& \hspace{3.15cm} = \int\limits_{-\infty}^{+\infty} d \gamma\,
e^{\frac{2\pi i}{\omega_1\omega_2} x_2\left(\lambda_1+\lambda_2-\gamma\right)}\,
K_g(\lambda_1-\gamma)\,K_g(\lambda_2-\gamma)\,
e^{\frac{2\pi i}{\omega_1\omega_2} x_1 \gamma},\\[6pt]
\label{Psi2bo}
&[\hat{Q}_2(x|\bo)\,\Phi_{x_1, x_2}](\lambda_1, \lambda_2|g,\bo) =
2\, \hat{q}(x,x_2|\bo) \,\hat{q}(x,x_1|\bo)\,
\Phi_{x_1, x_2}(\lambda_1, \lambda_2|g,\bo).
\end{align}

Using the Fourier transformation it is possible to establish connections
between eigenfunctions of dual operators
\begin{align} 
&\Psi_{\lambda_1,\lambda_2}(x_1, x_2) = \Phi_{x_1, x_2}(\lambda_1, \lambda_2),\\[6pt]
\label{dual}
&\Phi_{x_1, x_2}(\lambda_1, \lambda_2|g,\bo) =
S^2(g^*)\,\Phi_{\lambda_1, \lambda_2}(x_1,x_2|g^*,\bo).
\end{align}
Indeed the relation \eqref{F1} between $K(z)$ and $\KK(\lambda)$ and
\begin{align*}
\int\limits_{-\infty}^{+\infty}
\frac{d \lambda}{2\pi}\,e^{-i \lambda z}\,\KK(\lambda) = K(z)
\end{align*}
allows to transform the integral representation \eqref{Psi2} to the
integral representation \eqref{Psi2h}
\begin{multline}\label{KK1}
\int\limits_{-\infty}^{+\infty} d t\,
e^{i\lambda_2\left(x_1+x_2-t\right)}\,K(x_1-t)\,K(x_2-t)\,
e^{i\lambda_1 t} \\
= \int\limits_{-\infty}^{+\infty}
\frac{d \gamma}{2\pi}\,e^{i x_1(\lambda_1+\lambda_2-\gamma)}\,
\KK(\lambda_1-\gamma)\,\KK(\lambda_2-\gamma)\,e^{i x_2 \gamma}.
\end{multline}
In the same way relation \eqref{betah} between $K_g(z)$ and $K_{g^*}(\lambda)$
\begin{align*}
\int\limits_{-\infty}^{+\infty}
d z\,e^{\frac{2\pi i x z}{\omega_1\omega_2}}\,
K_g(z)  = \sqrt{\omega_1\omega_2}\,S(g^*)\,\KKK_{g^*}(x)
\end{align*}
allows to prove similar relation
\begin{multline}\label{KK2}
\int\limits_{-\infty}^{+\infty} d t\,
e^{\frac{2\pi i}{\omega_1\omega_2}\lambda_2\left(x_1+x_2-t\right)}\,K_g(x_1-t)\,K_g(x_2-t)\,
e^{\frac{2\pi i}{\omega_1\omega_2}\lambda_1 t}  \\ 
= S^2(g^*)\int\limits_{-\infty}^{+\infty}
d \gamma\,e^{\frac{2\pi i}{\omega_1\omega_2} x_1(\lambda_1+\lambda_2-\gamma)}\,
K_{g^*}(\lambda_1-\gamma)\,K_{g^*}(\lambda_2-\gamma)\,e^{\frac{2\pi i}{\omega_1\omega_2} x_2 \gamma}.
\end{multline}

Consider the first relation \eqref{KK1}. In the left hand side of \eqref{KK1} substitute $K(x_k-t)$ with their Fourier representations,
then $t$-integral gives $\delta$-function, which can be simply integrated; after that change the remaining integration variable $\gamma_1 = \gamma-\lambda_1$ and use $\hat{K}(\lambda)=\hat{K}(-\lambda)$:
\begin{align*}
&\int\limits_{-\infty}^{+\infty} d t\,
e^{i\lambda_2\left(x_1+x_2-t\right)+i\lambda_1 t}
\int\limits_{-\infty}^{+\infty}
\frac{d \gamma_1}{2\pi}\,e^{-i \gamma_1 (x_1-t)}\,\KK(\gamma_1)\,
\int\limits_{-\infty}^{+\infty}
\frac{d \gamma_2}{2\pi}\,e^{-i \gamma_2 (x_2-t)}\,\KK(\gamma_2) \\[6pt]
& = \int\limits_{-\infty}^{+\infty}
\frac{d \gamma_1}{2\pi}\,\frac{d \gamma_2}{2\pi}\,
\KK(\gamma_1)\,\KK(\gamma_2)\,
e^{i x_1(\lambda_2-\gamma_1) + i x_2(\lambda_2-\gamma_2)}\,
2\pi\delta(\lambda_1-\lambda_2+\gamma_1+\gamma_2) \\[6pt]
& = \int\limits_{-\infty}^{+\infty}
\frac{d \gamma_1}{2\pi}\,
\KK(\gamma_1)\,\KK(\lambda_2-\lambda_1-\gamma_1)\,
e^{i x_1(\lambda_2-\gamma_1) + i x_2(\lambda_1+\gamma_1)} \\[6pt]
& = \int\limits_{-\infty}^{+\infty}
\frac{d \gamma}{2\pi}\,
\KK(\lambda_1-\gamma)\,\KK(\lambda_2-\gamma)\,
e^{i x_1(\lambda_1+\lambda_2-\gamma) + i x_2 \gamma}.
\end{align*}
The relation \eqref{KK2} can be proved in a similar way.

Let us summarize our results:
\begin{itemize}
\item
The relation $\Psi_{\lambda_1,\lambda_2}(x_1, x_2) = \Phi_{x_1, x_2}(\lambda_1, \lambda_2)$
states the equivalence of the Hallnas-Ruijsenaars and Mellin-Barnes representations for the eigenfunctions
\begin{align*}
\Psi_{\lambda_1,\lambda_2}(x_1, x_2) & =
\int\limits_{-\infty}^{+\infty} d t\,
\frac{e^{i\lambda_1\left(x_1+x_2-t\right)}\,
e^{i\lambda_2 t}}{\mathrm{ch}^g(x_1-t)\,\mathrm{ch}^g(x_2-t)} \\[6pt]
& = \int\limits_{-\infty}^{+\infty} \frac{d\gamma}{2\pi}\,
\frac{\Gamma\left(\frac{g+i(\lambda_1-\gamma)}{2}\right)
\Gamma\left(\frac{g+i(\gamma-\lambda_1)}{2}\right)}
{2^{1-g}\Gamma(g)}\, \\[6pt]
& \times \frac{\Gamma\left(\frac{g+i(\lambda_2-\gamma)}{2}\right)
\Gamma\left(\frac{g+i(\gamma-\lambda_2)}{2}\right)}
{2^{1-g}\Gamma(g)}\,e^{i x_1(\lambda_1+\lambda_2-\gamma)}\,e^{i x_2 \gamma}.
\end{align*}
\item The Hallnas-Ruijsenaars representation is evidently has symmetry $x_1 \rightleftarrows x_2$, and the Mellin-Barnes representation is invariant under 
$\lambda_1 \rightleftarrows\lambda_2$
$$
\Psi_{\lambda_1,\lambda_2}(x_1,x_2) = \Psi_{\lambda_1,\lambda_2}(x_2,x_1) = \Psi_{\lambda_2,\lambda_1}(x_1,x_2).
$$
\end{itemize}

\subsubsection{Mellin transform} \label{sec:mell}
In this subsection we demonstrate another proof of the relation
\begin{equation*}
\Psi_{\l_1, \l_2}(x_1, x_2) = \Phi_{x_1,x_2}(\l_1, \l_2)
\end{equation*}
using Mellin transform. First, we separate variables in both functions shifting integration variables $y \rightarrow y + (x_1 + x_2)/2$ and $\gamma \rightarrow \gamma + (\l_1 + \l_2)/2$ in the corresponding representations
\begin{align}\notag 
\Psi_{\l_1, \l_2}(x_1, x_2) =  e^{i \frac{\l_1 + \l_2}{2}(x_1 +x_2)} \int\limits_{-\infty}^\infty & dy \, K \Bigl(\frac{x_1 - x_2}{2} - y \Bigr) \\[6pt] \label{Psi-fact0}
&\times K \Bigl(\frac{x_2 - x_1}{2} - y \Bigr) \, e^{i (\l_1 - \l_2) y },\\[8pt] \notag
\Phi_{x_1,x_2}(\l_1, \l_2) = e^{i \frac{\l_1 + \l_2}{2}(x_1 +x_2)}  \int\limits_{-\infty}^\infty & \frac{d\gamma}{2\pi} \, \hat{K} \Bigl( \frac{\l_1 - \l_2}{2} - \gamma \Bigr) \\[6pt] \notag
&\times  \hat{K} \Bigl(\frac{\l_2 - \l_1}{2} - \gamma \Bigr) \, e^{i \gamma (x_1 - x_2)}.
\end{align}
The exponents behind the integrals coincide, so it is left to prove the equivalence of integrals. Let $x = x_1 - x_2$ and $\l = (\l_1 - \l_2)/2$. The second integral in explicit form
\begin{equation}\label{phi}
\begin{aligned}
\phi_x(\l) = \frac{2^{2g - 2}}{\Gamma^2(g)} \int\limits_{-\infty}^\infty &\frac{d\gamma}{2\pi} \, \Gamma \Bigl(\frac{i\l - i \gamma + g}{2}\Bigr) \, \Gamma \Bigl(\frac{-i\l + i \gamma + g}{2}\Bigr) \\[6pt]
&\times \Gamma \Bigl(\frac{-i\l - i \gamma + g}{2}\Bigr) \, \Gamma \Bigl(\frac{i\l + i \gamma + g}{2}\Bigr) \, e^{i \gamma x}.
\end{aligned}
\end{equation}
Next denote by $M[f](s)$ and $M'[F](z)$ direct and inverse Mellin transforms respectively
\beqq
 M[f](s)=\int\limits_0^\infty \frac{dz}{z} \, z^s \, f(z),\qquad M'[F](z)=\frac{1}{2\pi i}\int\limits_{c-i\infty}^{c+i\infty }ds \, z^{-s} \, F(s).
\eeqq
We also write these relations in the form
 \beqq
 f(z)\mellin M[f](s),\qquad F(s)\mellininverse M'[F](z).
 \eeqq
Then the integral \eqref{phi} after change of variable $s = i \gamma/2$ can be written as the following inverse Mellin transform
\begin{multline} \label{phi-M}
\phi_x(\l) = \frac{2^{2g - 1}}{\Gamma^2(g)} \, M'\biggl[ \Gamma \Bigl(\frac{i\l+ g}{2} - s\Bigr) \, \Gamma \Bigl(\frac{-i\l + g}{2} + s\Bigr) \\[6pt]
\times  \Gamma \Bigl(\frac{-i\l + g}{2} - s\Bigr) \, \Gamma \Bigl(\frac{i\l + g}{2} + s\Bigr) \biggr] (e^{-2x}).
\end{multline}
The integration contour lies in the strip $|\Re s|<g/2$.

Beta integral, written in a form
\beqq
\int\limits_0^\infty \frac{dz}{z} \, \frac{z^a}{(1+z)^{a+b}}=\frac{\Gamma(a)\Gamma(b)}{\Gamma(a+b)}
\eeqq
says that the Mellin transform	 	
of the function $\dfrac{z^a}{(1+z)^{a+b}}$ is
\beqq \dfrac{z^a}{(1+z)^{a+b}}\mellin \frac{1}{\Gamma(a+b)}\Gamma(a-s)\Gamma(b+s),
\eeqq	
so that the inverse formula reads as
\beq \label{1}	
\frac{1}{\Gamma(a+b)} \Gamma(a-s)\Gamma(b+s)\mellininverse \dfrac{z^a}{(1+z)^{a+b}}.
\eeq	 	
The integration contour lies in the strip $-\Re b< \Re s<\Re a$.

Now we use the property
\beq \notag \label{4}f*g(z)\mellin F[s]\cdot G[s] \eeq
where
$$f*g(z)=\int\limits_0^\infty \frac{dt}{t} \, f(t)g(z/t).$$
Due to \eqref{phi-M} and  \rf{1} it says that
\begin{equation*}
\phi_x(\l) = 2^{2g - 1} \, \frac{z^{\frac{i\l+g}{2}}}{(1+z)^g} \ast \frac{z^{\frac{-i\l+g}{2}}}{(1+z)^g} \, (e^{-2x}).
\end{equation*}
Let us write the last formula in explicit form
\begin{equation*}
\phi_x(\l) = 2^{2g - 1} \int\limits_0^\infty \frac{dt}{t} \, \frac{t^{\frac{i\l+g}{2}}}{(1+t)^g} \, \frac{(e^{-2x}/ t)^{\frac{-i\l+g}{2}}}{(1+e^{-2x}/ t)^g}.
\end{equation*}
After the change of integration variable $t = e^{2y - x}$ this integral takes the form
\begin{equation*}
\phi_x(\l) = \int\limits_{-\infty}^\infty dy \, \frac{e^{2 i \l y}}{\ch^g(x/2 - y) \, \ch^g(x/2 + y)}
\end{equation*}
which coincides with the integral in \eqref{Psi-fact0} after identifying $x = x_1 - x_2$, $\l = (\l_1 - \l_2)/2$.

\subsubsection{Equivalence through dual $Q$-operators}

There is yet another way to establish the equivalence of two integral representations
\begin{equation*}
\Psi_{\l_1, \l_2}(x_1, x_2) = \Lambda_2(\l_2) \, e^{i \lambda_1 x_1} = \hat{\Lambda}_2(x_2) \, e^{i \lambda_1 x_1}
\end{equation*}
using both dual $Q$-operators. This particular way can be generalized to the case of $n$ particles, this is done for relativistic case in \cite{BDKK2}. First, note that the raising and Baxter operators are connected as
\begin{equation*}
\Lambda_2(\l_2) = e^{i \l_2 x_2} \, Q_1(\l_2) \, K(x_2 - x_1), \qquad  \hat{\Lambda}_2(x_2) = e^{i \l_2 x_2} \, \hat{Q}_1(x_2) \, \hat{K}(\l_2 - \l_1).
\end{equation*}
Recall also how one-particle $Q$-operators act on plane waves
\begin{equation*}
\hat{Q}_1(x_2) \, e^{i \l_1 x_1} = K(x_2 - x_1) \, e^{i \l_1 x_1}, \qquad Q_1(\l_2) \, e^{i \l_1 x_1} = \hat{K}(\l_2 - \l_1) \, e^{i \l_1 x_1}.
\end{equation*}
Hence, the first integral representation can be written as
\begin{equation*}
\begin{aligned}
\Lambda_2(\l_2) \, e^{i \lambda_1 x_1} &= e^{i \l_2 x_2} \, Q_1(\l_2) \, K(x_2 - x_1) \, e^{i \lambda_1 x_1} \\[6pt]
&= e^{i \l_2 x_2} \, Q_1(\l_2) \, \hat{Q}_1(x_2) \, e^{i \lambda_1 x_1}.
\end{aligned}
\end{equation*}
Since dual $Q$-operators act on different variables, they can be interchanged (the corresponding double integral is absolutely convergent). After that a similar chain of equations leads to the claim
\begin{equation*}
\begin{aligned}
\Lambda_2(\l_2) \, e^{i \lambda_1 x_1} &= e^{i \l_2 x_2} \,  \hat{Q}_1(x_2) \, Q_1(\l_2) \, e^{i \lambda_1 x_1} \\[6pt]
&= e^{i \l_2 x_2} \,  \hat{Q}_1(x_2) \, \hat{K}(\l_2 - \l_1) \, e^{i \lambda_1 x_1} = \hat{\Lambda}_2(x_2) \, e^{i \lambda_1 x_1}.
\end{aligned}
\end{equation*}

\subsection{Scalar product}

As it was outlined in introduction, if we trasform $Q$-operator's eigenfunction $\Psi_{\l_1, \l_2}$ \eqref{Psi2}
\begin{equation*}
\Psi^S_{\l_1, \l_2}(x_1,x_2) = \sh^{g}|x_1 - x_2| \, \Psi_{\l_1, \l_2}(x_1,x_2)
\end{equation*}
it becomes an eigenfunction of the Sutherland Hamiltonian $H_S$
\begin{equation*}
H_S \, \Psi^S_{\l_1, \l_2} = (\l_1^2 + \l_2^2) \, \Psi^S_{\l_1, \l_2}, \qquad H_S = -\partial_{x_1}^2 - \partial_{x_2}^2 + \frac{2g(g - 1)}{\sh^2(x_1 - x_2)},
\end{equation*}
which is symmetric with respect to the measure $dx_1 dx_2$. Therefore, the scalar product between $Q$-operator's eigenfunctions contains a nontrivial measure $\sh^{2g}|x_1 - x_2| dx_1 dx_2$
\begin{multline}\label{Psi-scalar}
\int\limits_{-\infty}^{+\infty}
d x_1 d x_2\, \overline{\Psi^S_{\lambda_1,\lambda_2}(x_1, x_2)}\,\Psi^S_{\rho_1,\rho_2}(x_1, x_2) \\
= \int\limits_{-\infty}^{+\infty}
d x_1 d x_2\, \mathrm{sh}^{2g}|x_1-x_2|\,
\overline{\Psi_{\lambda_1,\lambda_2}(x_1, x_2)}\,\Psi_{\rho_1,\rho_2}(x_1, x_2).
\end{multline}
The goal of this section is to calculate this scalar product by two methods: the standard method from textbooks and using the $Q$-operator. Furthermore, using dual operators $\hat{Q}(x)$ and $\hat{Q}(x|\bo)$ we calculate the corresponding scalar products between their eigenfunctions $\Phi_{x_1, x_2}(\l_1, \l_2)$ and $\Phi_{x_1, x_2}(\l_1, \l_2| g, \bo)$ from the previous section.

\subsubsection{Standard quantum mechanical calculation}
In this subsection we calculate the scalar product between the transformed functions $\Psi^S_{\l_1, \l_2}$ following the way written in textbooks (for example see \cite{FY}, \S36). It says that once we have two  functions $\psi_1(x)$ and $\psi_2(x)$,
which are the eigenfunctions  of the Hamiltonian on the line with real potential
\beqq\label{scal1}H=-\frac{d^2}{dx^2}+V(x),\qquad H\psi_1=k_1^2\psi_1,\qquad H\psi_2=k_2^2\psi_2,\eeqq
then multpying the first equation in the system
\beq \notag
\begin{aligned}
	-\bar{\psi}_1'' + V \bar{\psi}_1 &= k_1^2 \bar{\psi}_1,  \\[6pt]
	-\psi_2'' + V \psi_2 &= k_1^2 \psi_2 ,
\end{aligned}
\eeq
by $\psi_2$, the second by $\bar\psi_1$ and subtracting them we get the equality
$$(k_1^2-k_2^2)\bar\psi_1 \psi_2=\bar\psi_1 {\psi}''_2-\bar\psi''_1 {\psi}_2=(\bar\psi_1 {\psi}'_2-\bar\psi'_1 {\psi}_2)'.$$
Integrating it we arrive at the scalar product of eigenfunctions
\beq \label{scal3} \int\limits_{-\infty}^\infty dx \, \bar\psi_1{\psi}_2 =\lim_{x\to\infty}
\frac{W(\bar\psi_1,{\psi}_2)\big\vert_{-x}^{x}}{k_1^2-k_2^2}
\eeq
where we introduced the Wronskian
\beq \notag W(\bar\psi_1,{\psi}_2)=\bar\psi_1 {\psi}'_2-\bar\psi'_1 {\psi}_2.\eeq
Calculation of the scalar product therefore reduces to the calculation of Wronskian asymptotics.

Returning to our case, first, we use Mellin-Barnes representation of the eigenfunction and factor out the ``center of mass'' part changing the integration variable $\gamma \rightarrow \gamma + ( \l_1 + \l_2 ) / 2$
\begin{equation*}
\begin{aligned}
& \Psi^S_{\l_1, \l_2}(x_1, x_2) = \sh^{g}|x_1 - x_2| \, \int\limits_{-\infty}^\infty \frac{d\gamma}{2\pi} \, \hat{K}(\l_1 - \gamma) \, \hat{K}(\l_2 - \gamma) \, e^{i (\l_1 + \l_2 - \gamma) x_2 + i \gamma x_1} \\[6pt]
& = e^{i \frac{\l_1 + \l_2}{2}(x_1 +x_2)} \, \sh^{g}|x_1 - x_2|  \int\limits_{-\infty}^\infty \frac{d\gamma}{2\pi} \, \hat{K} \Bigl( \frac{\l_1 - \l_2}{2} - \gamma \Bigr) \\[6pt]
& \hspace{5.9cm} \times \hat{K} \Bigl(\frac{\l_2 - \l_1}{2} - \gamma \Bigr) \, e^{i \gamma (x_1 - x_2)}.
\end{aligned}
\end{equation*}
Introducing
\begin{equation*}
\psi_\l(x) = \sh^g |x| \int\limits_{-\infty}^\infty \frac{d\gamma}{2\pi} \, \hat{K}(\l - \gamma) \, \hat{K}(- \l - \gamma) \, e^{i \gamma x}
\end{equation*}
we separate the variables in the eigenfunction
\begin{equation}\label{Psi-fact}
\Psi^S_{\l_1, \l_2} (x_1, x_2)  = e^{i \frac{\l_1 + \l_2}{2}(x_1 +x_2)} \, \psi_{\frac{\l_1 - \l_2}{2}}(x_1 - x_2).
\end{equation}
The scalar product between plane waves is known. Hence, the scalar product between $\Psi^S_{\l_1,\l_2}$ reduces to the product between functions $\psi_\l$, which solve the equation
\begin{equation*}
\biggl( -\partial^2_x + \frac{g(g - 1)}{\sh^2 x} \biggr) \psi_\l(x) = \l^2 \psi_\l(x).
\end{equation*}
By \eqref{scal3} the latter product can be evaluated through the Wronskian asymptotics. Note that $\psi_\l(x)$ is an even function of $x$ so that we only need its asymptotics as $x \rightarrow \infty$
\begin{equation*}
\int\limits_{-\infty}^\infty dx \, \overline{\psi_\l (x)} \, \psi_\rho(x) = 2 \lim_{x \rightarrow \infty} \frac{W(\psi_\l(x), \psi_\rho(x))}{\l^2 - \rho^2}.
\end{equation*}
In the last expression we also used the fact that the eigenfunction is real $\overline{\psi_\l} = \psi_\l$.

The asymptotic of $\psi_\l(x)$ as $x \rightarrow \infty$ can be deduced from the asymptotic of two-particle eigenfunction \eqref{Psi-as} calculated in introduction. It is given by
\begin{equation*}
\psi_\l(x) = \frac{2^{g - 1}}{\Gamma(g)} \, \Bigl[ \, \Gamma(i \l) \, \Gamma(g - i \l) \, e^{i \l x} +  \Gamma(-i \l) \, \Gamma(g + i \l)\, e^{- i \l x} \,\Bigr](1 + O(e^{-2x})).
\end{equation*}
Consequently, its derivative has asymptotic
\begin{equation*}
\psi'_\l(x) = \frac{2^{g - 1}}{\Gamma(g)} \, i \l \Bigl[\,  \Gamma(i \l) \, \Gamma(g - i \l) \, e^{i \l x} -  \Gamma(-i \l) \, \Gamma(g + i \l) \, e^{- i \l x} \, \Bigr](1 + O(e^{-2x})).
\end{equation*}
Combining them we calculate the asymptotic of the Wronskian
\begin{equation*}
\begin{aligned}
2\frac{W(\psi_\l(x), \psi_\rho(x))}{\l^2 - \rho^2} &= \frac{2^{2g-1}}{i\Gamma^2(g)} \, \Gamma(i \l) \, \Gamma(-i\l) \, \Gamma(g + i \l) \, \Gamma(g - i\l) \\[6pt]
&\times \Biggl( \frac{e^{i(\l - \rho)x} - e^{i(\rho - \l)x}}{\l - \rho} + \frac{e^{i(\l + \rho)x} - e^{-i(\l + \rho)x}}{\l + \rho} + o(x) \Biggr)
\end{aligned}
\end{equation*}
where by $o(x)$ we mean terms which tend to zero in a sense of distributions of $\l, \rho$ as $x \rightarrow \infty$. Using the well-known identity
\begin{equation*}
\lim_{x \rightarrow \infty} \frac{\sin kx}{k} = \pi \delta(k)
\end{equation*}
we arrive at the expression for the scalar product
\begin{equation*}
\int\limits_{-\infty}^\infty dx \, \overline{\psi_\l (x)} \, \psi_\rho(x) = \frac{2^{2g}\pi}{\Gamma^2(g)} \, \Gamma(i \l) \, \Gamma(-i\l) \, \Gamma(g + i \l) \, \Gamma(g - i\l) \, \bigl( \delta(\l - \rho) + \delta(\l + \rho) \bigr).
\end{equation*}
Finally, using it together with the factorization \eqref{Psi-fact} we evaluate the scalar product between the original eigenfunctions
\begin{multline}\label{scal-qm}
\int\limits_{-\infty}^\infty dx_1  dx_2 \, \overline{\Psi^S_{\l_1, \l_2}(x_1, x_2)} \Psi^S_{\rho_1, \rho_2}(x_1, x_2) \\
= C\, \bigl( \delta(\l_1 - \rho_1) \delta(\l_2 - \rho_2) + \delta(\l_1 - \rho_2) \delta(\l_2 - \rho_1)\bigr)
\end{multline}
where the coefficient is given by
\begin{equation*}
C = \frac{2^{2g+1} \pi^2}{\Gamma^2(g)} \, \Gamma \Bigl( \frac{i \l_1 - i \l_2}{2} \Bigr) \, \Gamma \Bigl( g + \frac{i \l_1 - i \l_2}{2} \Bigr) \, \Gamma \Bigl( \frac{i \l_2 - i \l_1}{2} \Bigr)  \, \Gamma \Bigl( g + \frac{i \l_2 - i \l_1}{2} \Bigr).
\end{equation*}

\subsubsection{Eigenfunctions of the operator $Q(\l)$}

In this subsection we calculate the same scalar product \eqref{Psi-scalar} using the fact that $\Psi_{\l_1, \l_2}$ is an eigenfunction of the operator $Q_2(\l)$. This calculation is universal in two ways. Firstly, translation to the operators $\hat{Q}_2(x)$, $\hat{Q}_2(x|\bo)$ and scalar products between their eigenfunctions reduces to different expressions for the main building blocks --- the kernel and measure functions $K$ and $\mu$. Secondly, this calculation can be generalized to the general case of $n$ particles, as we will show in our future work.

Expressions for the eigenfunction and conjugated eigenfunction have the form
\begin{align*}
\Psi_{\rho_1,\rho_2}(x_1, x_2) &=
\int\limits_{-\infty}^{+\infty} d t\,
e^{i\rho_1\left(x_1+x_2-t\right)}\,e^{i\rho_2 t}
\,K(x_1-t)\,K(x_2-t), \\[6pt]
\overline{\Psi_{\lambda_1,\lambda_2}(x_1, x_2)} &=
\int\limits_{-\infty}^{+\infty} d t\,
e^{-i\lambda_1\left(x_1+x_2-t\right)}\,e^{-i\lambda_2 t}\,
K(x_1-t)\,K(x_2-t).
\end{align*}
and integral over $t$ absolutely converges for $g > 0 $.
Denote $x_{12} = x_1 - x_2$. The scalar product is
\begin{multline*}
\langle\Psi_{\lambda_1,\lambda_2}|\Psi_{\rho_1,\rho_2}\rangle =
\int\limits_{-\infty}^{+\infty}
d x_1 d x_2\, \mathrm{sh}^{2g}|x_{12}|\,
\overline{\Psi_{\lambda_1,\lambda_2}(x_1, x_2)}\,\Psi_{\rho_1,\rho_2}(x_1, x_2) \\
= \int\limits_{-\infty}^{+\infty}
d x_1 d x_2\, \mathrm{sh}^{2g}|x_{12}|\,
\int\limits_{-\infty}^{+\infty} d t_1\,
\int\limits_{-\infty}^{+\infty} d t_2\,
e^{-i\lambda_1\left(x_1+x_2-t_1\right)} \\
\times e^{-i\lambda_2 t_1}
\,e^{i\rho_1\left(x_1+x_2-t_2\right)}\,e^{i\rho_2 t_2}
\prod_{i,j=1}^{2} K(x_i-t_j).
\end{multline*}
Note that the ordering of integrals is fixed from the very beginning:
at the first step we integrate over $t_1$ and $t_2$
(absolutely convergent integrals) and then integrate over $x_1,x_2$.
The $x_1,x_2$-integral separately
\begin{align*}
\int\limits_{-\infty}^{+\infty}
d x_1 d x_2\,
\frac{e^{i(\rho_1-\lambda_1)\left(x_1+x_2\right)}\sh^{2g}|x_1-x_2|}
{\ch^g(x_1-t_1)\ch^g(x_2-t_1)\ch^g(x_1-t_2)\ch^g(x_2-t_2)}
\end{align*}
doesn't converge.

Let us use regularize it adding external
point $t_0$ and small $\varepsilon >0 $ in exponent 
\begin{multline*}
\int\limits_{-\infty}^{+\infty}
d x_1 d x_2\,
e^{i(\rho_1-\lambda_1)\left(x_1+x_2\right)}\,
\mathrm{sh}^{2g}|x_{12}|\,
\prod_{i,j=1}^{2} K(x_i-t_j) \\
= \frac{1}{2^{2g}}\lim _{t_0 \to +\infty}\,
\lim _{\varepsilon \to 0}\,
\int\limits_{-\infty}^{+\infty}
d x_1 d x_2\,\mathrm{sh}^{2g}|x_{12}|\,
e^{-g(x_1-t_0)-g(x_2-t_0)} \\
\times e^{\varepsilon\left(x_1+x_2\right)}\,\,
e^{i(\rho_1-\lambda_1)\left(x_1+x_2\right)}\,
\prod_{i=1}^{2}\prod_{j=0}^{2} K(x_i-t_j)\,.
\end{multline*}
To prove that in the limit $t_0\to +\infty$ we reproduce
the initial integrand use the following asymptotic
\begin{align*}
\frac{1}{2^{2g}}\,K(x_1-t_0)\,K(x_2-t_0) \to
e^{g(x_1-t_0)+g(x_2-t_0)}, \qquad t_0 \rightarrow +\infty.
\end{align*}
The regularized $x_1,x_2$-integral is already convergent.
Furthermore, the chosen regularization completes the $\Lambda_2$-operator to
the corresponding $Q_2$-operator and this allows to calculate everything
in a closed form.
To do this we convert $t_2$-integral back to the eigenfunction
$\Psi_{\rho_1,\rho_2}(x_1, x_2)$ and rearrange exponents
\begin{multline*}
\langle\Psi_{\lambda_1\lambda_2}|\Psi_{\rho_1\rho_2}\rangle =
\frac{1}{2^{2g}}\lim _{t_0 \to +\infty}\,
\lim _{\varepsilon \to 0}\,e^{-i(\lambda_1+ig+i\varepsilon)t_0}\,
\int\limits_{-\infty}^{+\infty} d t_1\,
e^{-i(\lambda_2-ig+i\varepsilon)t_1}\,\\
\times \int\limits_{-\infty}^{+\infty}
d x_1 d x_2\,
e^{i(\lambda_1-ig+i\varepsilon)\left(t_1+t_0-x_1-x_2\right)}\,
\mathrm{sh}^{2g}|x_{12}| \\
\times \prod_{i=1}^{2} K(x_i-t_1)K(x_i-t_0)\Psi_{\rho_1,\rho_2}(x_1, x_2).
\end{multline*}
Now in the second line it is easy to recognize the action
of the $Q$-operator on the eigenfunction
\begin{multline*}
[Q_2(\lambda_1-ig+i\varepsilon)\Psi_{\rho_1,\rho_2}](t_1,t_0)  \\[6pt]
= \int\limits_{-\infty}^{+\infty}
d x_1 d x_2\,
e^{i(\lambda_1-ig+i\varepsilon)\left(t_1+t_0-x_1-x_2\right)}\,
\mathrm{sh}^{2g}|x_{12}| \qquad \\
\qquad \times \prod_{i=1}^{2} K(x_i-t_1)K(x_i-t_0)\Psi_{\rho_1,\rho_2}(x_1, x_2) \\[6pt]
= 2\,q(\lambda_1-ig+i\varepsilon,\rho_1)\,
q(\lambda_1-ig+i\varepsilon,\rho_2)\,
\Psi_{\rho_1,\rho_2}(t_1, t_0).
\end{multline*}
So, we have managed to calculate $x_1,x_2$-integrals in explicit form and
it remains to calculate $t_1,t_2$-integrals
\begin{multline*}
e^{-i(\lambda_1+ig+i\varepsilon)t_0}\,
\int\limits_{-\infty}^{+\infty} d t_1\,
e^{-i(\lambda_2-ig+i\varepsilon)t_1}\,\Psi_{\rho_1,\rho_2}(t_1, t_0) \\
= e^{-i(\lambda_1+ig+i\varepsilon)t_0}\,
\int\limits_{-\infty}^{+\infty} d t_1\,
e^{-i(\lambda_2-ig+i\varepsilon)t_1} \\
\times \int\limits_{-\infty}^{+\infty} d t_2\,
e^{i\rho_1(t_1+t_0-t_2)}\,
K(t_1-t_2) K(t_0-t_2)\, e^{i\rho_2 t_2}.
\end{multline*}
The $t_1$-integral reproduces the action
of the $Q$-operator on the eigenfunction $\Psi_{\rho_1}(t_1) = e^{i\rho_1 t_1}$
and can be calculated explicitly
\begin{align*}
[Q_1(\lambda_2-ig+i\varepsilon)\Psi_{\rho_1}](t_2) &=
\int\limits_{-\infty}^{+\infty}
d t_1\,
e^{i(\lambda_2-ig+i\varepsilon)\left(t_2-t_1\right)}\,
K(t_2-t_1)\,e^{i\rho_1 t_1} \\
&= q(\lambda_2-ig+i\varepsilon,\rho_1)\,e^{i\rho_1 t_2}.
\end{align*}
The last step --- calculation of the $t_2$-integral.
After simple rearrangements of the exponents we obtain
\begin{align*}
e^{-i(\lambda_1+\lambda_2-\rho_1+2 i \varepsilon)t_0}\,
\int\limits_{-\infty}^{+\infty} d t_2\,
e^{i(\lambda_2-ig+i\varepsilon)(t_0-t_2)}\,
K(t_0-t_2)\, e^{i\rho_2 t_2},
\end{align*}
so that the $t_2$-integral also gives the action
of the $Q$-operator on eigenfunction $\Psi_{\rho_2}(t_2) = e^{i\rho_2 t_2}$
\begin{align*}
\left[Q_1(\lambda_2-ig+i\varepsilon)\,\Psi_{\rho_2}\right](t_0) &=
\int\limits_{-\infty}^{+\infty} d t_2\,
e^{i(\lambda_2-ig+i\varepsilon)(t_0-t_2)}\,
K(t_0-t_2)\, e^{i\rho_2 t_2} \\
&= q(\lambda_2-ig+i\varepsilon,\rho_2)\,e^{i\rho_2 t_0}.
\end{align*}
Collecting everything together we obtain the following expression
for the regularized scalar product
\begin{equation*}
\langle\Psi_{\lambda_1,\lambda_2}|\Psi_{\rho_1,\rho_2}\rangle =
2^{1-2g}\lim _{t_0 \to +\infty}\,
\lim _{\varepsilon \to 0}\,
e^{-i(\lambda_1+\lambda_2-\rho_1-\rho_2+2 i \varepsilon)t_0}
\prod_{k,j=1}^{2} q(\lambda_k-ig+i\varepsilon,\rho_j).
\end{equation*}
It is left to show that the function on the right is a delta-sequence. Let us write it explicitly and factor out a singular part
\begin{align*}
&\langle\Psi_{\lambda_1,\lambda_2}|\Psi_{\rho_1,\rho_2}\rangle = 2^{1 - 2g}\lim _{t_0 \to +\infty}\,\lim _{\varepsilon \to 0}\,
\,e^{i(\rho_1+\rho_2-\lambda_1-\lambda_2) t_0} \\[8pt]
& \hspace{2.5cm} \times \prod_{i,k=1,2}
\frac{\Gamma\left(g+\frac{i(\lambda_i-\rho_k)-\varepsilon}{2}\right)
\,\Gamma\left(\frac{i(\rho_k-\lambda_i)+\varepsilon}{2}\right)}
{2^{1-g}\Gamma(g)} \\[6pt]
&= \frac{2^{2g - 3}}{\Gamma^4(g)} \prod_{i,k=1,2} \Gamma\left(g+\frac{i(\lambda_i-\rho_k)}{2}\right)  \lim _{t_0 \to +\infty}\,\lim _{\varepsilon \to 0}\,
e^{i(\rho_1+\rho_2-\lambda_1-\lambda_2)t_0} \\[8pt]
& \hspace{2.5cm} \times \prod_{i,k=1,2}
\Gamma\left(\frac{i(\rho_k-\lambda_i)+\varepsilon}{2}\right) \\[8pt]
&= \frac{2^{2g - 3}}{\Gamma^4(g)}\,
\prod_{i,k=1,2} \Gamma\left(g+\frac{i(\lambda_i-\rho_k)}{2}\right)
\Gamma\left(1+\frac{i(\rho_k-\lambda_i)}{2}\right) \\[8pt]
& \hspace{2.5cm} \times \lim _{t_0 \to +\infty}\,\lim _{\varepsilon \to 0}\,
\frac{e^{i(\rho_1+\rho_2-\lambda_1-\lambda_2)t_0}}
{\prod_{i,k=1,2}\frac{i(\rho_k-\lambda_i)+\varepsilon}{2}}.
\end{align*}
Next we transform the last expression to the more simple form
\begin{align*}
\lim _{t_0 \to +\infty}\,\lim _{\varepsilon \to 0}\,
\frac{e^{i(\rho_1+\rho_2-\lambda_1-\lambda_2)t_0}}
{\prod_{i,k=1,2}\frac{i(\rho_k-\lambda_i)+\varepsilon}{2}} =
\left(\frac{2}{i}\right)^4
\lim _{t_0 \to +\infty}\,\lim _{\varepsilon \to 0}\,
\frac{e^{i(\rho_1+\rho_2-\lambda_1-\lambda_2)t_0}}
{\prod_{i,k=1,2}(\rho_k-\lambda_i-i\varepsilon)}
\end{align*}
and use relation proven in Appendix \ref{App-delta}
\begin{multline}\label{n=2}
\lim _{t_0 \to +\infty}\,\lim _{\varepsilon \to 0}\,
\frac{e^{i(\rho_1+\rho_2-\lambda_1-\lambda_2)t_0}}
{\prod_{i,k=1,2}(\rho_k-\lambda_i-i\varepsilon)} \\[6pt]
= \frac{(2\pi)^{2}}{\left(\lambda_{1}-\lambda_{2}\right)^2}\,
\Bigl[\delta(\lambda_1-\rho_1)\,\delta(\lambda_2-\rho_2)+
\delta(\lambda_1-\rho_2)\,\delta(\lambda_2-\rho_1)\Bigl].
\end{multline}
Thus, for the scalar product we obtain
\begin{align*}
&\langle\Psi_{\lambda_1,\lambda_2}|\Psi_{\rho_1,\rho_2}\rangle  =
\frac{2^{2g - 3}}{\Gamma^4(g)} \Gamma^2(g)\,
\Gamma\left(g\pm\frac{i\lambda_{12}}{2}\right)
\Gamma\left(1\pm\frac{i\lambda_{12}}{2}\right)
\frac{2^4(2\pi)^{2}}{\lambda_{12}^2} \\[8pt]
&\qquad \qquad \qquad \times \Bigl[\delta(\lambda_1-\rho_1)\,\delta(\lambda_2-\rho_2)+
\delta(\lambda_1-\rho_2)\,\delta(\lambda_2-\rho_1)\Bigl]  \\[8pt]
&=\frac{2^{2g +1} \pi^2}{\Gamma^2(g)}\,
\Gamma\left(g\pm\frac{i\lambda_{12}}{2}\right)
\Gamma\left(\pm\frac{i\lambda_{12}}{2}\right)\,
 \\[6pt]
& \qquad \qquad \qquad \times \Bigl[\delta(\lambda_1-\rho_1)\delta(\lambda_2-\rho_2) + \delta(\lambda_1-\rho_2)\,\delta(\lambda_2-\rho_1)\Bigl].
\end{align*}
This result coincides with the one obtained by the standard method \eqref{scal-qm}.

\subsubsection{Eigenfunctions of the operator $\hat{Q}(x)$}
Now we perform all calculations from previous subsection
in the case of the dual $\QQ$-operators.
Expressions for the eigenfunction and conjugated eigenfunction have the form
\begin{align*}
&\Phi_{x_1, x_2}(\lambda_1, \lambda_2) =
\int\limits_{-\infty}^{+\infty} \frac{d \gamma}{2\pi}\,
e^{i x_1\left(\lambda_1+\lambda_2-\gamma\right)}\,e^{i x_2 \gamma}
\,\KK (\lambda_1-\gamma)\,\KK (\lambda_2-\gamma), \\
&\overline{\Phi_{y_1,y_2}(\lambda_1, \lambda_2)} =
\int\limits_{-\infty}^{+\infty} \frac{d \gamma}{2\pi}\,
e^{-iy_1\left(\lambda_1+\lambda_2-\gamma\right)}\,e^{-iy_2 \gamma}\,
\KK (\lambda_1-\gamma)\,\KK (\lambda_2-\gamma).
\end{align*}
The integral over $\gamma$ converges for $g > 0 $.
The scalar product is
\begin{multline*}
\langle\Phi_{y_1, y_2}|\Phi_{x_1, x_2}\rangle =
\int\limits_{-\infty}^{+\infty}
\frac{d \lambda_1}{2\pi} \frac{d \lambda_2}{2\pi}\, \mu(\lambda_1, \lambda_2)\,
\overline{\Phi_{y_1, y_2}(\lambda_1, \lambda_2)}\,
\Phi_{x_1, x_2}(\lambda_1, \lambda_2) \\[6pt]
= \int\limits_{-\infty}^{+\infty}
\frac{d \lambda_1}{2\pi} \frac{d \lambda_2}{2\pi}\, \mu(\lambda_1, \lambda_2)\,
\int\limits_{-\infty}^{+\infty} \frac{d \gamma_1}{2\pi}\,
\int\limits_{-\infty}^{+\infty} \frac{d \gamma_2}{2\pi}\,
e^{-i y_1\left(\lambda_1 + \lambda_2-\gamma_1\right)}\,e^{-i y_2 \gamma_1} \\[6pt]
\times e^{i x_1\left(\lambda_1 + \lambda_2-\gamma_2\right)}\,e^{i x_2 \gamma_2}
\,\prod_{i,j=1}^{2} \KK (\lambda_i-\gamma_j).
\end{multline*}
The integration measure \eqref{mu} 
\begin{equation*}
\mu(\l_1,\l_2) = \frac{[2^{1-g}\Gamma(g)]^2}
{\Gamma\left(g\pm\frac{i(\l_1-\l_2)}{2}\right)\Gamma\left(\pm\frac{i(\l_1-\l_2)}{2}\right)}.
\end{equation*}
Again $\gamma_1,\gamma_2$-integral is absolutely convergent, but
$\lambda_1,\lambda_2$-integral diverges
\begin{align*}
\int\limits_{-\infty}^{+\infty}
\frac{d \lambda_1}{2\pi} \frac{d \lambda_2}{2\pi}\,
e^{i(x_1-y_1)\left(\lambda_1+\lambda_2\right)}\,
\mu(\lambda_1\,,\lambda_2)\,
\prod_{i,j=1}^{2} \KK (\lambda_i-\gamma_j).
\end{align*}
We regularize it adding external
point $\gamma_0$ and small $\varepsilon >0 $ in exponent
\begin{multline*}
\int\limits_{-\infty}^{+\infty}
\frac{d \lambda_1}{2\pi} \frac{d \lambda_2}{2\pi}\,
e^{i(x_1-y_1)\left(\lambda_1+\lambda_2\right)}\,
\mu(\lambda_1,\lambda_2)\,
\prod_{i,j=1}^{2} \KK (\lambda_i-\gamma_j) \\
= \lim _{\gamma_0 \to +\infty}\,
\lim _{\varepsilon \to 0}\, \left[\frac{2\pi}{\Gamma(g)}\,\gamma_0^{g-1}\right]^{-2} \int\limits_{-\infty}^{+\infty}
\frac{d \lambda_1}{2\pi} \frac{d \lambda_2}{2\pi}\,\mu(\lambda_1,\lambda_2)\,
e^{-\frac{\pi}{2}(\lambda_1-\gamma_0)-\frac{\pi}{2}(\lambda_2-\gamma_0)} \\
\times e^{\varepsilon\left(\lambda_1+\lambda_2\right)} \, e^{i(x_1-y_1)\left(\lambda_1+\lambda_2\right)}\,
\prod_{i=1}^{2}\prod_{j=0}^{2} \KK (\lambda_i-\gamma_j).
\end{multline*}
In the limit $\gamma_0\to +\infty$ we reproduce
the initial integrand, because
\begin{align*}
&\KK (\lambda-\gamma_0) \to \frac{2\pi}{\Gamma(g)}\,
\gamma_0^{g-1}\,e^{\frac{\pi}{2}(\lambda-\gamma_0)}, \qquad \gamma_0 \rightarrow + \infty,
\end{align*}
so that
\begin{align*}
&\left[\frac{2\pi}{\Gamma(g)}\,\gamma_0^{g-1}\right]^{-2}
\,\KK (\lambda_1-\gamma_0)\,\KK (\lambda_2-\gamma_0) \to
e^{\frac{\pi}{2}(\lambda_1-\gamma_0)+\frac{\pi}{2}(\lambda_2-\gamma_0)}, \qquad \gamma_0 \rightarrow + \infty.
\end{align*}
Th regularized integral is absolutely convergent. Next we convert $\gamma_2$-integral back to the eigenfunction $\Phi_{x_1, x_2}(\lambda_1, \lambda_2)$ and rearrange some factors 
\begin{multline*}
\langle\Phi_{y_1, y_2}|\Phi_{x_1, x_2}\rangle =
\lim _{\gamma_0 \to +\infty}\,
\lim _{\varepsilon \to 0}\,
\left[\frac{2\pi}{\Gamma(g)}\,\gamma_0^{g-1}\right]^{-2}
\,e^{-i(y_1+\frac{i\pi}{2}+i\varepsilon)\gamma_0} \\[6pt]
\times \int\limits_{-\infty}^{+\infty} \frac{d \gamma_1}{2\pi}\,
e^{-i(y_2-\frac{i\pi}{2}+i\varepsilon)\gamma_1}\,
\int\limits_{-\infty}^{+\infty}
\frac{d \lambda_1}{2\pi} \frac{d \lambda_2}{2\pi}\,\mu(\lambda_1,\lambda_2)\,
e^{i(y_1-\frac{i\pi}{2}+i\varepsilon)\left(\gamma_1+\gamma_0-\lambda_1-\lambda_2\right)} \\
\times \prod_{i=1,2} \KK (\lambda_i-\gamma_1)\KK (\lambda_i-\gamma_0)\,
\Phi_{x_1, x_2}(\lambda_1, \lambda_2)
\end{multline*}
In the last line it is easy to recognize the action
of the $\QQ$-operator on eigenfunction $\Phi_{x_1 x_2}(\lambda_1\,, \lambda_2)$
\begin{multline*}
[\QQ_2(y_1-\textstyle\frac{i\pi}{2}+i\varepsilon)
\Phi_{x_1, x_2}](\gamma_1,\gamma_0)  \\[6pt]
= \int\limits_{-\infty}^{+\infty}
\frac{d \lambda_1}{2\pi} \frac{d \lambda_2}{2\pi}\,
e^{i(y_1-\frac{i\pi}{2}+i\varepsilon)\left(\gamma_1+\gamma_0-\lambda_1-\lambda_2\right)}\,
\mu(\lambda_1, \lambda_2) \qquad  \\[6pt]
\qquad \times \prod_{i=1,2} \KK (\lambda_i-\gamma_1)\KK (\lambda_i-\gamma_0)
\Phi_{x_1, x_2}(\lambda_1, \lambda_2) \\[6pt]
= \textstyle 2\,\hat{q}(y_1-\frac{i\pi}{2}+i\varepsilon,x_1)\,
\hat{q}(y_1-\frac{i\pi}{2}+i\varepsilon,x_2)\,
\Phi_{x_1, x_2}(\gamma_1, \gamma_0)\,.
\end{multline*}
Thus, we have managed to calculate $\lambda_1,\lambda_2$-integrals in explicit form and
it remains to calculate $\gamma_1,\gamma_2$-integrals
\begin{multline*}
e^{-i(y_1+\frac{i\pi}{2}+i\varepsilon)\gamma_0}\,
\int\limits_{-\infty}^{+\infty} \frac{d \gamma_1}{2\pi}\,
e^{-i(y_2-\frac{i\pi}{2}+i\varepsilon)\gamma_1}\,
\Phi_{x_1 ,x_2}(\gamma_1, \gamma_0) \\
= e^{-i(y_1+\frac{i\pi}{2}+i\varepsilon)\gamma_0}\,
\int\limits_{-\infty}^{+\infty} \frac{d \gamma_1}{2\pi}\,
e^{-i(y_2-\frac{i\pi}{2}+i\varepsilon)\gamma_1} \\
\times \int\limits_{-\infty}^{+\infty} \frac{d \gamma_2}{2\pi}\,
e^{i x_1(\gamma_1+\gamma_0-\gamma_2)}  \KK (\gamma_1-\gamma_2) \KK (\gamma_0-\gamma_2)\, e^{i x_2 \gamma_2}.
\end{multline*}
The $\gamma_1$-integral represents action
of the $\QQ$-operator on the eigenfunction $\Phi_{x_1}(\gamma_1)=e^{ix_1 \gamma_1}$
\begin{align*}
[\QQ_1(y_2-{\textstyle\frac{i\pi}{2}+i\varepsilon})\Phi_{x_1}](\gamma_2) & =
\int\limits_{-\infty}^{+\infty}
\frac{d \gamma_1}{2\pi}\,
e^{i(y_2-\frac{i\pi}{2}+i\varepsilon)\left(\gamma_2-\gamma_1\right)}\,
\KK (\gamma_2-\gamma_1)\,e^{ix_1 \gamma_1} \\
& = \hat{q}(y_2-\textstyle\frac{i\pi}{2}+i\varepsilon,x_1)\,e^{ix_1 \gamma_2}
\end{align*}
The last step is to calculate the $\gamma_2$-integral.
After simple rearrangements of the exponents we obtain
\begin{align*}
e^{-i(y_1+y_2-x_1+2 i \varepsilon)\gamma_0}\,
\int\limits_{-\infty}^{+\infty} \frac{d \gamma_2}{2\pi}\,
e^{i(y_2-\frac{i\pi}{2}+i\varepsilon)(\gamma_0-\gamma_2)}\,
\KK (\gamma_0-\gamma_2)\, e^{i x_2 \gamma_2}
\end{align*}
and it is again action of the $\QQ$-operator on
eigenfunction $\Phi_{x_2}(\gamma_2)=e^{ix_2 \gamma_2}$
\begin{align*}
[\QQ_1(y_2-{\textstyle\frac{i\pi}{2}+i\varepsilon})\Phi_{x_2}](\gamma_0) &=
\int\limits_{-\infty}^{+\infty} \frac{d \gamma_2}{2\pi}\,
e^{i(y_2-\frac{i\pi}{2}+i\varepsilon)(\gamma_0-\gamma_2)}\,
\KK (\gamma_0-\gamma_2)\, e^{i x_2 \gamma_2} \\
&= \hat{q}(y_2-\textstyle\frac{i\pi}{2}+i\varepsilon,x_2)\,e^{ix_1 \gamma_0}.
\end{align*}
Collecting everything together we obtain the following expression
for the scalar product
\begin{align*}
\langle\Phi_{y_1, y_2}|\Phi_{x_1, x_2}\rangle &=
\lim _{\gamma_0 \to +\infty}\,\lim _{\varepsilon \to 0}\,2
\left[\frac{2\pi}{\Gamma(g)}\,\gamma_0^{g-1}\right]^{-2} \\[6pt]
& \times e^{-i(y_1+y_2-x_1-x_2+2 i \varepsilon)\gamma_0}
\prod_{k,j=1}^{2} \hat{q}(y_k-\textstyle\frac{i\pi}{2}+i\varepsilon,x_j).
\end{align*}
Let us write it in explicit form and factor out a singular part
\begin{align*}
\langle\Phi_{y_1, y_2}|\Phi_{x_1, x_2}\rangle =2\,\frac{\Gamma^2(g)}{(2\pi)^2}\,e^{-2 \pi i g}\,
\lim_{\gamma_0 \to +\infty}\,\lim _{\varepsilon \to 0}\,
\frac{\gamma_0^{2(1-g)}\,e^{i(x_1+x_2-y_1-y_2)\gamma_0}}
{\prod_{k,j=1}^{2} \mathrm{sh}^g(x_j-y_k-i\varepsilon)} \\[6pt]
= 2\,\frac{\Gamma^2(g)}{(2\pi)^2}\,e^{-2 \pi i g}\,
\lim _{\varepsilon \to 0}\,
\prod_{k,j=1}^{2} \frac{(x_j-y_k-i\varepsilon)^g}
{\mathrm{sh}^g(x_j-y_k-i\varepsilon)} \\[6pt]
\times \lim_{\gamma_0 \to +\infty}\,\lim _{\varepsilon \to 0}\,
\frac{\gamma_0^{2(1-g)}\,e^{i(x_1+x_2-y_1-y_2)\gamma_0}}
{\prod_{k,j=1}^{2} (x_j-y_k-i\varepsilon)^g}
\end{align*}
where we used
\begin{align*}
\hat{q}(y-{\textstyle\frac{i\pi}{2}}+i\varepsilon,x) =
\frac{1}{\mathrm{ch}^g(y-x-\frac{i\pi}{2}+i\varepsilon)} =
\frac{e^{-i\frac{\pi}{2}g}}{\mathrm{sh}^g(x-y-i\varepsilon)}
\end{align*}
and the fact that function $z^{-1}\,\mathrm{sh}(z)$
is regular at the point $z=0$.

Now we have to use analog of the formulas \eqref{g} and \eqref{n=2}
(we postpone the proof to the next paper)
\begin{multline*}
\lim _{\gamma_0 \to +\infty}\,\lim _{\varepsilon \to 0}\,
\frac{\gamma_0^{2(1-g)}\,e^{i(x_1+x_2-y_1-y_2)\gamma_0}}
{\prod_{i,k=1,2}(x_i-y_k-i\varepsilon)^g} \\[6pt]
= \frac{(2\pi)^{2}}{\Gamma^2(g)}\,e^{2\pi i g}
\frac{1}{\left|x_{1}-x_{2}\right|^{2g}}\,
\Bigl[\delta(x_1-y_1)\,\delta(x_2-y_2)+
\delta(x_1-y_2)\,\delta(x_2-y_1)\Bigl].
\end{multline*}
Then we have
\begin{align*}
&\lim_{\varepsilon \to 0}\,
\prod_{k,j=1}^{2} \frac{(x_k-y_j-i\varepsilon)^g}
{\mathrm{sh}^g(x_k-y_j-i\varepsilon)}\,\delta(x_1-y_1)\,\delta(x_2-y_2)  \\[6pt]
&= \lim _{\varepsilon \to 0}\,
\frac{(x_1-y_1-i\varepsilon)^g}{\mathrm{sh}^g(x_1-y_1-i\varepsilon)}\,
\frac{(x_1-y_2-i\varepsilon)^g}{\mathrm{sh}^g(x_1-y_2-i\varepsilon)}\,
\frac{(x_2-y_1-i\varepsilon)^g}{\mathrm{sh}^g(x_2-y_1-i\varepsilon)}  \\[6pt]
&\times \frac{(x_2-y_2-i\varepsilon)^g}{\mathrm{sh}^g(x_2-y_2-i\varepsilon)}\,
\delta(x_1-y_1)\,\delta(x_2-y_2) \\[6pt]
& = \lim _{\varepsilon \to 0}\,
\frac{(-i\varepsilon)^g}{\mathrm{sh}^g(-i\varepsilon)}\,
\frac{(x_1-x_2-i\varepsilon)^g}{\mathrm{sh}^g(x_1-x_2-i\varepsilon)}\,
\frac{(x_2-x_1-i\varepsilon)^g}{\mathrm{sh}^g(x_2-x_1-i\varepsilon)}  \\[6pt]
& \times \frac{(-i\varepsilon)^g}{\mathrm{sh}^g(-i\varepsilon)}\,
\delta(x_1-y_1)\,\delta(x_2-y_2)  = \left(\frac{x_{12}}{\mathrm{sh}\, x_{12}}\right)^{2g}\,
\delta(x_1-y_1)\,\delta(x_2-y_2).
\end{align*}
The second contribution with $\delta(x_1-y_2)\,\delta(x_2-y_1)$
produces the same coefficient, so that collecting everything
together we obtain
\begin{align*}
\langle\Phi_{y_1, y_2}|\Phi_{x_1, x_2}\rangle & =
2\,\frac{\Gamma^2(g)}{(2\pi)^2}\,
\left(\frac{x_{12}}{\mathrm{sh}\, x_{12}}\right)^{2g}\,
\frac{(2\pi)^{2}}{\Gamma^2(g)}\,
\frac{1}{|x_{12}|^{2g}} \\[6pt]
& \times \Bigl[\delta(x_1-y_1)\,\delta(x_2-y_2)+
\delta(x_1-y_2)\,\delta(x_2-y_1)\Bigl] \\[6pt]
& = \frac{2}{\mathrm{sh}^{2g} |x_{12}|}\,
\Bigl[\delta(x_1-y_1)\,\delta(x_2-y_2)+
\delta(x_1-y_2)\,\delta(x_2-y_1)\Bigl].
\end{align*}

\subsubsection{Eigenfunctions of the operators $Q(\lambda|\vect{\omega})$ and $\QQ(x | \vect{\omega})$}

In this subsection we repeat the same calculation in the relativistic case. Due to the relation \eqref{dual} the transition to the dual $Q$-operator is reduced to change $g \to g^*$ and renaming of variables $x \rightleftarrows \lambda$. In this section we shall work with the eigenfunctions of the operator $Q_2(\lambda|\bo)$ \eqref{Q2bod}.

Expressions for the eigenfunction and conjugated eigenfunction have the form
\begin{align*}
&\Phi_{\rho_1,\rho_2}(x_1, x_2) =
\int\limits_{-\infty}^{+\infty} d t\,
e^{\frac{2\pi i}{\omega_1\omega_2}\rho_1\left(x_1+x_2-t\right)}\,
e^{\frac{2\pi i}{\omega_1\omega_2}\rho_2 t}
\,K_{g^*}(x_1-t)\,K_{g^*}(x_2-t), \\
&\overline{\Phi_{\lambda_1,\lambda_2}(x_1, x_2)} =
\int\limits_{-\infty}^{+\infty} d t\,
e^{-\frac{2\pi i}{\omega_1\omega_2}\lambda_1\left(x_1+x_2-t\right)}\,
e^{-\frac{2\pi i}{\omega_1\omega_2}\lambda_2 t}\,
K_{g^*}(x_1-t)\,K_{g^*}(x_2-t).
\end{align*}
The scalar product is
\begin{multline*}
\langle\Phi_{\lambda_1,\lambda_2}|\Phi_{\rho_1,\rho_2}\rangle =
\int\limits_{-\infty}^{+\infty}
d x_1 d x_2\, \mu_{g^*}(x_1,x_2)\,
\overline{\Phi_{\lambda_1,\lambda_2}(x_1, x_2)}\,\Phi_{\rho_1,\rho_2}(x_1, x_2) \\
= \int\limits_{-\infty}^{+\infty}
d x_1 d x_2\, \mu_{g^*}(x_1,x_2)\,\int\limits_{-\infty}^{+\infty} d t_1\,
\int\limits_{-\infty}^{+\infty} d t_2\,
e^{-\frac{2\pi i}{\omega_1\omega_2}\lambda_1\left(x_1+x_2-t_1\right)}\,
e^{-\frac{2\pi i}{\omega_1\omega_2}\lambda_2 t_1} \\
\times e^{\frac{2\pi i}{\omega_1\omega_2}
\rho_1\left(x_1+x_2-t_2\right)}\,
e^{\frac{2\pi i}{\omega_1\omega_2}\rho_2 t_2}\
\prod_{i,j=1}^{2} K_{g^*}(x_i-t_j)
\end{multline*}
where
\begin{align*}
\mu_{g^*}(x_1,x_2) = S(\pm ix_{12})S(\pm ix_{12}+{g^*}).
\end{align*}
The $x_1\,,x_2$-integrals diverge
\begin{align*}
\int\limits_{-\infty}^{+\infty}
d x_1 d x_2\,
e^{\frac{2\pi i}{\omega_1\omega_2}
(\rho_1-\lambda_1)\left(x_1+x_2\right)}\,
\mu_{g^*}(x_1\,,x_2)\,
\prod_{i,j=1}^{2} K_{g^*}(x_i-t_j)\,,
\end{align*}
so we regularize them adding external
point $t_0$ and small $\varepsilon >0 $ in exponent
\begin{multline*}
\int\limits_{-\infty}^{+\infty}
d x_1 d x_2\,
e^{\frac{2\pi i}{\omega_1\omega_2}(\rho_1-\lambda_1)
\left(x_1+x_2\right)}\,
\mu_{g^*}(x_1,x_2)\,
\prod_{i,j=1}^{2} K_{g^*}(x_i-t_j) \\
= \lim _{t_0 \to +\infty}\,
\lim _{\varepsilon \to 0}\,
\int\limits_{-\infty}^{+\infty}
d x_1 d x_2\,\mu_{g^*}(x_1,x_2) 
e^{-\frac{\pi g}{\omega_1\omega_2}\left(x_1-t_0\right)-
\frac{\pi g}{\omega_1\omega_2}\left(x_2-t_0\right)} \\ 
\times e^{\frac{2\pi}{\omega_1\omega_2}\varepsilon\left(x_1+x_2\right)}\,\,
e^{\frac{2\pi i}{\omega_1\omega_2}(\rho_1-\lambda_1)
\left(x_1+x_2\right)}\,
\prod_{i=1}^{2}\prod_{j=0}^{2} K_{g^*}(x_i-t_j)\,.
\end{multline*}
To prove that in the limit $t_0\to +\infty$ we reproduce
the initial integrand use the following asymptotic
\begin{align}
K_{g^*}(x_1-t_0)\,K_{g^*}(x_2-t_0) \to
e^{\frac{\pi g}{\omega_1\omega_2}\left(x_1-t_0\right)+
\frac{\pi g}{\omega_1\omega_2}\left(x_2-t_0\right)}, \qquad t_0 \rightarrow + \infty.
\end{align}
Next we  convert the $t_2$-integral back to the function $\Phi_{\rho_1,\rho_2}(x_1, x_2)$ and
rearrange exponents
\begin{multline*}
\langle\Phi_{\lambda_1,\lambda_2}|\Phi_{\rho_1,\rho_2}\rangle =
\lim _{t_0 \to +\infty}\,
\lim _{\varepsilon \to 0}\,
e^{-\frac{2\pi i}{\omega_1\omega_2}
(\lambda_1+\frac{i}{2}g+i\varepsilon)t_0}\,
\int\limits_{-\infty}^{+\infty} d t_1\,
e^{-i(\lambda_2-\frac{i}{2} g+i\varepsilon)t_1}\,\\
\int\limits_{-\infty}^{+\infty}
d x_1 d x_2\,
e^{\frac{2\pi i}{\omega_1\omega_2}(\lambda_1-\frac{i}{2} g+i\varepsilon)\left(t_1+t_0-x_1-x_2\right)}\,
\mu_{g^*}(x_1,x_2) \\
\times \prod_{i=1,2} K_{g^*}(x_i-t_1)K_{g^*}(x_i-t_0)\,\Phi_{\rho_1,\rho_2}(x_1, x_2).
\end{multline*}
In the last line the $x_1,x_2$-integral represents the action
of the $Q$-operator on the eigenfunction $\Phi_{\rho_1,\rho_2}(x_1, x_2)$
\begin{multline*}
[Q_2(\lambda_1-\textstyle\frac{i}{2}g+i\varepsilon|\bo)\Phi_{\rho_1,\rho_2}](t_1,t_0) \\[6pt]
= \int\limits_{-\infty}^{+\infty}
d x_1 d x_2\,
e^{\frac{2\pi i}{\omega_1\omega_2}(\lambda_1-\frac{i}{2} g+i\varepsilon)\left(t_1+t_0-x_1-x_2\right)}\,
\mu_{g^*}(x_1,x_2) \\[6pt]
\times \prod_{i=1,2} K_{g^*}(x_i-t_1)K_{g^*}(x_i-t_0)\Psi_{\rho_1,\rho_2}(x_1, x_2) \\[6pt]
= 2\,q(\lambda_1-\tfrac{i}{2} g+i\varepsilon,\rho_1|\bo)\,
q(\lambda_1-\tfrac{i}{2} g+i\varepsilon,\rho_2|\bo)\,
\Phi_{\rho_1,\rho_2}(t_1, t_0).
\end{multline*}
where
\begin{align}\label{eig}
q(\lambda,\rho|\bo) = \sqrt{\omega_1 \omega_2} S(g)
K_{g}(\lambda - \rho) = \frac{\sqrt{\omega_1\omega_2}\,S(g)}
{S\left(\frac{g}{2}+i(\lambda-\rho)\right)
S\left(\frac{g}{2}-i(\lambda-\rho)\right)}.
\end{align}
In fact we have managed to calculate $x_1,x_2$-integrals in explicit form and
it remains to calculate $t_1,t_2$-integrals
\begin{multline*}
e^{-\frac{2\pi i}{\omega_1\omega_2}(\lambda_1+\frac{i}{2} g+i\varepsilon)t_0}\,
\int\limits_{-\infty}^{+\infty} d t_1\,
e^{-\frac{2\pi i}{\omega_1\omega_2}(\lambda_2-\frac{i}{2} g+i\varepsilon)t_1}
\,\Phi_{\rho_1,\rho_2}(t_1, t_0) \\
= e^{-\frac{2\pi i}{\omega_1\omega_2}(\lambda_1+\frac{i}{2} g+i\varepsilon)t_0}\,
\int\limits_{-\infty}^{+\infty} d t_1\,
e^{-\frac{2\pi i}{\omega_1\omega_2}(\lambda_2-\frac{i}{2} g+i\varepsilon)t_1} \\
\times \int\limits_{-\infty}^{+\infty} d t_2\,
e^{\frac{2\pi i}{\omega_1\omega_2}\rho_1(t_1+t_0-t_2)}\,
K_{g^*}(t_1-t_2)\,K_{g^*}(t_0-t_2)\, e^{\frac{2\pi i}{\omega_1\omega_2}\rho_2 t_2}.
\end{multline*}
The $t_1$-integral represents the action
of $Q$-operator on eigenfunction
$\Phi_{\rho_1}(t_1) = e^{\frac{2\pi i}{\omega_1\omega_2}\rho_1 t_1}$
\begin{multline*}
[Q_1(\lambda_2-\textstyle\frac{i}{2} g+i\varepsilon|\bo)\Phi_{\rho_1}](t_2) \\
=\int\limits_{-\infty}^{+\infty}
d t_1\,
e^{\frac{2\pi i}{\omega_1\omega_2}(\lambda_2-\frac{i}{2} g +i\varepsilon)
\left(t_2-t_1\right)}\,
K_{g^*}(t_2-t_1)\,e^{\frac{2\pi i}{\omega_1\omega_2}\rho_1 t_1} \\
= q(\lambda_2-\textstyle\frac{i}{2} g+i\varepsilon,\rho_1|\bo)\,
e^{\frac{2\pi i}{\omega_1\omega_2}\rho_1 t_2}
\end{multline*}
After simple rearrangements of the exponents
\begin{align*}
e^{-\frac{2\pi i}{\omega_1\omega_2}(\lambda_1+\lambda_2-\rho_1+2 i \varepsilon)t_0}\,
\int\limits_{-\infty}^{+\infty} d t_2\,
e^{\frac{2\pi i}{\omega_1\omega_2}(\lambda_2-\frac{i}{2} g+i\varepsilon)(t_0-t_2)}\,
K_{g^*}(t_0-t_2)\, e^{\frac{2\pi i}{\omega_1\omega_2}\rho_2 t_2}
\end{align*}
we see that the $t_2$-integral also represents the action of the
$Q$-operator on eigenfunction
$\Phi_{\rho_2}(t_2) = e^{\frac{2\pi i}{\omega_1\omega_2}\rho_2 t_2}$
\begin{multline*}
[Q_1(\lambda_2-\textstyle\frac{i}{2} g+i\varepsilon|\bo)\Phi_{\rho_2}](t_0) \\
= \int\limits_{-\infty}^{+\infty} d t_2\,
e^{\frac{2\pi i}{\omega_1\omega_2}(\lambda_2-\frac{i}{2} g+i\varepsilon)(t_0-t_2)}\,
K_{g^*}(t_0-t_2)\, e^{\frac{2\pi i}{\omega_1\omega_2}\rho_2 t_2} \\
= q(\lambda_2-\textstyle\frac{i}{2} g+i\varepsilon,\rho_2|\bo)\,
e^{\frac{2\pi i}{\omega_1\omega_2}\rho_2 t_0}.
\end{multline*}
Collecting everything together and using exact
representation for the eigenvalue \eqref{eig}
we obtain the following expression
for the regularized scalar product
\begin{multline*}
\langle\Phi_{\lambda_1,\lambda_2}|\Phi_{\rho_1,\rho_2}\rangle =
2\,\lim _{t_0 \to +\infty}\,
\lim _{\varepsilon \to 0}\,
e^{-\frac{2\pi i}{\omega_1\omega_2}(\lambda_1+\lambda_2-\rho_1-\rho_2+2 i \varepsilon)t_0} \\
\times \prod_{k,j=1}^{2} q(\lambda_k-\textstyle\frac{i}{2} g^*+i\varepsilon,\rho_j|\bo),
\end{multline*}
or explicitly
\begin{align*}
&\langle\Phi_{\lambda_1,\lambda_2}|\Phi_{\rho_1,\rho_2}\rangle = \lim _{t_0 \to +\infty}\,\lim _{\varepsilon \to 0}\,
2\,e^{\frac{2\pi i}{\omega_1\omega_2}(\rho_1+\rho_2-\lambda_1-\lambda_2) t_0} \\
&  \hspace{2.5cm} \times \prod_{k,j=1,2}
\frac{\sqrt{\omega_1 \omega_2} S(g)}
{S\left(i(\lambda_k - \rho_j +i\varepsilon)+g\right)
S\left(i(\rho_j-\lambda_k - i\varepsilon)\right)}   \\[6pt]
& = 2\,\lim _{\varepsilon \to 0}\,
\prod_{k,j=1,2}
\frac{\sqrt{\omega_1 \omega_2} S(g)\,(\rho_j-\lambda_k - i\varepsilon)}
{S\left(i(\lambda_k - \rho_j +i\varepsilon)+g\right)
S\left(i(\rho_j-\lambda_k - i\varepsilon)\right)} \\
&  \hspace{2.5cm} \times \lim _{t_0 \to +\infty}\,\lim _{\varepsilon \to 0}\,
\frac{e^{\frac{2\pi i}{\omega_1\omega_2}(\rho_1+\rho_2-\lambda_1-\lambda_2)t_0}}
{\prod_{j,k=1,2}(\rho_j-\lambda_k - i\varepsilon)}.
\end{align*}
In the last line we used the fact that function $z^{-1}\,S_2(z)$
is regular at the point $z=0$ and extracted the singular part arising
at coinciding arguments $\rho_j =\lambda_k$.
Next we use formula \eqref{n=2}
\begin{multline*}
\lim_{t_0 \to +\infty}\,\lim _{\varepsilon \to 0}\,
\frac{e^{\frac{2\pi i}{\omega_1\omega_2}(\rho_1+\rho_2-\lambda_1-\lambda_2)t_0}}
{\prod_{j,k=1,2}(\rho_j-\lambda_k-i\varepsilon)} \\[6pt]
= \frac{(2\pi)^{2}}{\lambda_{12}^2}\,
\Bigl[\delta(\lambda_1-\rho_1)\,\delta(\lambda_2-\rho_2)+
\delta(\lambda_1-\rho_2)\,\delta(\lambda_2-\rho_1)\Bigr]
\end{multline*}
and calculate the contribution arising from the first term $\delta(\lambda_1-\rho_1)\,\delta(\lambda_2-\rho_2)$
\begin{multline*}
 \lim _{\varepsilon \to 0}\,
\prod_{k,j=1,2}
\frac{\sqrt{\omega_1 \omega_2} S(g)\,(\rho_j-\lambda_k - i\varepsilon)}
{S\left(i(\lambda_k - \rho_j +i\varepsilon)+g\right)
S\left(i(\rho_j-\lambda_k - i\varepsilon)\right)}\,
\delta(\lambda_1-\rho_1)\,\delta(\lambda_2-\rho_2) \\[6pt]
 = \frac{(\o_1 \o_2)^3 S^2(g)}{(2\pi)^2} \,\frac{\lambda^2_{12}}
{S\left(\pm i\lambda_{12}+g\right)
S\left(\pm i\lambda_{12}\right) }\,
\delta(\lambda_1-\rho_1)\,\delta(\lambda_2-\rho_2)
\end{multline*}
where we used the formula (see Appendix \ref{AppA})
\begin{align*}
\lim_{z\to 0} z^{-1}\,S(z) = \frac{2\pi}{\sqrt{\omega_1\omega_2}}.
\end{align*}
The second contribution with
$\delta(\lambda_1-\rho_2)\,\delta(\lambda_2-\rho_1)$ enters with
the same coefficient so that finally one obtains
\begin{multline*}
\langle\Phi_{\lambda_1,\lambda_2}|\Phi_{\rho_1,\rho_2}\rangle  =
\frac{2\,\left(\omega_1 \omega_2\right)^3 S^2(g)}
{S\left(\pm i\lambda_{12}+g\right)
S\left(\pm i\lambda_{12}\right)} \\[6pt]
\times \Bigl[\delta(\lambda_1-\rho_1)\,\delta(\lambda_2-\rho_2)+
\delta(\lambda_1-\rho_2)\,\delta(\lambda_2-\rho_1)\Bigl].
\end{multline*}
The scalar product for the eigenfunctions 
$\Phi_{x_1, x_2}(\l_1, \l_2| g, \bo)$  of the operator $\QQ(x|\bo)$ 
is obtained from the previous formula by change $g \rightleftarrows g^*$ and renaming variables
\begin{multline*}
\langle\Phi_{x_1,x_2}|\Phi_{y_1,y_2}\rangle  =
\frac{2\,\left(\omega_1 \omega_2\right)^3 S^2(g^*)}
{S\left(\pm i x_{12}+g^*\right)
S\left(\pm i x_{12}\right)} \\[6pt]
\times \Bigl[\delta(x_1-y_1)\,\delta(x_2-y_2)+
\delta(x_1-y_2)\,\delta(x_2-y_1)\Bigl].
\end{multline*}

\subsubsection{Orthogonality and completeness}

In previous sections we have established orthogonality 
relations for all sets of eigenfunctions. Due to the duality properties 
\begin{align*}
&\Psi_{\lambda_1,\lambda_2}(x_1, x_2) = \Phi_{x_1, x_2}(\lambda_1, \lambda_2),\\[6pt]
&\Phi_{x_1, x_2}(\lambda_1, \lambda_2|g,\bo) =
S^2(g^*)\,\Phi_{\lambda_1, \lambda_2}(x_1,x_2|g^*,\bo).
\end{align*}
the completeness relations of these eigenfunctions coincide with orthogonality relations for the dual ones.

Let us list the corresponding formulas in explicit form using the exact 
expressions for the measures.
We have orthogonality relation for eigenfunctions $\Psi_{\lambda_1,\lambda_2}(x_1, x_2)$ 
of the operator $Q_2(\lambda)$
\begin{multline*}
\int\limits_{-\infty}^{+\infty}
d x_1 d x_2\, \sh^{2g}|x_1-x_2|\,
\overline{\Psi_{\lambda_1,\lambda_2}(x_1, x_2)}\,
\Psi_{\rho_1,\rho_2}(x_1, x_2)  \\
= \frac{2^{2g +1} \pi^2}{\Gamma^2(g)}\,
\Gamma\left(g\pm\frac{i\lambda_{12}}{2}\right)
\Gamma\left(\pm\frac{i\lambda_{12}}{2}\right) \\[6pt]
\times \Bigl[\delta(\lambda_1-\rho_1)\,\delta(\lambda_2-\rho_2)+
\delta(\lambda_1-\rho_2)\,\delta(\lambda_2-\rho_1)\Bigl]
\end{multline*}
and due to relation $\Psi_{\lambda_1,\lambda_2}(x_1, x_2) = \Phi_{x_1, x_2}(\lambda_1, \lambda_2)$
it is possible to rewrite orthogonality relations for the 
eigenfunctions $\Phi_{x_1, x_2}(\lambda_1, \lambda_2)$ of the operator $\QQ(x)$ in terms of functions $\Psi_{\lambda_1,\lambda_2}(x_1, x_2)$. 
In this way one obtains the needed completeness relation  
\begin{multline*}
\int\limits_{-\infty}^{+\infty}
\frac{d \lambda_1}{2\pi} \frac{d \lambda_2}{2\pi}\, 
\frac{[2^{1-g}\Gamma(g)]^2}
{\Gamma\left(g\pm\frac{i(\lambda_1-\lambda_2)}{2}\right)
\Gamma\left(\pm\frac{i(\lambda_1-\lambda_2)}{2}\right)}\,
\overline{\Psi_{\lambda_1, \lambda_2}(y_1, y_2)}\,
\Psi_{\lambda_1, \lambda_2}(x_1, x_2) \\[6pt]
= \frac{2}{\mathrm{sh}^{2g}| x_{1} - x_2|}\,
\Bigl[\delta(x_1-y_1)\,\delta(x_2-y_2)+
\delta(x_1-y_2)\,\delta(x_2-y_1)\Bigl].
\end{multline*}
These relations are compatible with each other. To verify it integrate the first one over $\rho_1, \rho_2$ with the eigenfunction $\overline{\Psi_{\rho_1, \rho_2}(y_1,y_2)}$ and the measure $\mu(\rho_1, \rho_2)$ from the second relation. Then using the second relation together with eigenfunction symmetry with respect to variables $\rho_1, \rho_2$ and $y_1, y_2$ we obtain obvious statement $4\overline{\Psi_{\l_1, \l_2}(y_1,y_2)} = 4\overline{\Psi_{\l_1, \l_2}(y_1,y_2)}$. 

In relativistic case we have orthogonality relations for the eigenfunctions 
$\Phi_{\lambda_1, \lambda_2}(x_1, x_2) = \Phi_{\lambda_1, \lambda_2}(x_1, x_2|g^*\,,\bo)$ of the operator $Q(\lambda|\bo)$
\begin{multline*}
\int\limits_{-\infty}^{+\infty}
d x_1 d x_2\, S(\pm ix_{12})S(\pm ix_{12}+{g^*})\,
\overline{\Phi_{\lambda_1,\lambda_2}(x_1, x_2)}\,\Phi_{\rho_1,\rho_2}(x_1, x_2) =\\
\frac{2\,\left(\omega_1 \omega_2\right)^3 S^2(g)}
{S\left(\pm i\lambda_{12}+g\right)
S\left(\pm i\lambda_{12}\right)}
\Bigl[\delta(\lambda_1-\rho_1)\,\delta(\lambda_2-\rho_2)+
\delta(\lambda_1-\rho_2)\,\delta(\lambda_2-\rho_1)\Bigl].
\end{multline*}
The scalar product for the eigenfunctions
$\Phi_{x_1, x_2}(\l_1, \l_2| g, \bo)$  of the operator $\QQ(x|\bo)$
is obtained from the previous formula by change $g \rightleftarrows g^*$ and renaming variables
\begin{multline*}
\int\limits_{-\infty}^{+\infty}
d \lambda_1 d \lambda_2\, S(\pm i\lambda_{12})S(\pm i\lambda_{12}+{g})\,
\overline{\Phi_{x_1,x_2}(\lambda_1, \lambda_2)}\,
\Phi_{y_1,y_2}(\lambda_1, \lambda_2)  = \\
\frac{2\,\left(\omega_1 \omega_2\right)^3 S^2(g^*)}
{S\left(\pm i x_{12}+g^*\right)
S\left(\pm i x_{12}\right)}
\Bigl[\delta(x_1-y_1)\,\delta(x_2-y_2)+
\delta(x_1-y_2)\,\delta(x_2-y_1)\Bigl].
\end{multline*}
The relation $\Phi_{x_1, x_2}(\lambda_1, \lambda_2|g,\bo) =
S^2(g^*)\,\Phi_{\lambda_1, \lambda_2}(x_1,x_2|g^*,\bo)$ 
allows to rewrite this formula as completeness relation for eigenfunctions $\Phi_{\lambda_1, \lambda_2}(x_1, x_2)$
\begin{multline*}
\int\limits_{-\infty}^{+\infty}
d \lambda_1 d \lambda_2\, S(\pm i\lambda_{12})S(\pm i\lambda_{12}+{g})\,
\overline{\Phi_{\lambda_1 \lambda_2}(x_1\,,x_2)}\,
\Phi_{\lambda_1 \lambda_2}(y_1\,,y_2)  = \\
\frac{2\,\left(\omega_1 \omega_2\right)^3 S^2(g)}
{S\left(\pm i x_{12}+g^*\right)
S\left(\pm i x_{12}\right)}
\Bigl[\delta(x_1-y_1)\,\delta(x_2-y_2)+
\delta(x_1-y_2)\,\delta(x_2-y_1)\Bigl].
\end{multline*}

\section*{Acknowledgments}

The work of N. Belousov and S. Derkachov was supported by Russian Science Foundation, project No 23-11-00311, used for the proof of statements of Section 3. The work of S. Khoroshkin was supported by Russian Science Foundation, project No 23-11-00150, used for the proof of statements of Sections 2.1--2.3. The work of S. Kharchev was supported by Russian Science Foundation, project No 20-12-00195, used for the proof of statements of Section 2.4 and Appendices A, B, C. The authors also thank Euler International Mathematical Institute for hospitality during the PDMI and HSE joint workshop on quantum integrability, where they got a possibility to discuss many subtle points of this work.

\setcounter{equation}{0}

\section*{Appendix}
\appendix
\section{The double sine function} \label{AppA}
The  double sine  function $S(z):=S_2(z):=S_2(z|\bo)$, see \cite{Ku} and references therein, is a meromorphic function that satisfies two functional relations
\beq \notag \frac{S_2(z)}{S_2(z+\o_1)}=2\sin \frac{\pi z}{\o_2},\qquad \frac{S_2(z)}{S_2(z+\o_2)}=2\sin \frac{\pi z}{\o_1}
\eeq
and inversion relation
\beq \notag S_2(z)S_2(-z)=-4\sin\frac{\pi z}{\o_1}\sin\frac{\pi z}{\o_2},\eeq
or equivalently
\beq \notag S_2(z)S_2(\o_1+\o_2-z)=1. \eeq
The function $S_2(z)$ has poles at the points
\beq \notag
z = m \o_1 + k\o_2, \qquad m,k\geq 1
\eeq
and zeros at
\beq\notag
z=-m\o_1-k\o_2,\qquad m,k\geq 0.
\eeq
For $\o_1 / \o_2 \not\in \mathbb{Q}$ all poles and zeros are simple.
In the analytic region $ \Re z \in ( 0, \Re(\omega_1 + \omega_2) )$ we have the following integral representation for the logarithm of $S_2(z)$
\begin{equation*}\label{S2-int}
	\ln S_2 (z) = \int_0^\infty \frac{dt}{2t} \left( \frac{\sh \left[ (2z - \omega_1 - \omega_2)t \right]}{ \sh (\omega_1 t) \sh (\omega_2 t) } - \frac{ 2z - \omega_1 - \omega_2 }{ \omega_1 \omega_ 2 t } \right).
\end{equation*}
It is clear from this representation that the double sine function is homogeneous
\beq\notag
S_2( \gamma z | \gamma\o_1, \gamma \o_2 ) = S_2(z|\o_1, \o_2), \qquad \gamma \in (0, \infty)
\eeq
and invariant under permutation of periods
\beq\notag
S_2(z| \o_1, \o_2) = S_2(z | \o_2, \o_1).
\eeq
The double sine function can be expressed through the Barnes double Gamma function $\Gamma_2(z|\bo)$ \cite{B},
\beq \notag
S_2(z|\bo)=\Gamma_2(\o_1+\o_2-z|\bo)\Gamma_2^{-1}(z|\bo),
\eeq
and its properties follow from the corresponding properties of the double Gamma function.
It is also connected to the Ruijsenaars hyperbolic Gamma function $G(z|\bo)$ \cite{R1}
\beq \notag
G(z|\bo) = S_2\Bigl(\imath z + \frac{\o_1 + \o_2}{2} \,\Big|\, \bo \Bigr)
\eeq
and to the Faddeev quantum dilogarithm $\gamma(z|\bo)$ \cite{F}
\beq \notag
\gamma(z|\bo) = S_2\Bigl(-\imath z + \frac{\o_1+\o_2}{2}\, \Big|\, \bo\Bigr) \exp \Bigl( \frac{\imath \pi}{2\o_1 \o_2} \Bigl[z^2 + \frac{\o_1^2+\o_2^2}{12} \Bigr]\Bigr).
\eeq
Both $G(z|\bo)$ and $\gamma(z|\bo)$ were investigated independently.

\subsection{Asymptotics with respect to periods}

The asymptotics of double sine function as $\o_2\rightarrow\infty$ \cite[Prop III.6]{R1}
\begin{equation}
\label{as1}
S_2(u|\bo)=\sqrt{2\pi}
\left(\frac{2\pi\o_1}{\omega_2}\right)^{\frac{1}{2}-\frac{u}{\o_1}}
\Gamma^{-1}\Big(\frac{u}{\o_1}\Big)\Big\{1+O(\o_2^{-1})\Big\}.
\end{equation}
We also use another limiting formula:
\begin{equation*}
\lim_{\o_2\to\infty}\frac{S_2(\frac{\o_2}{\pi}u+h|\bo)}{S_2(\frac{\o_2}{\pi}u+g|\bo)}
=(2\sin u)^{\frac{g-h}{\o_1}}.
\end{equation*}
It agrees with \cite[Prop III.7]{R1}:
\begin{equation}
\label{as3}
\lim_{\o_2\to0}\frac{S_2(u+h\o_2|\bo)}{S_2(u+g\o_2|\bo)}=
\Big(2\sin\frac{\pi u}{\o_1}\Big)^{g-h}
\end{equation}
due to automorphy property
$S_2(u|\o_1,\o_2)=S_2(\l u|\l\o_1,\l\o_2)$, for $\l\in(0,\infty)$ one has
$$
S_2\Bigl(\frac{\o_2}{\pi}u+g\Big|\bo\Bigr)=S_2\Bigl(u+\frac{g}{\o_1}\cdot\frac{\pi\o_1}{\o_2}\Big|
\pi,\frac{\pi\o_1}{\o_2}\Bigr).
$$

\subsection{Beta integrals}

Here we show that integrals \eqref{beta0} and \eqref{beta00} can be obtained by reduction from the
higher level beta integral
\begin{align}\label{beta}
\int\limits_{-\infty}^{+\infty}
d z\, e^{\frac{2\pi i x z}{\omega_1\omega_2}}\,
\frac{S\left(\frac{\omega_1+\omega_2}{2}+iz+\frac{g}{2}\right)}
{S\left(\frac{\omega_1+\omega_2}{2}+iz-\frac{g}{2}\right)} =
\frac{\sqrt{\omega_1\omega_2}\,S(g)}{S\left(\frac{g}{2}+ix\right)
S\left(\frac{g}{2}-ix\right)}.
\end{align}

The first reduction: we substitute $g \to \omega_2 g$ and $x \to \omega_2 x$ and
then send $\omega_2 \to 0$ using in the left hand side the following
formula obtained from \eqref{as3}
\begin{align}\label{ch}
\frac{S\left(\frac{\omega_1+\omega_2}{2}+iz+\frac{g\omega_2}{2}\right)}
{S\left(\frac{\omega_1+\omega_2}{2}+iz-\frac{g\omega_2}{2}\right)} \to
\frac{1}{2^g\cos^g\left(\frac{i\pi z}{\omega_1}\right)}.
\end{align}
In the right hand side we need the following asymptotic for $\omega_2 \to 0$
\begin{align*}
S\left(x\omega_2\right) \to \frac{\sqrt{2\pi}\left(\frac{2\pi\omega_2}{\omega_1}\right)^{\frac{1}{2}-x}}
{\Gamma(x)}.
\end{align*}
To derive the last formula we use the following relation which can be get from integral representation of the double sine function(see \eqref{as1})
\begin{align}\label{Gamma}
\lim_{\omega_2\to \infty} \sqrt{2\pi}\left(\frac{2\pi\omega_1}{\omega_2}\right)^{\frac{1}{2} -\frac{x}{\omega_1}}\,S_2^{-1}(x|\omega_1,\omega_2) = \Gamma\left(\frac{z}{\omega_1}\right).
\end{align}
We use the modular property
\begin{align*}
S_2(x\omega_2|\omega_1,\omega_2) =
S_2\left(x|\omega_1/\omega_2,1\right) =
S_2\left(x|1,\omega_1/\omega_2\right)
\end{align*}
and after that we are ready to apply \eqref{Gamma} but now the role of $\omega_1$ in \eqref{Gamma} plays $\omega_1 \to 1$ and the role of $\omega_2$ in \eqref{Gamma} plays $\omega_2 \to \omega_1/\omega_2$.

Performing needed reductions \eqref{ch} in the left hand side and \eqref{Gamma}
in the right hand side of \eqref{beta} we obtain
\begin{multline*}
\int\limits_{-\infty}^{+\infty}
d z\, e^{\frac{2\pi i x z}{\omega_1}}\,
\frac{1}{2^g\cos^g\left(\frac{i\pi z}{\omega_1}\right)} \\[6pt]
= \frac{\sqrt{\omega_1\omega_2}\,
\frac{\sqrt{2\pi}\left(\frac{2\pi\omega_2}{\omega_1}\right)^{\frac{1}{2}-g}}
{\Gamma(g)}}
{\frac{\sqrt{2\pi}\left(\frac{2\pi\omega_2}{\omega_1}\right)^{\frac{1}{2}-\frac{g}{2}-ix}}
{\Gamma(\frac{g}{2}+ix)} \frac{\sqrt{2\pi}\left(\frac{2\pi\omega_2}{\omega_1}\right)^{\frac{1}{2}-\frac{g}{2}+ix}}
{\Gamma(\frac{g}{2}-ix)}}  = \frac{\omega_1}{2\pi}\frac{\Gamma(\frac{g}{2}-ix)\Gamma(\frac{g}{2}+ix)}{\Gamma(g)}.
\end{multline*}
Next we rescale $z \to \frac{\omega_1}{\pi} z$ and $x\to \frac{x}{2}$ and obtain
relation
\begin{align*}
\int\limits_{-\infty}^{+\infty}
d z\, e^{i x z}\,
\frac{1}{2^g\mathrm{ch}^g z} =
\frac{1}{2}\frac{\Gamma(\frac{g}{2}-\frac{ix}{2})
\Gamma(\frac{g}{2}+\frac{ix}{2})}{\Gamma(g)}
\end{align*}
which coincides exactly with \eqref{beta0}.

The second reduction: first of all we switch to the dual coupling constant
$g \to g^* =\omega_1+\omega_2-g$ in \eqref{beta}
\begin{align}\label{beta*}
\int\limits_{-\infty}^{+\infty}
d z\, e^{\frac{2\pi i x z}{\omega_1\omega_2}}\,
\frac{S\left(\omega_1+\omega_2+iz-\frac{g}{2}\right)}
{S\left(iz+\frac{g}{2}\right)} =
\frac{\sqrt{\omega_1\omega_2}\,S(\omega_1+\omega_2-g)}
{S\left(\frac{\omega_1+\omega_2}{2}-\frac{g}{2}+ix\right)
S\left(\frac{\omega_1+\omega_2}{2}-\frac{g}{2}-ix\right)}
\end{align}
then rescale $g \to g\omega_2$ and $z \to z\omega_2$ and
transform integrand in a following way
\begin{multline*}
\frac{S\left(\omega_1+\omega_2(1+iz-\frac{g}{2})\right)}
{S\left(\omega_2(iz+\frac{g}{2})\right)} =
\frac{1}{2\sin\pi(1+iz-\frac{g}{2})}\,
\frac{S\left(\omega_2(1+iz-\frac{g}{2})\right)}
{S\left(\omega_2(iz+\frac{g}{2})\right)} \\[8pt]
\xrightarrow{\omega_2 \to 0} \frac{1}{2\sin\pi(1+iz-\frac{g}{2})}\,\frac{\frac{\sqrt{2\pi}
\left(\frac{2\pi\omega_2}{\omega_1}\right)^{\frac{1}{2}-(1+iz-\frac{g}{2})}}
{\Gamma(1+iz-\frac{g}{2})}}
{\frac{\sqrt{2\pi}
\left(\frac{2\pi\omega_2}{\omega_1}\right)^{\frac{1}{2}-(iz+\frac{g}{2})}}
{\Gamma(iz+\frac{g}{2})}} \\[8pt]
= \frac{1}{2\sin\pi(1+iz-\frac{g}{2})}
\frac{\Gamma(iz+\frac{g}{2})}{\Gamma(1+iz-\frac{g}{2})}\,
\left(\frac{2\pi\omega_2}{\omega_1}\right)^{g-1} \\[8pt]
= \frac{1}{2\pi}
\Gamma\left(\frac{g}{2}+iz\right)\Gamma\left(\frac{g}{2}-iz\right)\,
\left(\frac{2\pi\omega_2}{\omega_1}\right)^{g-1}.
\end{multline*}
In the right hand side we have
\begin{multline*}
\frac{\sqrt{\omega_1\omega_2}\,S(\omega_1+\omega_2(1-g))}
{S\left(\frac{\omega_1+\omega_2}{2}-\frac{g\omega_2}{2}+ix\right)
S\left(\frac{\omega_1+\omega_2}{2}-\frac{g\omega_2}{2}-ix\right)} \\[8pt]
= \frac{\sqrt{\omega_1\omega_2}\,S(\omega_2(1-g))}{2\sin\pi(1-g)}
\frac{S\left(\frac{\omega_1+\omega_2}{2}+\frac{g\omega_2}{2}+ix\right)}
{S\left(\frac{\omega_1+\omega_2}{2}-\frac{g\omega_2}{2}+ix\right)} \\[8pt]
 \xrightarrow{\omega_2 \to 0} \frac{\sqrt{\omega_1\omega_2}\,
\frac{\sqrt{2\pi}\left(\frac{2\pi\omega_2}{\omega_1}\right)^{\frac{1}{2}-(1-g)}}
{\Gamma(1-g)}}{2\sin\pi(1-g)}
\frac{1}{2^g\cos^g(\frac{i\pi x}{\omega_1})} = \left(\frac{2\pi\omega_2}{\omega_1}\right)^{g-1}
\frac{\omega_2\,\Gamma(g)}{2^g\cos^g(\frac{i\pi x}{\omega_1})}.
\end{multline*}
Factor $\left(\frac{2\pi\omega_2}{\omega_1}\right)^{g-1}$ appears
in both sides of our relations and the same with
factor $\omega_2$ (it appears in integral due to the change
of variables $z \to z\omega_2$)
so that both factors can be cancelled and we obtain
\begin{align*}
\int\limits_{-\infty}^{+\infty}
d z\, e^{\frac{2\pi i x z}{\omega_1}}\,
\frac{1}{2\pi}
\Gamma\left(\frac{g}{2}+iz\right)\Gamma\left(\frac{g}{2}-iz\right) =
\frac{\Gamma(g)}{2^g\cos^g(\frac{i\pi x}{\omega_1})}.
\end{align*}
This relation is reduced to the \eqref{beta00} by $z \to \frac{z}{2}$ and $x \to \frac{\omega_1 x}{\pi}$.

Finally, we have checked that integral \eqref{beta} reproduced in appropriate limits all integral relations \eqref{beta0} and \eqref{beta00}.

\subsection{Asymptotics for large argument}\label{AppA-as2}
Let us consider asymptotic of needed functions for large arguments.
The key formula is
\begin{align*}
S(z|\bo) \to e^{\pm\frac{i\pi}{2} B_{2,2}(z|\bo)}
\end{align*}
for $\pm \Im(z)>0$ and $|z| \to \infty$.
The polynomial $B_{2,2}(z|\bo)$ is given by the formula
\begin{align*}
B_{2,2}(z|\bo) = \frac{z^2}{\omega_1\omega_2} -
\frac{\omega_1+\omega_2}{\omega_1\omega_2}\,z +
\frac{\omega^2_1+3\omega_1\omega_2+\omega^2_2}{6\omega_1\omega_2}.
\end{align*}
We have for $\lambda \to \pm \infty$
\begin{multline*}
K_g(\lambda) = \frac{S\left(\omega_1+\omega_2+i\lambda-\frac{g}{2}\right)}
{S\left(i\lambda+\frac{g}{2}\right)} \\[6pt]
\rightarrow
e^{\pm\frac{i\pi}{2}
\left( B_{2,2}(\omega_1+\omega_2+i\lambda-\frac{g}{2}|\omega_1,\omega_2) -
B_{2,2}(i\lambda+\frac{g}{2}|\omega_1,\omega_2)\right)} =
e^{\pm\frac{2\pi i}{\omega_1\omega_2}\,\lambda\,\frac{i g^{*}}{2}}
\end{multline*}
where $g^{*} = \omega_1+\omega_2-g$.

The leading asymptotic of $\KK (\gamma-\mu)$ for
$\mu \to +\infty$ has the following form
\begin{multline}\label{K^*as}
\KK (\gamma-\mu) =
\frac{\Gamma\left(\frac{g+i(\gamma-\mu)}{2}\right)
\Gamma\left(\frac{g-i(\gamma-\mu)}{2}\right)}{2^{1-g}\Gamma(g)} \\[6pt]
\rightarrow \frac{2\pi i \,e^{-\frac{i\pi g}{2}}}{2^{1-g}\Gamma(g)}\,
\left(\frac{i\mu}{2}\right)^{g-1}e^{\frac{\pi}{2}(\gamma-\mu)} =
\frac{2\pi}{\Gamma(g)}\,\mu^{g-1}\,e^{\frac{\pi}{2}(\gamma-\mu)}.
\end{multline}
To derive this asymptotic we start from useful textbook formula
\begin{align*}
\frac{\Gamma(\Lambda+a)}{\Gamma(\Lambda+b)} \xrightarrow{\Lambda \to \infty} \Lambda^{a-b}.
\end{align*}
To derive needed asymptotic we shall use reflection formula also
\begin{multline*}
\Gamma\left(\frac{i\mu}{2} +a\right)\Gamma\left(-\frac{i\mu}{2} +b\right) =
\frac{\Gamma(\frac{i\mu}{2} +a)}{\Gamma(\frac{i\mu}{2} + 1-b)}
\frac{\pi}{\sin\pi(b-\frac{i\mu}{2})} \\[6pt]
\rightarrow \left(\frac{i\mu}{2}\right)^{a+b-1} 2\pi i\, e^{-\frac{\mu \pi}{2}}e^{-i\pi b}.
\end{multline*}
In our case $a = \frac{g-i\lambda}{2}$ and $b=\frac{g-i\lambda}{2}$ so that we
obtain \eqref{K^*as}.

\section{Commutativity of $Q$-operators} \label{App-g=1}
\setcounter{equation}0

Let us formulate the relation of commutativity of $Q$-operators
as some integral relation which we have to prove.
The $Q$-operator is the integral operator
\begin{align*}
[Q(u)\Psi](x_1\ldots x_n) =
\int\limits_{-\infty}^{+\infty} \prod_{i=1}^n d t_i \prod_{i<k} \mathrm{sh}^{2g}(t_i-t_k)\,
\frac{e^{i u \sum_{i=1}^n \left(x_i-t_i\right)}}{\prod_{i,k=1}^n \mathrm{ch}^g(x_i-t_k)}\,
\Psi(t_1\ldots t_n)
\end{align*}
We shall rewrite all in new variables and the reason is very simple:
it seems that work with usual rational functions instead of hyperbolic
functions is simpler but of course it is just matter of habit.
We have for $x = \ln z_1$ and $y = \ln z_2$
\begin{align*}
\ch(x-y) = \frac{e^{x-y}+e^{y-x}}{2} = \frac{z^2_1+z^2_2}{2z_1z_2}, \quad
\sh(x-y) = \frac{e^{x-y}-e^{y-x}}{2} = \frac{z^2_1-z^2_2}{2z_1z_2}
\end{align*}
so that for rationalization of our integral we perform the change of variables
\begin{align*}
x_k = \ln\sqrt{z_k}, \quad t_k = \ln\sqrt{s_k}, \quad d t_k = \frac{d s_k}{2 s_k}
\end{align*}
and switch to new function $\Psi(\ln\sqrt{s_1}\ldots \ln\sqrt{s_n})  = \Phi(s_1\ldots s_n)$
\begin{multline*}
[Q(u)\Phi](z_1\ldots z_n) =
\int\limits_{0}^{+\infty} \prod_{k=1}^n \frac{d s_k}{2 s_k} \prod_{i<k}
\left(\frac{s_i-s_k}{2\sqrt{s_i s_k}}\right)^{2g} \\[6pt]
\times \prod_{i=1}^n\,\frac{z_i^{\frac{i u}{2}}}{s_i^{\frac{i u}{2}}} \prod_{i,k=1}^n
\frac{\left(2\sqrt{z_i s_k}\right)^g}{(z_i+s_k)^g}\,
\Phi(s_1\ldots s_n) \\[6pt]
= 2^{n(g-1)}\,\prod_{i=1}^n\,z_i^{\frac{i u+gn}{2}} \,
\int\limits_{0}^{+\infty} \prod_{k=1}^n d s_k\, s_k^{g-1-\frac{i u+ng}{2}}
\frac{\prod_{i<k} \left(s_i-s_k\right)^{2g}}
{\prod_{i,k=1}^n(z_i+s_k)^g}\,
\Phi(s_1\ldots s_n).
\end{multline*}
We are going to consider the commutativity relation $Q(u)Q(v) = Q(v)Q(u)$ as integral relation for the kernels of $Q$-operators. We have
\begin{multline*}
[Q(u)Q(v)\Phi](z_1\ldots z_n) =
2^{n(g-1)}\,\prod_{i=1}^n\,z_i^{\frac{iu+gn}{2}} \,
\int\limits_{0}^{+\infty} \prod_{k=1}^n d s_k\, s_k^{g-1-\frac{iu+ng}{2}} \\[6pt]
\times \frac{\prod_{i<k} \left(s_i-s_k\right)^{2g}}
{\prod_{i,k=1}^n(z_i+s_k)^g}\, 2^{n(g-1)}  \prod_{i=1}^n\,s_i^{\frac{iv+gn}{2}} \,
\int\limits_{0}^{+\infty} \prod_{k=1}^n d t_k\, t_k^{g-1-\frac{iv+ng}{2}} \\[6pt]
\times \frac{\prod_{i<k} \left(t_i-t_k\right)^{2g}}
{\prod_{i,k=1}^n(s_i+t_k)^g}\,
\Phi(t_1\ldots t_n)  \\[6pt]
= 2^{2n(g-1)}\,\prod_{i=1}^n\,z_i^{\frac{iu+gn}{2}} \,
\int\limits_{0}^{+\infty} \prod_{k=1}^n d s_k\, s_k^{g-1-i\frac{u-v}{2}}
\frac{\prod_{i<k} \left(s_i-s_k\right)^{2g}}
{\prod_{i,k=1}^n(z_i+s_k)^g(t_i+s_k)^g}\, \\[6pt]
\times \int\limits_{0}^{+\infty} \prod_{k=1}^n d t_k\, t_k^{g-1-\frac{iv+ng}{2}}
\prod_{i<k} \left(t_i-t_k\right)^{2g}\,
\Phi(t_1\ldots t_n)
\end{multline*}
so that the commutativity relation $Q(u)Q(v) = Q(v)Q(u)$ is equivalent
to the following integral relation
\begin{multline*}
\prod_{i=1}^n\,z_i^{i\frac{u-v}{2}} \,
\int\limits_{0}^{+\infty} \prod_{i=1}^n d s_k\, s_k^{g-1-i\frac{u-v}{2}}
\frac{\prod_{i<k} \left(s_i-s_k\right)^{2g}}
{\prod_{i,k=1}^n(z_i+s_k)^g(t_i+s_k)^g} \\
= \prod_{k=1}^n t_k^{i\frac{v-u}{2}}\,\int\limits_{0}^{+\infty} \prod_{i=1}^n d s_k\, s_k^{g-1-i\frac{v-u}{2}}
\frac{\prod_{i<k} \left(s_i-s_k\right)^{2g}}
{\prod_{i,k=1}^n(z_i+s_k)^g(t_i+s_k)^g}.
\end{multline*}
Let us use notation $\lambda = \frac{u-v}{2}$ for simplicity
so that main relation has the form
\begin{multline*}
\left(z_1\cdots z_n\right)^{i\lambda} \,
\int\limits_{0}^{+\infty} \prod_{i=1}^n d s_k\, s_k^{g-1-i\lambda}
\frac{\prod_{i<k} \left(s_i-s_k\right)^{2g}}
{\prod_{i,k=1}^n(z_i+s_k)^g(t_i+s_k)^g} \\
= \left(t_1\cdots t_n\right)^{-i\lambda}\,\int\limits_{0}^{+\infty} \prod_{i=1}^n d s_k\, s_k^{g-1+i\lambda}
\frac{\prod_{i<k} \left(s_i-s_k\right)^{2g}}
{\prod_{i,k=1}^n(z_i+s_k)^g(t_i+s_k)^g}.
\end{multline*}

\subsection{$n=1$ and arbitrary $g$}

Of course it is natural to start from the simplest  case $n=1$ and to check everything
\begin{align}\label{n=1}
z^{i\lambda}\,
\int\limits_{0}^{+\infty} d s\, s^{g-1-i\lambda}
\frac{1}
{(z+s)^g(t+s)^g} =
t^{-i\lambda}\,\int\limits_{0}^{+\infty} d s\, s^{g-1+i\lambda}
\frac{1}{(z+s)^g(t+s)^g}.
\end{align}
Here main steps are more or less evident -- we should use inversion $s \to \frac{1}{s}$ and
dilatation $s \to \frac{s}{tz}$
\begin{align*}
z^{i\lambda}\,
\int\limits_{0}^{+\infty} d s\, s^{g-1-i\lambda}
\frac{1}
{(z+s)^g(t+s)^g} = z^{i\lambda}\,
\int\limits_{0}^{+\infty} \frac{d s}{s^2}\, s^{-g+1+i\lambda}
\frac{s^{2g}}
{(sz+1)^g(st+1)^g} \\
= z^{i\lambda}\,
\int\limits_{0}^{+\infty} d s \, s^{g-1+i\lambda}
\frac{1}{(sz+1)^g(st+1)^g} =
z^{i\lambda}\,
(zt)^{-g-i\lambda}\int\limits_{0}^{+\infty} d s \, s^{g-1+i\lambda}
\frac{(zt)^g}{(s+t)^g(s+z)^g} \\
= t^{-i\lambda}\,\int\limits_{0}^{+\infty} d s\, s^{g-1+i\lambda}
\frac{1}{(z+s)^g(t+s)^g}.
\end{align*}

\subsection{$g=1$ and arbitrary $n$}

Now we are going to reformulate everything in some determinant form.
To do that we use Cauchy determinant identity
($z_{kj}=z_k-z_j$ and so on )
\begin{align*}
\frac{\prod_{k<j}z_{kj}\,s_{kj}}{\prod_{k,j=1}^{n}(z_i+s_k)}
= \det\left(\frac{1}{z_i+s_k}\right), \quad  \frac{\prod_{k<j}t_{kj}\,s_{kj}}{\prod_{k,j=1}^{n}(t_i+s_k)}
= \det\left(\frac{1}{t_i+s_k}\right)
\end{align*}
and rewrite the main relation in a very suggestive form
\begin{multline*}
\left(z_1\cdots z_n\right)^{i\lambda} \,
\int\limits_{0}^{+\infty} \prod_{i=1}^n d s_k\, s_k^{g-1-i\lambda}
\left(\det\left(\frac{1}{z_i+s_k}\right)
\det\left(\frac{1}{t_i+s_k}\right)\right)^g \\
= \left(t_1\cdots t_n\right)^{-i\lambda}\,\int\limits_{0}^{+\infty} \prod_{i=1}^n d s_k\, s_k^{g-1+i\lambda} \left(\det\left(\frac{1}{z_i+s_k}\right)\det\left(\frac{1}{t_i+s_k}\right)\right)^g.
\end{multline*}
In the case $g=1$ this reformulation in fact solves the problem.
Now it is possible to convert each $n$-fold integral to the determinant of
the matrix constructed from the one dimensional integrals
\begin{multline*}
\int\limits_{0}^{+\infty} \prod_{i=1}^n d s_k\, s_k^{-i\lambda}
\det\left(\frac{1}{z_i+s_k}\right)
\det\left(\frac{1}{t_i+s_k}\right) \\
= n!\,\det\left(\int\limits_{0}^{+\infty} d s\, s^{-i\lambda} \frac{1}{(z_i+s)(s+t_k)}\right)
\end{multline*}
and then everything is reduced to the $n=1$ case. Indeed we have
\begin{multline*}
\left(z_1\cdots z_n\right)^{i\lambda}\,
\int\limits_{0}^{+\infty} \prod_{i=1}^n d s_k\, s_k^{-i\lambda}
\det\left(\frac{1}{z_i+s_k}\right)
\det\left(\frac{1}{t_i+s_k}\right) \\
= n!\,\det\left(\int\limits_{0}^{+\infty} d s\, s^{-i\lambda}
\frac{z_i^{i\lambda}}{(z_i+s)(s+t_k)}\right) \\
= n!\,\det\left(\int\limits_{0}^{+\infty} d s\, s^{i\lambda}
\frac{t_k^{-\lambda}}{(z_i+s)(s+t_k)}\right) \\
= \left(t_1\cdots t_n\right)^{-i\lambda}\,
\int\limits_{0}^{+\infty} \prod_{i=1}^n d s_k\, s_k^{i\lambda}
\det\left(\frac{1}{z_i+s_k}\right)
\det\left(\frac{1}{t_i+s_k}\right)
\end{multline*}
where we used the identity \eqref{n=1} for one-dimensional integrals
(for $g=1$)
\begin{align*}
\int\limits_{0}^{+\infty} d s\, s^{-i\lambda} \frac{z_i^{i\lambda}}{(z_i+s)(s+t_k)} = \int\limits_{0}^{+\infty} d s\, s^{i\lambda} \frac{t_k^{-i\lambda}}{(z_i+s)(s+t_k)}.
\end{align*}

\setcounter{equation}{0}
\section{Delta-sequence}\label{App-delta}

We are going to show that in the sense of distributions the following identity holds
\begin{align}\label{id}
\lim_{\lambda\rightarrow\infty}\lim_{\epsilon\rightarrow 0^+}
\frac{e^{i \lambda \sum_{a=1}^{n}(x_a-y_a)}}{
\prod_{a,b=1}^{n} \left(x_a-y_b -i\varepsilon\right)}
= \frac{(-1)^{\frac{n(n-1)}{2}}(2\pi i )^{n} n!}{ \prod_{a<b}^{n} \left(x_{a}-x_{b}\right)^2}\,
\delta\big(\bm{x}_n,\bm{y}_n \big),
\end{align}
where
\begin{align*}
\delta\big(\bm{x}_n,\bm{y}_n \big) = \frac1{n!}\,
\sum_{w\in S_n} \prod_{k=1}^n\delta(x_k-y_{w(k)}).
\end{align*}
This identity is written in a compact formal way and should be understood
in the following sense:
for any test function $f(x_1,\ldots,x_n)$ we have
\begin{multline}\label{lem}
\lim_{\lambda\rightarrow\infty}\lim_{\epsilon\rightarrow 0^+}
\int dx_1\cdots dx_n \prod_{a<b}^{n} (x_{a}-x_{b})^2\,f(x_1,\ldots,x_n)\,
\frac{e^{i \lambda \sum_{a=1}^{n}(x_a-y_a)}}{
\prod_{a,b=1}^{n} \left(x_a-y_b -i\varepsilon\right)}
\\ = (-1)^{\frac{n(n-1)}{2}}(2\pi i )^{n}\,
\sum_{w\in S_n} f\big(y_{w(1)},\ldots,y_{w(n)}\big).
\end{multline}
First of all we are going to prove the equivalent identity
\begin{multline}\label{0}
\lim_{\lambda\rightarrow\infty}\lim_{\epsilon\rightarrow 0^+}
e^{i \lambda \sum_{a=1}^{n}(x_a-y_a)}\,\frac{
\prod_{a<b}^{n} \left(x_{a}-x_{b}\right)\left(y_{b}-y_{a}\right)}{
\prod_{a,b=1}^{n} \left(x_a-y_b -i\varepsilon\right)}
\\
= (2\pi i )^{n}\,
\sum_{w\in S_n} (-1)^{s(w)}\prod_{k=1}^n\,\delta(x_k-y_{w(k)})
\end{multline}
where $s(w)$ is the sign of the permutation $w$.
Let us start from the simplest example $n=1$. We have to prove that
\begin{align*}
\lim_{\lambda\rightarrow\infty}\lim_{\epsilon\rightarrow 0^+}
\frac{e^{i \lambda (x-y)}}{\left(x-y -i\varepsilon\right)}
= 2\pi i\,\delta(x - y ),
\end{align*}
or equivalently
\begin{align*}
\lim_{\lambda\rightarrow\infty}\lim_{\epsilon\rightarrow 0^+}
\int_{\mathbb{R}} f(x)\,
\frac{e^{i \lambda (x-y)}}{x-y -i\varepsilon}\,dx = 2\pi i\,f(y).
\end{align*}
First of all we transform integral with the test function.
We divide integral on two parts: the first
integral can be calculated by residues and due
to cancelation of singularity at $x=y$ it is possible to put
$\varepsilon \to 0 $ in the second part
\begin{multline*}
\int_{\mathbb{R}} f(x)\,
\frac{e^{i\lambda(x-y)}}{x-y -i\varepsilon}\,dx =
f(y)\,\int_{\mathbb{R}}\,
\frac{e^{i\lambda(x-y)}}{x-y-i\varepsilon}\,dx  +
\int_{\mathbb{R}} \frac{f(x)-f(y)}
{x-y-i\varepsilon}\,e^{i\lambda(x-y)}\,dx \\[6pt]
= 2\pi i\,f(y)\,e^{-\varepsilon\lambda}  +
\int_{\mathbb{R}} \frac{f(x)-f(y)}
{x-y-i\varepsilon}\,e^{i\lambda(x-y)}\,dx \\[6pt]
\xrightarrow{\varepsilon\to 0} 2\pi i\,f(y)  +
\int_{\mathbb{R}} \frac{f(x)-f(y)}
{x-y}\,L^{ix-iy}\,dx.
\end{multline*}
Due to the Riemann-Lebesgue lemma the second contribution
tends to zero in the limit $L\to\infty$ so that we obtain
after removing $\varepsilon$-regularization and $\lambda\to\infty$
\begin{align*}
\lim_{\lambda\rightarrow\infty}\lim_{\epsilon\rightarrow 0^+}
\int_{\mathbb{R}} f(x)\,
\frac{e^{i\lambda(x-y)}}{x-y -i\varepsilon}\,dx = 2\pi i\,f(y).
\end{align*}
The whole consideration in the case $n=2$ is almost identical
to the case of general $n$.
We have to prove the following relation
\begin{multline*}
\lim_{\lambda\rightarrow\infty}\lim_{\epsilon\rightarrow 0^+}
\frac{e^{i\lambda(x_1+x_2-y_1-y_2)}\,x_{12}\,y_{21}}{
\prod_{a,b=1}^{2} \left(x_a-y_b -i\varepsilon\right)} \\
= (2\pi i)^{2}\,
\bigl[\delta(x_1-y_1)\,\delta(x_2-y_2)-\delta(x_1-y_2)\,\delta(x_2-y_1)
\bigl].
\end{multline*}
First of all we use Cauchy determinant identity
\begin{multline}\label{12}
\frac{x_{12}\,y_{21}}{
\prod_{a,b=1}^{2} \left(x_a-y_b -i\varepsilon\right)} \\
= \frac{1}{\left(x_1-y_1 -i\varepsilon\right)
\left(x_2-y_2 -i\varepsilon\right)} -
\frac{1}{\left(x_1-y_2 -i\varepsilon\right)
\left(x_2-y_1 -i\varepsilon\right)}.
\end{multline}
Let us consider the convolution of the first term with
the test function
\begin{align*}
\int dx_1\,dx_2\,f(x_1,x_2)\,
\frac{e^{i\lambda(x_1+x_2-y_1-y_2)}}{\left(x_1-y_1 -i\varepsilon\right)
\left(x_2-y_2 -i\varepsilon\right)}
\end{align*}
and introduce two commuting operators $X_1$ and $X_2$ acting on the test function
\begin{align*}
X_1 f(x_1\,,x_2) = f(y_1\,,x_2), \quad
X_2 f(x_1\,,x_2) = f(x_1\,,y_2).
\end{align*}
As a consequence of evident identity
\begin{multline*}
1 = (1-X_1+X_1)(1-X_2+X_2) \\
= (1-X_1)(1-X_2) + X_1(1-X_2) + X_2(1-X_1) + X_1 X_2
\end{multline*}
and explicit formulas
\begin{align*}
X_1(1-X_2)f(x_1\,,x_2) &= f(y_1\,,x_2) - f(y_1\,,y_2)\,;\\
X_2(1-X_1)f(x_1\,,x_2) &= f(x_1\,,y_2) - f(y_1\,,y_2)\,;\\
(1-X_1)(1-X_2)f(x_1\,,x_2) &=
(1-X_1)\left[f(x_1\,,x_2) - f(x_1\,,y_2)\right]  \\
& \hspace{-1cm}= f(x_1\,,x_2) - f(x_1\,,y_2) - f(y_1\,,x_2) + f(y_1\,,y_2)
\end{align*}
we obtain the following useful representation for the function $f(x_1\,,x_2)$
\begin{align*}
f(x_1\,,x_2) = f(y_1\,,y_2) + \left[f(y_1\,,x_2) - f(y_1\,,y_2)\right] +
\left[f(x_1\,,y_2) - f(y_1\,,y_2)\right] \\[6pt]
+ \left[f(x_1\,,x_2) - f(x_1\,,y_2) - f(y_1\,,x_2) + f(y_1\,,y_2)\right].
\end{align*}
Note that the first term does not depend on $x_1$ and $x_2$,
second term does not depend on $x_1$ and is equal to zero at the point $x_2=y_2$,
third term does not depend on $x_2$ and is equal to zero at the point $x_1=y_1$.
The Taylor expansion of the last term in vicinity of the point $x_1=y_1\,,x_2=y_2$ started from the contribution $\sim(x_1-y_1)(x_2-y_2)$ because it turns to zero at points $x_1=y_1$ and $x_2=y_2$ independently. In the first three terms the corresponding integrals can be calculated by residues
and we obtain
\begin{align*}
&\int dx_1\,dx_2\,
\frac{f(x_1\,,x_2)\,e^{i\lambda(x_1+x_2-y_1-y_2)}}{\left(x_1-y_1 -i\varepsilon\right)
\left(x_2-y_2 -i\varepsilon\right)} \\[6pt]
&= f(y_1\,,y_2)\,(2\pi i)^2\,e^{-2\varepsilon\lambda} +
2\pi i\,e^{-\varepsilon\lambda}\,
\int dx_2\,
\frac{\left[f(y_1\,,x_2) - f(y_1\,,y_2)\right]\,e^{i\lambda(x_2-y_2)}}{x_2-y_2 -i\varepsilon}
\\[6pt]
&+ 2\pi i\,e^{-\varepsilon\lambda}\,
\int dx_1\,
\frac{\left[f(x_1\,,y_2) - f(y_1\,,y_2)\right]\,
e^{i\lambda(x_1-y_1)}}{x_1-y_1 -i\varepsilon} 
\\[6pt]
&+ \int dx_1\,dx_2\,
\frac{\left[f(x_1\,,x_2) - f(x_1\,,y_2) - f(y_1\,,x_2) + f(y_1\,,y_2)\right]\,e^{i\lambda(x_1+x_2-y_1-y_2)}}{\left(x_1-y_1 -i\varepsilon\right)
\left(x_2-y_2 -i\varepsilon\right)}.
\end{align*}
Inside of remaining integrals all singularities of integrand
are cancelled so that it is possible to perform the limit $\varepsilon \to 0$.
Due to the Riemann-Lebesgue lemma all contributions with integrals
tend to zero in the limit $\lambda\to\infty$ and we have
after removing $\varepsilon$-regularization and $\lambda\to\infty$
\begin{align*}
\lim_{\lambda\rightarrow\infty}\lim_{\epsilon\rightarrow 0^+}
\int dx_1\,dx_2\,
\frac{f(x_1\,,x_2)\,e^{i\lambda(x_1+x_2-y_1-y_2)}}{\left(x_1-y_1 -i\varepsilon\right)
\left(x_2-y_2 -i\varepsilon\right)} = (2\pi i)^2\,f(y_1\,,y_2).
\end{align*}
The second term in \eqref{12} is obtained by $y_1\rightleftarrows y_2$
so that finally one obtains the stated result
\begin{multline*}
\hspace{-0.3cm} \lim_{\lambda\rightarrow\infty}\lim_{\epsilon\rightarrow 0^+}
\int dx_1\,dx_2\,
\frac{f(x_1\,,x_2)\,e^{i\lambda(x_1+x_2-y_1-y_2)}x_{12}y_{21}}
{\left(x_1-y_1 -i\varepsilon\right)\left(x_1-y_2 -i\varepsilon\right)
\left(x_2-y_1 -i\varepsilon\right)
\left(x_2-y_2 -i\varepsilon\right)} \\[6pt]
= (2\pi i)^2\,\left[ f(y_1\,,y_2)- f(y_2\,,y_1) \right].
\end{multline*}
It is evident that the symmetric part of the function $f(x_1,x_2)$ does not contribute so that
the nontrivial contribution is due to antisymmetric part of the function $f(x_1,x_2)$.
Antisymmetric part of the test function $f(x_1,x_2)$ should be zero at $x_1=x_2$ and
without loss of generality it is possible to use representation
$f(x_1,x_2) = (x_1-x_2)\phi(x_1,x_2)$, where $\phi(x_1,x_2)$ can be generic because
antisymmetric part of the function $\phi(x_1,x_2)$ does not contribute.
Finally one obtains the formula \eqref{lem} in the case $n=2$
\begin{multline*}
\hspace{-0.3cm} \lim_{\lambda\rightarrow\infty}\lim_{\epsilon\rightarrow 0^+}
\int dx_1\,dx_2\,
\frac{x^2_{12}\,\phi(x_1\,,x_2)\,e^{i\lambda(x_1+x_2-y_1-y_2)}}
{\left(x_1-y_1 -i\varepsilon\right)\left(x_1-y_2 -i\varepsilon\right)
\left(x_2-y_1 -i\varepsilon\right)
\left(x_2-y_2 -i\varepsilon\right)} \\[6pt]
= (2\pi i)^2\,
\left[ \phi(y_1\,,y_2) + \phi(y_2\,,y_1) \right].
\end{multline*}
In general case we again use Cauchy determinant identity in the form
\begin{align*}
\frac{\prod_{k<j}x_{kj}\,y_{jk}}{\prod_{k,j=1}^{n}(x_k-y_j-i\varepsilon)} 
&= \det\left(\frac{1}{x_k-y_{j}-i\varepsilon}\right) \\
&= \sum_{\sigma\in S_{n}} (-1)^{s(\sigma)}
\prod_{k=1}^{n}\frac{1}{x_k-y_{\sigma(k)}-i\varepsilon}.
\end{align*}
In analogy with $n=2$ we shall prove that
\begin{align*}
\lim_{\lambda\rightarrow\infty}\lim_{\epsilon\rightarrow 0^+}
\int dx_1\,\cdots dx_n\,
\frac{f(x_1,\ldots ,x_n)\,e^{i\lambda\sum_k(x_k-y_k)}}{
\prod_k\left(x_k-y_k -i\varepsilon\right)} = (2\pi i)^n\,f(y_1,\ldots,y_n)
\end{align*}
and then use the same identity with evident permutations.
We introduce the natural generalization of the operators $X_k$
\begin{align*}
X_k f(x_1,\ldots ,x_k,\ldots ,x_n) = f(x_1,\ldots ,y_k,\ldots ,x_n)
\end{align*}
and the main expansion
\begin{multline*}
1 = \prod_{k=1}^N (1-X_k+X_k) = \prod_{k=1}^N (1-X_k) \\
+\sum_{k=1} X_k \prod_{i\neq k}^N (1-X_i)+
\sum_{k,p=1} X_k X_p \prod_{i\neq k,p}^N (1-X_i)+ \ldots + \prod_{k=1}^N X_k.
\end{multline*}
Due to the Riemann-Lebesgue lemma all contributions containing
$\prod_{i} (1-X_i)\,f(x_1,\ldots,x_n)$ in integrand are regular
at corresponding points so that corresponding integrals
tend to zero in the limit $L\to\infty$. In the needed limit only one term
$\prod_{k=1}^n X_k$ survives and produce $(2\pi i)^n\,f(y_1,\ldots,y_n)$ in the full anagoly with the case $n=2$.
Then for the whole sum we obtain
\begin{multline}\label{lem0}
\lim_{\lambda\rightarrow\infty}\lim_{\epsilon\rightarrow 0^+}
\int dx_1\cdots dx_n\, f(x_1,\ldots,x_n)\,
\frac{\prod_{k<j}x_{kj}\,y_{jk}\,e^{i\lambda\sum_{a=1}^{n}(x_a-y_a)}}{
\prod_{a,b=1}^{n} \left(x_a-y_b -i\varepsilon\right)}
\\
=  (-1)^{\frac{n(n-1)}{2}}(2\pi i )^{n}\,
\sum_{w\in S_n} (-1)^{s(w)}\,
f\big(y_{w(1)},\ldots,y_{w(n)}\big)
\end{multline}
and this identity is equivalent to \eqref{0}.
Next step is very similar to the case $n=2$.
Indeed, only the antisymmetric part of the test function
$f(x_1,\ldots,x_n)$ give nontrivial contribution so that without
loss of generality it is possible to use the following
representation for the test function
$f(x_1,\ldots,x_n) = \Delta(x_1,\ldots,x_n)\,\phi(x_1,\ldots,x_n)$,
where $\Delta(x_1,\ldots,x_n) = \prod_{k<j} x_{kj}$.
We have evident relation
\begin{align*}
\Delta\big(x_{w(1)},\ldots,x_{w(n)}\big) = (-1)^{s(w)}\,
\Delta\big(x_{1},\ldots,x_{n}\big)
\end{align*}
and as consequence one obtains \eqref{lem}
\begin{multline*}
\lim_{\lambda\rightarrow\infty}\lim_{\epsilon\rightarrow 0^+}
\int dx_1\cdots dx_n\, \phi(x_1,\ldots,x_n)\,
\frac{\prod_{k<j}x^2_{kj}\,e^{i\lambda\sum_{a=1}^{n}(x_a-y_a)}}{
\prod_{a,b=1}^{n} \left(x_a-y_b -i\varepsilon\right)}
\\
= (-1)^{\frac{n(n-1)}{2}}(2\pi i )^{n}\,
\sum_{w\in S_n} \,
\phi\big(y_{w(1)},\ldots,y_{w(n)}\big).
\end{multline*}

\end{document}